\documentclass[useAMS,usenatbib,usegraphicx]{mn2e}

			\usepackage{amsmath}
			\usepackage{amssymb}
		
			\usepackage{times}

			\usepackage{natbib}
			
			\usepackage[usenames,dvipsnames]{color}
		
		\newcommand{\Mc}[1]{#1} 
		\newcommand{\Mn}[1]{} 
			
		\newcommand{\Ac}[1]{#1} 
		\newcommand{\An}[1]{} 
		\newcommand{\Ann}[1]{} 

		\newcommand{\At}[1]{} 
			
		\newcommand{\Acc}[1]{#1} 

			\usepackage{url}	
		
			\newcommand{\chandra}{\textit{Chandra}}
			\newcommand{\xmm}{XMM-\textit{Newton}}
			
			\newcommand{\PoSp}{power spectrum}
			\newcommand{\PoSpa}{power spectra}

			\newcommand{\OHT}{one-halo term}
			\newcommand{\THT}{two-halo term}

			\newcommand{\rES}{resolved extended source}
			\newcommand{\rPS}{resolved point source}
			
			\newcommand{\eSoft}{\ensuremath{0.5-2.0\,\mathrm{keV}}}
			
			\newcommand{\eGal}{\ensuremath{0.5-0.7\,\mathrm{keV}}}
			\newcommand{\eExgal}{\ensuremath{1.0-2.0\,\mathrm{keV}}}
			\newcommand{\eFlare}{\ensuremath{2.3-7.3\,\mathrm{keV}}}
			\newcommand{\eFull}{\ensuremath{0.5-7.0\,\mathrm{keV}}}
			\newcommand{\ePart}{\ensuremath{9.5-12.0\,\mathrm{keV}}}
			
			\newcommand{\ePartB}{\ePart{} band}
			\newcommand{\eSoftB}{\eSoft{} band}

			\newcommand{\eExgalB}{\eExgal{} band}
			
			\newcommand{\Cref}[1]{Chapter~\ref{#1}}
			\newcommand{\Sref}[1]{Sect.~\ref{#1}}
			\newcommand{\Fref}[1]{Fig.~\ref{#1}}
			\newcommand{\Tref}[1]{Table~\ref{#1}}
			\newcommand{\Eref}[1]{Eq.~\ref{#1}}
			\newcommand{\Eqref}[1]{Eq.~\eqref{#1}}
			\newcommand{\Aref}[1]{App.~\ref{#1}}

			\newcommand{\VEC}[1]{\ensuremath{\mathbf{#1}}}
			\renewcommand{\d}{\ensuremath{\mathrm{d}}} 
			
			\newcommand{\InvArcSec}{\ensuremath{\,\mathrm{arcsec^{-1}}}}
			
			\newcommand{\LogNLogS}{\ensuremath{\log N - \log S}}
			
			\newcommand{\NH}{\ensuremath{N_\mathrm{H}}}
			
			\newcommand{\APEC}{\texttt{APEC}}

			\newcommand{\GCG}{clusters and groups of galaxies}
			
			\newcommand{\acisi}{\mbox{ACIS-I}}

\begin{document}

	
		\defcitealias{Kenter2005}{K05}
		\newcommand{\Ken}{\citetalias{Kenter2005}}
		
		\defcitealias{Hickox2006}{H06}
		\newcommand{\HickI}{\citetalias{Hickox2006}}

		\defcitealias{Hasinger2005}{H05}
		\newcommand{\HMS}{\citetalias{Hasinger2005}}
		
		\defcitealias{Kim2007}{K07}
		\newcommand{\KIM}{\citetalias{Kim2007}}
		
		\defcitealias{Georgakakis2008}{G08}
		\newcommand{\GEO}{\citetalias{Georgakakis2008}}
		
		\defcitealias{Aird2010}{A10}
		\newcommand{\AIRD}{\citetalias{Aird2010}}
		
		\defcitealias{Lehmer2012}{L12}
		\newcommand{\Leh}{\citetalias{Lehmer2012}}
		
		\defcitealias{eROSITA.SB}{SB}
		\newcommand{\SB}{\citetalias{eROSITA.SB}}
	

		\title[Can AGN and clusters explain the CXB fluctuations?]{
			Can AGN and galaxy clusters explain the surface brightness fluctuations of the cosmic X-ray background?
		}
		
		\author[A. Kolodzig et al.]{
			Alexander~Kolodzig$^{1,2}$,
			Marat~Gilfanov$^{2,3}$,
			Gert~H\"{u}tsi$^{2,4}$,
			Rashid~Sunyaev$^{2,3}$ \\
			$^1$Kavli Institute for Astronomy and Astrophysics (KIAA), Peking University, 100871 Beijing, China
			-- KIAA fellow, \url{http://kiaa.pku.edu.cn} \\
			$^2$Max-Planck-Institut f\"{u}r Astrophysik (MPA), Karl-Schwarzschild-Str. 1, D-85741 Garching, Germany \\
			$^3$Space Research Institute (IKI), Russian Academy of Sciences, Profsoyuznaya ul. 84/32, Moscow, 117997 Russia \\
			$^4$Tartu Observatory, T\~oravere 61602, Estonia \\
		}
		
		\date{Accepted 2O!6 Xxx XX. Received 2O!6 Xxx XX; in original form 2O!6 Xxx XX}
		
		\pagerange{\pageref{firstpage}--\pageref{lastpage}} \pubyear{2O!6}
		
		\maketitle
		\label{firstpage}

		\begin{abstract} 
			Fluctuations of the surface brightness of  cosmic X-ray background (CXB)  carry  unique information about  faint and low luminosity source populations, which is inaccessible for conventional large-scale structure (LSS) studies based on resolved sources.
			We  used \chandra{} data of the XBOOTES field ($\sim9$~deg$^2$) to conduct the most accurate measurement to date of the power spectrum of fluctuations of the unresolved CXB on the angular scales of $3\arcsec-17\arcmin$.
			We find that at sub-arcmin angular scales, the power spectrum is  
			consistent with the AGN shot noise,
			without much need for any significant contribution  from their one-halo term.
			This is consistent with the theoretical expectation that low-luminosity AGN reside alone in their dark matter halos.
			However, at larger angular scales  we detect a significant LSS signal above the AGN shot noise.
			Its power spectrum, obtained  after subtracting the AGN shot noise, follows a power law with
			the slope of $-0.8\pm0.1$
			and its amplitude is  much larger than what can be plausibly explained by the two-halo term of AGN.
			We demonstrate that the detected LSS signal is produced by unresolved clusters and groups of galaxies. For the flux limit of the XBOOTES survey, their flux-weighted mean redshift  equals $\left<z\right>\sim0.3$, and the mean  temperature of their intracluster medium (ICM), $\left<T\right>\approx 1.4$~keV, corresponds to the  mass of $M_{500}\sim 10^{13.5}~M_\odot$.  
			The power spectrum of CXB fluctuations carries information about the redshift distribution of these objects and  the spatial structure of their ICM on the linear scales of up to $\sim$Mpc, i.e. of the order of the virial radius. 
		\end{abstract}

		

	
	\section{Introduction} \label{sec:intro}
		
		Since the discovery of the cosmic X-ray background (CXB) about half a century ago \citep{Giacconi1962},
		understanding of its origin has been one of the major drivers for the development of X-ray astronomy
		and most X-ray space telescopes, such as the currently active missions: \xmm{}, \chandra{}, and NuSTAR \citep[e.g.][]{Fabian1992,Giacconi2013,Tanaka2013}.
		Thanks to the many, in particular deep X-ray surveys of \chandra{} \citep[e.g.][]{Brandt2005,Alexander2013,Brandt2015},
		we now know for certain that the CXB is dominated by extragalactic discrete sources, with Active Galactic Nuclei (AGN) leading the way \citep[e.g.][]{Comastri1995,Moretti2003,Hickox2006,Hickox2007,Gilli2007,Moretti2012,Lehmer2012}.
		This makes the CXB the prefect window to study the accretion history of the Universe up to high redshift ($z\sim5$) \citep[e.g.][]{Hasinger2005,Gilli2007,Aird2010,Ueda2014,Miyaji2015},
		which is an essential base to understand galaxy evolution \citep[e.g.][]{Hopkins2006,Hickox2009,Alexander2012}.
		
		Since the first X-ray surveys, angular correlation studies of the CXB had two major applications.
		They are used to disentangle the components of the CXB and at the same time to perform large-scale structure (LSS) studies \citep[e.g.][]{Scheuer1974,Hamilton1987,Shafer1983,Barcons1988,Soltan1994,Vikhlinin1995b,Miyaji2002}.
		The advantage  of such studies is that one can analyze the CXB beyond the survey sensitivity limit, since one does not require any source identification or/and redshift information.
		Thanks to these studies it has been long known that the CXB must be dominated by point-sources with a redshift distribution similar to optical QSOs but somewhat higher clustering strength.
		
		These results were confirmed in the last $\sim$two decades by very deep pencil beam surveys \citep[e.g.][]{Hickox2006,Hickox2007,Lehmer2012} and LSS studies with resolved samples of X-ray-selected AGN from wide but more shallow  surveys \citep[see reviews of][]{Cappelluti2012,Krumpe2013}.
		This became  possible thanks to the high-angular resolution of the current generation of X-ray telescopes, complemented with optical spectroscopic redshift surveys of sufficient size and depth.
		Due to this clustering measurements with resolved AGN developed in the last decade to an important branch of LSS studies in general. 
		It led to major advances in understanding how AGN activity is triggered and how does it depend on its environment, such as the host galaxy and dark matter halo (DMH) properties, and how do supermassive black holes (SMBH) grow and co-evolve with their DMH over cosmic time, which are essential questions in the field of galaxy evolution \citep[e.g.][]{Cappelluti2012,Krumpe2013}.
		In the future, it will become  possible to use AGN as a cosmological probe via baryon acoustic oscillation measurements \citep[for details see e.g.][]{Kolodzig2013,Huetsi2013} with the $\sim3$~million AGN to be detected in the upcoming SRG/eROSITA all-sky survey \citep[for details see][]{eROSITA,eROSITA.SB,Kolodzig2012}.
		
		Due the focus on resolved AGN, the current knowledge of AGN clustering properties and its implications for AGN and galaxy evolution are biased towards objects  of $L_{\eSoft}>10^{42}\,\mathrm{erg\;s^{-1}}$, in particular for higher redshifts ($z>0.5$), due to the luminosity cut from the AGN identification process and the signal-to-noise ratio (S/N) cut for the spectroscopic redshift \citep[e.g.][]{Allevato2011,Allevato2012,Allevato2014,Krumpe2010,Krumpe2012,Miyaji2011,Krumpe2015}.
		An important question to ask is if we are able to extrapolate these clustering properties to less luminous AGN, which trace galaxies at an earlier evolutionary stage with a less massive SMBH and/or smaller accretion rate than luminous AGN? 
		A significant step towards answering this question is to study the surface brightness fluctuations of the unresolved CXB measured with the current generation of X-ray telescopes, which allows us to measure angular fluctuations on small scales down to the arc-second regime.
		
		This type of clustering measurement offers us a great window to the small-scale clustering regime ($<1\;\mathrm{Mpc\,h^{-1}}$).
		Clustering studies of spatially resolved AGN samples have difficulties to access this regime,
		because of the low spatial density of AGN in general, and because multiobject spectroscopy surveys are typically limited to an angular separation of $\sim1\arcmin$ \citep[e.g.][]{Blanton2003,Dawson2013}.
		Therefore, our best spatially resolved measurement of the small-scale clustering regime comes from the use of dedicated catalogs of close AGN pairs \citep[e.g. SDSS Quasar Lens Search,][]{Kayo2012} or the direct measurement of the halo occupation distribution (HOD) of AGN from galaxy groups \citep[e.g.][]{Allevato2012}.
		Both types of measurement require an extensive amount of multi-wavelength survey data.
		In terms of standard clustering studies, the best results come from studies of optically-selected AGN thanks to the sufficient size of the available survey data \citep[e.g.][]{Hennawi2006,Kayo2012,Richardson2012,Shen2012}.
		For X-ray-selected AGN, the situation is more difficult due to the so far rather limited survey data \citep[e.g.][]{Allevato2012,Richardson2013}.
		Here, non-spatially resolved studies, such as the brightness fluctuations of the unresolved CXB, may offer a true alternative for small-scale clustering measurements, which have not been fully utilized yet.
		
		Due to their scientific focus, the only two existing  studies of the brightness fluctuations of the unresolved CXB at these angular scales used very deep surveys  \citep[e.g.][]{Cappelluti2013,Helgason2014}.
		However, this also implies a very small sky coverage of these surveys ($\sim0.1\,\mathrm{deg^2}$).
		
		In our study, we aim to conduct the most accurate measurement to date of the brightness fluctuations of the unresolved CXB on angular scales below $\sim17\arcmin$.
		We are able to achieve this by using the XBOOTES survey \citep[hereafter \Ken{}]{Murray2005,Kenter2005}, the currently largest available continuous \chandra{} survey, with a surface area of $\sim9\,\mathrm{deg^2}$.
		The advantage in comparison to previous studies is that a higher S/N makes any comparison with current clustering models from known source populations much more meaningful
		and it enables us to do clustering measurements in an energy resolved manner in order to separate different source populations.
		In this first study we present our measurement of the brightness fluctuations of the unresolved CXB with angular scales up to $\sim17\arcmin$, and make novel tests for systematic uncertainties such as the brightness fluctuations of the instrumental background. %

		Covering the angular scales from the arc-second to arc-minute regime may allow us to study the clustering properties of AGN within the same DMH (one-halo-term) and AGN of different DMHs (two-halo-term) of low-luminosity AGN ($L_{\eSoft}<10^{42}\,\mathrm{erg\;s^{-1}}$) and redshifts of $z>0.5$. 
		This  parameter regime is inaccessible for conventional  clustering studies of the resolved CXB with current X-ray surveys \citep[e.g.][]{Cappelluti2012}. 
		
		Diffuse emission from the intracluster medium (ICM) of \GCG{} and the associated warm-hot intergalactic medium (WHIM) also contributes to the CXB \citep[e.g.][]{Rosati2002,Hickox2007,Kravtsov2012,Roncarelli2012}.
		Since \GCG{} are more difficult to detect and an order of magnitude more sparse than AGN, our knowledge about their population,  in particular at low fluxes ($\lesssim10^{-16}\,\mathrm{erg\,cm^{-2}\,s^{-1}}$) is less certain \citep[e.g.][]{Finoguenov2007,Finoguenov2010,Finoguenov2015,Clerc2012,Boehringer2014}.
		Thanks to cosmological hydrodynamical simulations \citep[e.g.][]{Roncarelli2006,Roncarelli2007,Roncarelli2012,Ursino2011,Ursino2014} and analytical studies \citep[e.g.][]{Diego2003,Cheng2004} we have nevertheless some reasonable understanding of their clustering properties. 
		\Mc{As we will demonstrate in this paper,  angular correlation studies of  CXB fluctuations  can help to dramatically improve the situation from the observational side.} 
		
		This paper is organized as following:
		In \Sref{sec:DataProc} we explain our data processing procedure, 
		in \Sref{sec:Espec} we show the energy spectrum of the unresolved CXB of XBOOTES and estimate the contribution by different components of the CXB,
		in \Sref{sec:Ana} we present our measurement of the surface brightness fluctuations of the unresolved CXB,
		and in \Sref{sec:Excess} we \Ac{study the origin of the detected LSS signal at large angular scales.}		
		In the Appendixes  we present results of tests \Mc{for various systematic effects and investigate the impact} of the instrumental background on our measurements.

		For the work we assume a flat $\Lambda$CDM cosmology with the following parameters:  
		$H_0 = 70\,\mathrm{km\,s^{-1}\,Mpc^{-1}}$ ($h=0.70$),
		$\Omega_\mathrm{m} = 0.30$ ($\Omega_\Lambda = 0.70$),
		$\Omega_\mathrm{b} = 0.05$, 
		$\sigma_8=0.8$.
		The values for $H_0$ and $\Omega_\mathrm{m}$ were chosen to match the values assumed in the X-ray luminosity function studies, which we use in our calculations (e.g.\  \Sref{ss:UnresoAGN} or \ref{ss:AGNmodel}),
		and $\Omega_\mathrm{b}$ and $\sigma_8$ are derived from the cosmic microwave background (CMB) study of WMAP\footnote{\url{http://map.gsfc.nasa.gov}} \citep{Komatsu2011}.
		We note that the results of this work are not very sensitive to the exact values of the cosmological parameters and if we used the recently published, more precise cosmological parameters of the CMB study by PLANCK\footnote{\url{http://www.cosmos.esa.int/web/planck}} \citep{Planck_CosPara} our results would not change.

	\section{Data preparation and processing} \label{sec:DataProc}
		
		For our analysis of the surface brightness fluctuations of the unresolved CXB we are using the XBOOTES survey \citep[\Ken]{Murray2005}, which is currently the largest available continuous \chandra{} survey with a surface area of $\sim9\,\mathrm{deg^2}$.
		It consists of 126 individual, contiguous \chandra{} \acisi{} observations.
		In order to avoid unnecessary complication in our analysis, we exclude eight of them.
		The six observations with the ObsIDs 3601, 3607, 3617, 3625, 3641 \& 3657 are excluded because they all show much higher background count rate than the average.
		The observations with the ObsIDs 4228 and 4224 are excluded because they contain a very bright point- and extended source, respectively.
		Therefore, when referring to the ``XBOOTES survey'', we mean from now on the 118 remaining observations \citep[for a full list of observations see Table~1 of][]{Murray2005}.
		
		The average exposure time of an XBOOTES observation is $\sim5$~ksec and the combined exposure time of 118 observations is almost $0.6$~Msec.
		By excluding $8$ observations the surface area of XBOOTES reduces from originally $\sim9.3\,\mathrm{deg^2}$ to $\sim8.7\,\mathrm{deg^2}$.
 		Those values will further decrease after processing the observations. 
	
		For processing the observations of the XBOOTES survey we are using \chandra{}'s data analysis system CIAO \citep[v4.7, CALDB v4.6.9,][]{CIAO} and follow their standard analysis threads, unless stated otherwise.
		Since the observations were performed in the very faint mode (\texttt{\uppercase{vfaint}}),
		we are able to make use of CIAO's most strict filtering method%
			\footnote{
				For details see \url{http://cxc.harvard.edu/cal/Acis/Cal_prods/vfbkgrnd}.
				We activate it in the data processing script \texttt{chandra\_repro} with \texttt{check\_vf\_pha = yes}.
			}
		of background events in \acisi{} data.

		Unless otherwise stated, we use throughout the paper for the Galactic absorption a hydrogen column density of $\NH=10^{20}\mathrm{cm^{-2}}$ as determined for the XBOOTES survey and we convert the flux of extragalactic sources between different energy bands and between physical and instrumental units assuming an absorbed powerlaw with a photon-index of $\Gamma=1.70$ (\Ken{}, Sect.3.3).

				\begin{figure}
					\resizebox{\hsize}{!}{\includegraphics{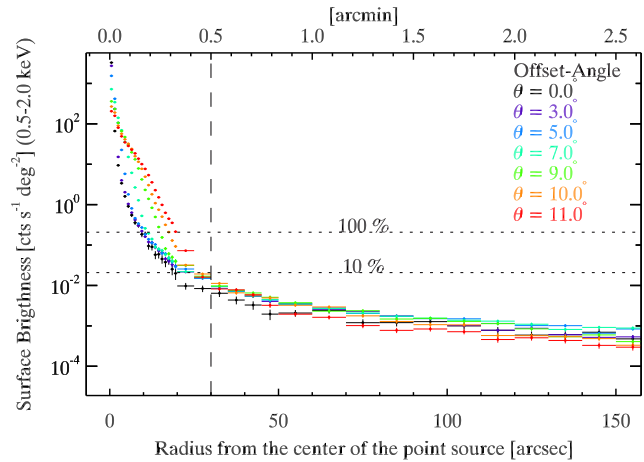}}  
					\caption{\label{fig:PSF_shape}%
						Shape of the PSF for \chandra{} \acisi{} (averaged over 4 azimuthal angles) for different offset angles ($\theta$) for the source flux of $0.63\times10^{-14}\,\mathrm{erg\,cm^{-2}\,s^{-1}\,deg^{-2}}$ (\eSoft{}).
						The vertical dashed line shows the radius of the circular exclusion area for this flux group (\Tref{tab:PS_Area}).
						The horizontal dotted lines show levels corresponding to $100\,\%$ and $10\,\%$ of the surface brightness of unresolved AGN (\Sref{ss:Espec_Exgal}).
					}
				\end{figure}

		\subsection{Exposure map and mask} \label{ss:MSK}
			
			For the following data processing and analysis, we will need the exposure map $\mathbf{E}$ [seconds] and mask $\mathbf{M}$, which we describe here.
			
			We use the exposure map $\mathbf{E}$ to convert our count maps $\mathbf{C}$ [counts] into flux maps $\mathbf{F}$ [cts s$^{-1}$] but also to take the vignetting into account. 
			For creating the exposure map $\mathbf{E}$ we are using the spectral model and best-fit parameters of our spectral fit of the unresolved CXB (\Tref{tab:CXB_spec}).
			Note that the exposure map $\mathbf{E}$ is not very sensitive to the choice of the spectral model in the \eSoftB{}, because we only compute the exposure map in units of seconds and the vignetting is not very energy depended in this energy range.
			
			We use the mask $\mathbf{M}$ to excluded certain regions of \chandra{} maps from the analysis.
			It is set to be large enough ($2900\times2900$~pixels) to contain the entire \acisi{} FOV of an observation.
			Pixels of the mask are set to zero,
			when they are outside the FOV,
			when they are within the exclusion area of a resolved source (\Sref{ss:ResoScr}),
			when they have zero exposure time, which takes also bad pixels into account,
			or when the exposure time falls below $63\,\%$ of the peak value of $\mathbf{E}$.
			The latter targets low exposed pixels, which are predominantly located in the CCD gaps and edges of the \acisi{} and occur due to the dithering movement of \chandra{} during an observation.
			The threshold was chosen to be $63\,\%$, because we see a clear break of the pixel distribution of $\mathbf{E}$ around this value.
			The average field-of-view (FOV) solid angle of one observation after this filter step but before removing resolved sources is $\sim0.07\,\mathrm{deg^2}$.
			The solid angle of the mask is computed as 
				\begin{align} \label{eq:Omega}
					\Omega = (\Delta p)^2 \, \left(\Sigma_{i,j} M_{i,j}\right) \text{ ,}					
				\end{align}
			whereby $\Delta p$ is the size of a pixel. 
			Since for our analysis we use an image pixel binning of one, the size of a pixel%
				\footnote{\url{http://cxc.harvard.edu/proposer/POG/html/chap6.html\#tab:acis_char}}
			is $\Delta p = 0.492\arcsec$.

			Whenever we convert counts to count-rate, where we can not use the exposure map $\mathbf{E}$, we are using the average value of this map:
				\begin{align} \label{eq:AveETM}
					\langle E \rangle =\dfrac{\Sigma_{i,j} E_{i,j} \, M_{i,j}}{\Sigma_{i,j} M_{i,j}} \text{ ,}					
				\end{align}
			This for instance is the case for the energy spectrum in \Sref{sec:Espec} or for energy bands above $9\,\mathrm{keV}$, where the effective area of \chandra{} becomes neglectable.
		
		\subsection{Removing resolved sources} \label{ss:ResoScr}

				\begin{figure}
					\resizebox{\hsize}{!}{\includegraphics{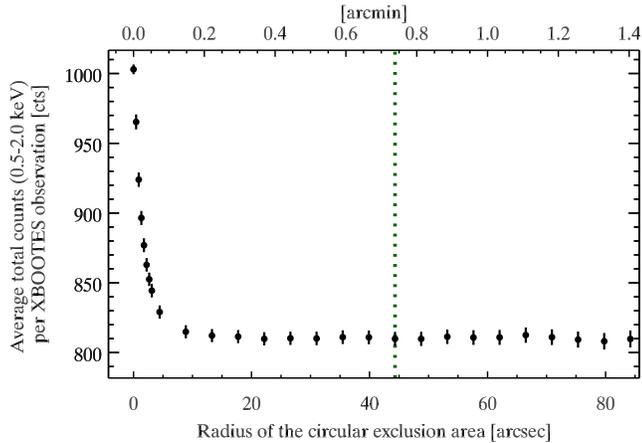}}   
					\caption{\label{fig:PS_radius}%
						Average total counts (\eSoft) per XBOOTES observation (instrumental background not removed) as a function of radius of the circular exclusion area of point sources.
						We change the radius for each flux group between $1\,\%$ to $200\,\%$ of its value in \Tref{tab:PS_Area}.
						The displayed radius is a weighted average based on the number of point sources per flux group.
						Vertical dotted line: Average radius for our definition in \Tref{tab:PS_Area}.
					}
				\end{figure}
			
			
				\begin{table*}
					\renewcommand\arraystretch{1.3}		
					\renewcommand\tabcolsep{10pt}
					\caption{Radius of the circular exclusion area of \rPS{}s in XBOOTES for different flux groups.} 
					\label{tab:PS_Area}
					\begin{center}     
						\begin{tabular}{c c c c c c c}	
							\hline
							\hline
							Flux  			 & \# of	& (b)	& Radius	& Depth$^{(\mathrm{c})}$ & ECF$^{(\mathrm{d})}$ & (e) \\
							Groups$^{(\mathrm{a})}$  & Sources	& [\%] 	&  [arcsec] 	& [\%] 			 & [\%] & [cts]  \\
							\hline
							$[0.47,0.63[$		& 1673		& $\sim51$	&  30	& $\sim10$		& $\sim98.4$		& $\sim0.06$ \\
							$[0.63,2.10[$		& 1328		& $\sim40$	&  55	& $\sim10$		& $\sim98.7$		& $\sim0.16$ \\
							$[2.10,9.00[$		&  268		& $\sim8$	&  80	& $\sim25$		& $\sim99.0$		& $\sim0.55$ \\
							$[9.0,47.0[$ 		&   23		& $\sim1$	& 140	& $\sim50$		& $\sim99.4$		& $\sim1.76$ \\
							\hline
						\end{tabular}
					\end{center}
					\begin{flushleft}
					(a) In $10^{-14}$ erg cm$^{-2}$ s$^{-1}$ (\eSoft).  \\
					(b) Number of sources in fraction of the total number of sources (3293). \\
					(c) The PSF  surface brightness in fraction of the surface brightness of unresolved AGN (\Sref{ss:Espec_Exgal}) for the brightest sources in the given flux group at the edge of the circular exclusion area. \\
					(d) Enclosed count fraction (based on the integration of our simulated PSF, averaged over all azimuthal and offset angles). \\
					(e) Residual counts per point-source outside the exclusion area for the brightest sources in the given flux group. Computed with the corresponding ECF and an exposure time of $4.3$~ksec (average value for XBOOTES, \Sref{ss:Flares}).
					\end{flushleft}
				\end{table*}

			In order to study the unresolved CXB, we need to remove the resolved (point-like and extended) sources to such a level that the residual counts of resolved sources contribute only insignificantly to the surface brightness of the unresolved CXB.
			For this purpose 
			we are using the two source catalogs of \Ken{} for point and extended sources.

			\subsubsection{Point sources} \label{ss:PointScr}
			
			The point-source catalog of \Ken{} includes 3293 sources with at least 4~counts in the \eFull{} band.
			The sky coverage of the XBOOTES survey as a function of the point-source detection sensitivity is shown in Fig.~12 of \Ken{} (for the \eFull{} band).
			The  average flux limit of the survey  in the \eSoftB{} is $\sim2.3\times10^{-15}\,\mathrm{erg\,cm^{-2}\,s^{-1}}$.
			It is defined as the flux level, which gives the same resolved CXB fraction as computed with the sky coverage vs. sensitivity distribution of the survey  (e.g.\  \Eref{eq:SurfBright}).
			To estimate the appropriate size of a circular exclusion area of a point source, we simulated the point-spread-function (PSF) shape for 13 offset angles ($\theta=0\arcmin-12\arcmin$, in $1\arcmin$-steps)
			and for four equally distributed azimuthal angles (from the aimpoint roughly along the diagonal of each CCD)
			with the \chandra{} Ray Tracer simulator%
				\footnote{\url{http://cxc.harvard.edu/chart}} 
			\citep{ChaRT} and the MARX software package%
				\footnote{\url{http://space.mit.edu/ASC/MARX}}
			(v5.0.0), as shown in \Fref{fig:PSF_shape} for the average over all azimuthal angles.
			Based on these simulations we define the circular exclusion area as presented in \Tref{tab:PS_Area}, where we split the point sources into different flux groups in order to make source removal more efficient.
			The shape of the PSF does not depend on the flux of a point source but the normalization of the PSF does.
			Hence, for each flux group the PSF normalization is defined by the upper-limit of its flux-interval (first column of \Tref{tab:PS_Area}).
			For the first two flux groups, which represent about $90\,\%$ of all point sources, the radius of the exclusion area is chosen in the way that both groups are removed at a same \emph{depth}.
			We quantify this depth with the surface brightness of the PSF at the edge of the exclusion area,
			which corresponds for both groups to $\sim10\,\%$ of the surface brightness of unresolved AGN (\Sref{ss:Espec_Exgal}).
			For the other two flux groups we make a compromise in the depth in order to keep the radius of the exclusion area in a reasonable regime.
			In our simulations we further find that for radii of $\gtrsim20\arcsec$ the PSF shape does not change very significantly as a function of azimuthal and offset angle (\Fref{fig:PSF_shape}).
			Therefore, we use in the following a PSF averaged over all four azimuthal angles and all offset angles. 
			With this we compute the enclosed count fraction (ECF) and residual counts of a point source for each flux-group, which are displayed as well in \Tref{tab:PS_Area}.
			
			In order to test that our definition of the exclusion area of point sources in \Tref{tab:PS_Area} removes sufficiently well the counts of resolved sources,
			we estimated how the average total counts per XBOOTES observation change as a function of radius of the exclusion area (\Fref{fig:PS_radius}).
			We create a list of evaluation steps, where we set the radius from $1\,\%$ to $200\,\%$ of its value in \Tref{tab:PS_Area} in each flux group.
			For presentation purposes (\Fref{fig:PS_radius}), we compute an average radius per evaluation step of all flux groups, where we weight the radius of each flux group by the corresponding number of point sources.
			The measured number of total counts per observation is normalized for each evaluation step to the surface area of the observation before sources were removed.
			To ensure a clean test without any bias due to our choice of removing the extended sources (\Sref{ss:ExteScr}), we take here only those observations into account (83 out of 118 observations), which do not contain extended sources.
			
			The result of this test is shown in \Fref{fig:PS_radius}.
			We can see for radii of $\gtrsim20\arcsec$ that the total counts do not change significantly.
			The rise in total counts for $\lesssim20\arcsec$ indicates that there is still a significant contamination by counts from \rPS{}s at these apertures.
			The average radius of our definition of the exclusion area in \Tref{tab:PS_Area} is $\approx44\arcsec$ (weighted with number of point sources per flux group) and is shown as vertical dotted line in  \Fref{fig:PS_radius}. 
			This figure demonstrates that our definition of the exclusion  regions for point sources  is rather conservative. 
			In average the FOV is reduce by $\sim17\,\%$ after removing all \rPS{}s with our definition.
			
			To estimate contamination by the residual source counts, we note the following.
			With our definition of the point-source exclusion area, the ECF averaged over all resolved point sources equals to  $\sim98.6\,\%$,
			i.e. about $\approx 1.4\%$ of the point source counts in the wings of the PSF  remain in the image.
			Using observations without resolved extended sources (83 of 118) we compute the average number of counts in the exclusion regions,
			$\approx 181$~counts per observation in the \eSoftB{}. 
			Therefore, there is about $\approx 2.6$ residual counts per image left from the resolved point sources.
			This should be compared with the total number of counts in the unresolved emission, $\approx601$
			per image, of which $\approx208$ are from unresolved CXB and $\approx393$ are due to the instrumental background.
			Thus, residual counts from resolved point sources constitute about $\approx 1\%$ of the total unresolved CXB counts,
			i.e. their contamination can be neglected.

			\subsubsection{Extended sources} \label{ss:ExteScr}
			There are 43 extended sources detected  with a detection limit of $\approx1\times10^{-14}\,\mathrm{erg\,cm^{-2}\,s^{-1}}$ (\eSoft{}) (\Ken, Sect.~3.2. \& Table~1).
			The extended sources in the XBOOTES catalog were fitted with a Gaussian model in order to estimate their size.
			We define the radius of the circular exclusion area as six times this size. 
			We tested circular exclusion areas between four and eight times the size and did not find any significant difference in the remaining source counts after we normalize to the same surface area.
			Therefore, we believe that this is a reasonable definition.
			We also note that the total source counts of the \rES{}s only accounts $\sim4\,\%$ to the total source counts of all resolved sources, based on the source catalogs of XBOOTES.

			\subsubsection{Summary}
			
			After removing all resolved sources the average FOV area is reduced by $\sim18\%$ down to $\sim0.0610\,\mathrm{deg^2}$.
			The average surface brightness is reduced by $\sim43\%$ from $1.42\pm0.01$ to $0.81\pm0.01\,\mathrm{cts\,s^{-1}deg^{-2}}$ in the \eSoftB{} (after removing the instrumental background, see \Sref{ss:Flares} and \ref{ss:QuiBKG}).

		\subsection{Removing background flares} \label{ss:Flares}
		
			In order to detect and remove time intervals of an observation, which are contaminated by background flares, we adopt the main concept of \citet[hereafter \HickI{}]{Hickox2006} and adjust them to the XBOOTES data.
			We analyze the light curve of each observation in the energy-band \eFlare{}.
			\HickI{} show that this band is the best choice for background flare detection, because of the different energy-spectra of background flares and the quiescent background (see their Fig.~3).
			
			Our de-flaring method consists of three consecutive steps of filtering the light curve:
			
			\textbf{(a)} We run a $3\sigma$-clipping with the CIAO tool \texttt{deflare}, which is a standard procedure and removes the most obvious flares.
			Hereby, we use bins of $\sim63$~sec (10 frames), which is large enough to assume a Gaussian error distribution in each time bin but small enough to not conceal short, strong flares.
			
			\textbf{(b)} We create a light curve with a binning of $\sim252$~sec (80 frames) and remove all bins, which are $30\,\%$ above the mean count rate of the $3\sigma$-clipped light curve from step (a).
			This step targets weaker and longer lasting flares with a maximum duration of the order of the bin size.
			In comparison to \HickI{}, we only remove positive deviations from the mean. 
			
			\textbf{(c)} We compute a light curve in bins of $\sim252$~sec (80 frames) of the ratio between the \eFlare{} and \ePartB{} and remove all bins, which are $40\,\%$ above the mean ratio of all considered XBOOTES observations.
			This method was introduced by \HickI{} and is best suited for weak flares. 
			It takes advantage of the fact that for a typical flare the flux-ratio of \eFlare{} to \ePartB{} will be larger than for the normal instrumental background (alias quiescent background) due to the different energy-spectrum shapes.
			We use the same threshold for all observations to ensure a constant energy-spectrum shape for all of them.

			The major difference between \HickI{} and our filtering arises due to the fact that our observations have exposure times of the order of kiloseconds, whereas \HickI{} use observations with more than one Megasecond.
			This leads in our case to much smaller bin sizes for the light curves and less restrictive thresholds for removing flare events for step (b) and (c).
			The light curves of all observations were visually inspected and the thresholds of (b) and (c) were tuned to removed any obvious feature of the light curve, which could be interpret as a background flare.
			
			For a typical observation, our de-flaring method removes on average $\sim190$~sec ($\sim4\,\%$) of the exposure time.
			After the de-flaring we have an average exposure time per observation of $\sim4.3$~ksec and a total exposure time is reduced to $\sim0.50$~Msec. We note that de-flaring does not significantly affect the \PoSp{} of the unresolved CXB,
			but it is necessary for accurate measurement of the absolute CXB flux (\Sref{sec:Espec}).

		\subsection{Instrumental background and background-subtracted map} \label{ss:QuiBKG}
		
			We estimate the contribution of the instrumental background with the method presented in \HickI{}.
			They show in their study with the \chandra{}'s \acisi{} stowed background data%
				\footnote{\url{http://cxc.harvard.edu/contrib/maxim/acisbg/}}
			that the shape of the energy spectrum of the instrumental background of \acisi{} from different observations is very stable over the course of five years, which includes the time when the XBOOTES observations were performed.
			Further, we know that all detected photons in the \ePartB{} are due to the instrumental background because the effective area of \chandra{} in this energy range is neglectable.
			With those two facts combined we can estimate the \emph{instrumental-background map} $\mathbf{C}^\mathrm{BKG}$ for an observation with the \emph{total-count map} $\mathbf{C}^\mathrm{Total}$ in the energy band $X$ by scaling the \acisi{} stowed-background map $\mathbf{C}^\mathrm{Stow}$ as following:
				\begin{align} \label{eq:BKG}
					\mathbf{C}_X^\mathrm{BKG} = \mathbf{M} \cdot \mathbf{C}_X^\mathrm{Stow} \cdot \left(\dfrac{\Sigma_{i,j} \, C_{\ePart}^\mathrm{Total} \, M }{\Sigma_{i,j} \, C_{\ePart}^\mathrm{Stow} \, M }\right)
					\text{ .}	
				\end{align}
			With this method we estimate an average background surface brightness of $1.55\pm0.01\,\mathrm{cts\,s^{-1}deg^{-2}}$ in the \eSoftB{}, which is consistent with the value from the spectral fit (see also \Tref{tab:PB_BkgSB}).
			This means that $\sim65\,\%$ of the total surface brightness of $2.37\pm0.01\,\mathrm{cts\,s^{-1}deg^{-2}}$ (after removing resolved sources, \Sref{ss:ResoScr}) is due to the instrumental background.
			
			The \emph{background-subtracted map} is then 
				\begin{align} \label{eq:NetCtsMap}
					\mathbf{C}_X^\mathrm{CXB} = \mathbf{C}_X^\mathrm{Total} \, \mathbf{M} - \mathbf{C}_X^\mathrm{BKG} \text{ .}	
				\end{align}
			We estimate from the background-subtracted map (after removing resolved sources) the average surface brightness of  the the unresolved CXB equal to $0.81\pm0.01\,\mathrm{cts\,s^{-1}deg^{-2}}$, which  in physical units corresponds to $7.9\pm0.1\,\times10^{-12}\,\mathrm{erg\,cm^{-2}\,s^{-1}\,deg^{-2}}$, using our spectral model of the unresolved CXB from \Sref{sec:Espec}.
			This value is consistent with $7.8\pm0.1\,\times10^{-12}\,\mathrm{erg\,cm^{-2}\,s^{-1}\,deg^{-2}}$ obtained from the  spectral fit  in \Sref{ss:Espec_Scr}.
			
			We show in \Aref{app:ss:Bkg} that in the \eSoftB{}, surface brightness fluctuations of the instrumental background are much smaller than fluctuations of unresolved CXB.
			Therefore subtraction of the instrumental background is unnecessary for the calculation of the \PoSp{} of CXB fluctuations.
			Accordingly, it is not performed in \Sref{sec:Ana} where total-count maps ($\mathbf{C}^\mathrm{Total}$) are used for construction of the \PoSpa{}.
			However, accurate account for the instrumental background is necessary for  computing the  CXB flux and its spectral analysis.

				\begin{figure*}
					\resizebox{15.5cm}{!}{\includegraphics[angle=270]{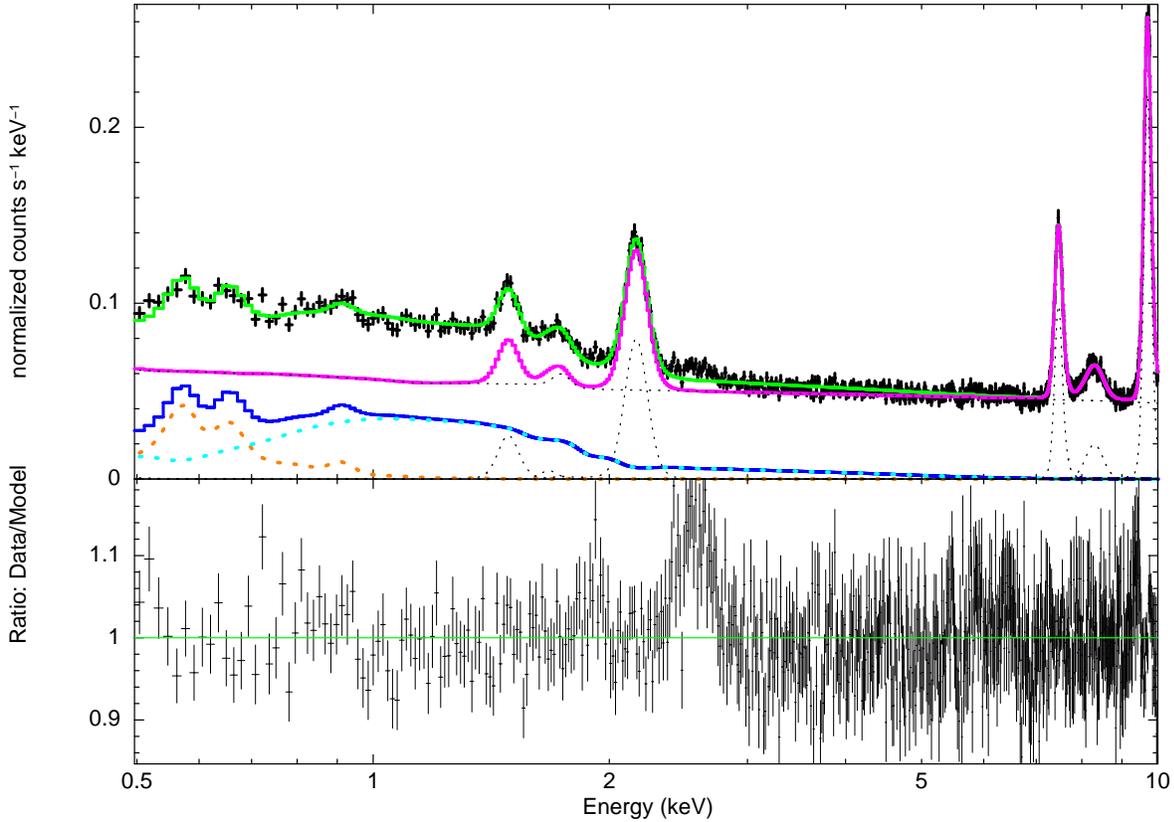}}   
					\caption{\label{fig:CXB_spec_full}%
						The 0.5--10 keV stacked energy spectrum of 118 XBOOTES observations after removing resolved sources (black crosses, \Sref{ss:ResoScr}).
						Note that for our brightness fluctuation analysis in \Sref{sec:Ana} we only use the \eSoftB{}.
						The lower panel shows the ratio of model and data.
						Green solid curve: Total model (CXB and background model).
						Dark blue solid curve: Total CXB model (\Tref{tab:CXB_spec}).
						Light blue dotted curve: absorbed powerlaw model (\texttt{phabs(powerlaw)}), representing the contribution of extragalactic emission.
						Orange dotted curve: \APEC{}, representing the contribution of Galactic emission.
						Pink solid curve: Instrumental background model (\Aref{app:SpecPB}).
						Black dotted curves: components of the instrumental background model.
					}	
				\end{figure*}

		\subsection{Flux and fluctuation maps} \label{ss:FlucMaps}
		
			The \emph{flux map} $\mathbf{F}_X$ of the energy band $X$ is computed as the ratio between the count and exposure map (for each pixel):
				\begin{align} \label{eq:F_x}
					\mathbf{F}_X= \mathbf{M} \cdot \dfrac{ \mathbf{C}_X}{\mathbf{E}} \text{ ,}	
				\end{align}	
			and is only computed  in instrumental units [cts s$^{-1}$].
			The average flux map $\langle \mathbf{F} \rangle$ is defined as: 
			\begin{align} \label{eq:MeanFlx}
				\langle \mathbf{F}_X \rangle = \mathbf{M} \cdot \left(\dfrac{\Sigma_{i,j} \, C_X \, M }{\Sigma_{i,j} \, E \, M }\right) \text{ .}
			\end{align}
			These definitions minimizes statistical errors, while taking the vignetting of the exposure map properly into account.
			Note, that in order to compute the average count map $\langle \mathbf{C} \rangle$, which also treats vignetting properly, one has to multiply $\langle \mathbf{F} \rangle$ with the exposure map:
			\begin{align} \label{eq:C_x} 
				\langle \mathbf{C}_X \rangle = \langle \mathbf{F}_X \rangle \cdot \mathbf{E} \text{ .}
			\end{align}
			
			For our analysis in \Sref{sec:Ana} we need the \emph{fluctuation map} $\delta\mathbf{F}$ in different energy bands for each observation. 
			We compute this map for an energy band $X$ as following:
				\begin{align} \label{eq:FlucMap}
					\delta\mathbf{F}_X =  \mathbf{F}_X - \left\langle \mathbf{F}_X \right\rangle \text{ .}
				\end{align}

%
	\section{Composition of the unresolved CXB} \label{sec:Espec}

	\subsection{Energy spectrum}
		
		The unresolved CXB%
			\footnote{We note that the term ``CXB'' is used ambiguously in the literature and some may use it exclusively for extragalactic emission.}
		consists of two components: the Galactic and extragalactic emission.
		We will use spectral analysis to separate their contributions to the CXB. 
		We create the energy spectrum by stacking the energy spectra of all 118 considered XBOOTES observations, after removing resolved sources (\Sref{ss:ResoScr}) and background flares (\Sref{ss:Flares}).
		The stacked energy spectrum has a total exposure time of $\sim0.50$~Msec and is based on a total surface area of $\sim7.2\,\mathrm{deg^2}$ (without taking overlaps into account).
		We fit%
			\footnote{
				With the X-Ray spectral fitting package XSPEC \citep[v12.8.2,][]{XSPEC}. 
			}
		it in the energy range of $0.5-10.0\,\mathrm{keV}$ with a model for the unresolved CXB (\Sref{ss:Espec_Scr}, blue curve in \Fref{fig:CXB_spec_full}) and an instrumental background model (\Aref{app:SpecPB}, pink curve in \Fref{fig:CXB_spec_full}).

			\subsubsection{Spectral model} \label{ss:Espec_Scr} 

				
				\begin{table*}
					\caption{Best-fit parameters of our spectral model (\texttt{APEC + phabs(powerlaw)}, \Sref{ss:Espec_Scr}) of the unresolved CXB of XBOOTES (\Fref{fig:CXB_spec_full}).}
					\label{tab:CXB_spec}
					\begin{center}     
						\begin{tabular}{l l c}	
							\hline
							\hline
							Model Component  & Parameter & Value  \\
							\hline
							\APEC{}		& Temperature ($T$)	& $0.164\pm0.003\;\mathrm{keV}$ \\
									& Normalization		& $1.9\pm0.1\;\times10^{-4}\;\mathrm{cm^{-5}}$ \\
									& Surface Brightness$^{(\mathrm{a})}$	& $3.2\pm0.1\;\times10^{-12}$ \\
							\hline
							\texttt{phabs(powerlaw)} & \NH{} (fixed) & $10^{20}\;\mathrm{cm^{-2}}$ \\
									& Photon Index ($\Gamma$)	& $1.74\pm0.03$ \\
									& Normalization$^{(\mathrm{b})}$		& $1.23\pm0.02\;\times10^{-4}$ \\
									& Surface Brightness$^{(\mathrm{a})}$	& $4.6\pm0.1\;\times10^{-12}$ \\
							\hline
						\end{tabular}
					\end{center}
					(a) $\mathrm{erg\,cm^{-2}\,s^{-1}\,deg^{-2}}$ (\eSoft);
					(b) $\mathrm{photons\,keV^{-1}\,cm^{-2}\,s^{-1}}$ at $1\;\mathrm{keV}$.
				\end{table*}

			Our spectral model for the unresolved CXB consists of an absorbed powerlaw (\texttt{phabs(powerlaw)}) with a fixed absorption column
			of $\NH=10^{20}\mathrm{cm^{-2}}$ \citep[\Ken]{Kalberla2005}  and of an unabsorbed \APEC{}%
				\footnote{
					A collisionally-ionized diffuse gas model, based on the atomic database ATOMDB (v2.0.2), \url{http://www.atomdb.org}.
					Other diffuse gas models, such as \texttt{RAYMOND} or \texttt{MEKAL} are also appropriate.
					We use the solar abundances of \citet{Anders1989}, since it was used in several previous CXB studies.
				}
			model, the former representing the extragalactic sources and the latter representing the Galactic diffuse emission.
			The spectrum and model fit are shown in \Fref{fig:CXB_spec_full} and the best-fit parameters are listed  in \Tref{tab:CXB_spec}.
			The CXB model gives a surface brightness of $7.8\pm0.1\,\times10^{-12}\,\mathrm{erg\,cm^{-2}\,s^{-1}\,deg^{-2}}$ ($0.81\pm0.01\,\mathrm{cts\,s^{-1}deg^{-2}}$) in the \eSoftB{}, which is in good agreement with the value from the flux maps of $7.9\pm0.1\,\times10^{-12}\,\mathrm{erg\,cm^{-2}\,s^{-1}\,deg^{-2}}$ (\Sref{ss:QuiBKG}).
			The individual components of our CXB model are discussed in the following sections.
			
			We note that there is a significant emission feature in the energy spectrum around $2.5\,\mathrm{keV}$ (close to the right wing of the third instrumental line, \Fref{fig:CXB_spec_full}). 
			We believe that it arises from the instrumental background, since we can see a similar feature in the spectrum of the latter (\Fref{fig:PB_spec}).
			This is further supported by the fact that
			we do not detect any excess continuum associated with this feature.
			However, we can not entirely exclude the possibility that it may be of astrophysical origin			
			As it is outside the energy range of our fluctuation analysis (\eSoft{}) we do not investigate it any further here.			
			
			\subsubsection{Galactic emission}  \label{ss:Espec_Gal}

			The \APEC{} model of our spectral model encapsulates all the Galactic emission, which is a superposition of various diffuse sources \citep[e.g.][]{Lumb2002,Hickox2006,Henley2013}:
			the Galactic halo emission and the foreground emission, which is a composite of emission from solar wind charge exchange (SWCX) and the local bubble.
			All of them have in common that they are anisotropically distributed over the sky.
			In \Fref{fig:CXB_spec_full} we can see that the Galactic emission dominates the soft part of the energy spectrum but above $\sim1\,\mathrm{keV}$ it becomes negligibly small  in comparison to the extragalactic component.
			
			The surface brightness of our Galactic emission model is $3.2\pm0.1\,\times10^{-12}$~erg cm$^{-2}$ s$^{-1}$ deg$^{-2}$ 
			for the \eSoftB{} (\Tref{tab:CXB_spec}).
			This is in reasonable agreement with the  measurements of \HickI{} (Table~2, $\sim(3-4)\times10^{-12}\,\mathrm{erg\,cm^{-2}\,s^{-1}\,deg^{-2}}$), which use \chandra{} Deep Field surveys (hereafter CDFs), and \citet[Table~3, $\sim3.8\times10^{-12}\,\mathrm{erg\,cm^{-2}\,s^{-1}\,deg^{-2}}$]{Lumb2002}, which use several deep \xmm{} observations.

			\subsubsection{Extragalactic emission} \label{ss:Espec_Exgal}

				\begin{figure} 
					\centering
					\resizebox{\hsize}{!}{\includegraphics{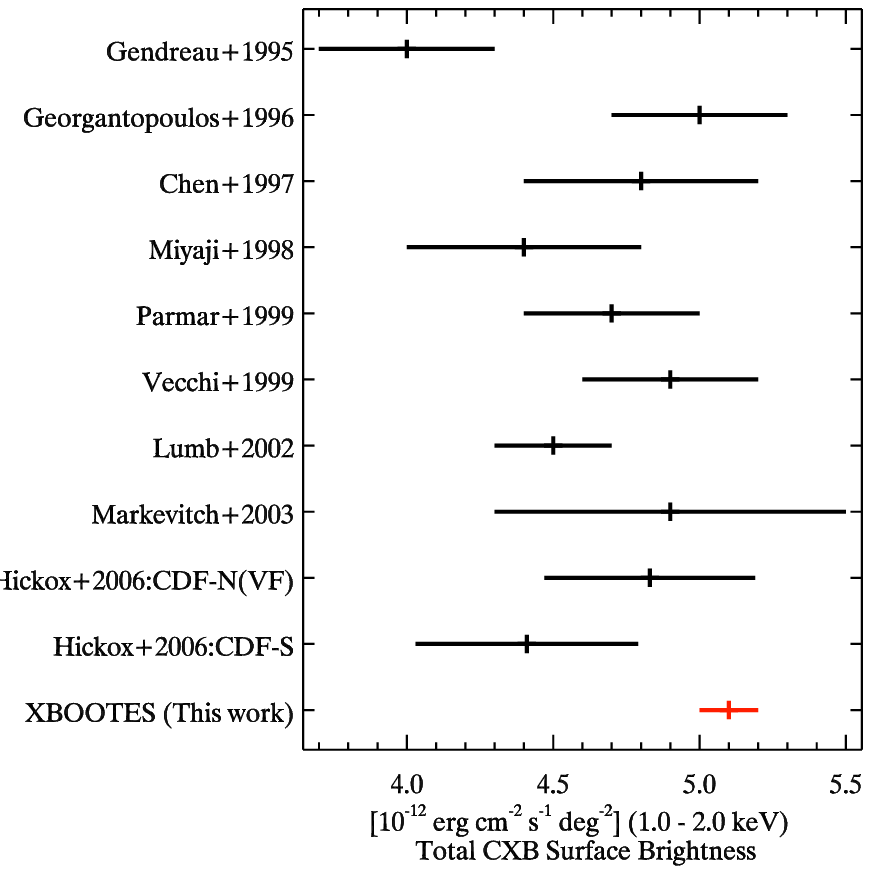}}   
					\caption{\label{fig:CXB_comp}%
						Comparison of recent surface brightness measurements of the CXB in the \eExgalB{}.
						See also Table~5 and Table~6 (last column) of \HickI{}.
					}	
				\end{figure}
				

			The absorbed powerlaw (\texttt{phabs(powerlaw)}) of our spectral model describes the extragalactic emission and its total surface brightness is $4.6\pm0.1\,\times10^{-12}\,\mathrm{erg\,cm^{-2}\,s^{-1}\,deg^{-2}}$ for the \eSoftB{}.
			Together with the emission of the resolved sources ($4.4\pm0.1\,\times10^{-12}\,\mathrm{erg\,cm^{-2}\,s^{-1}\,deg^{-2}}$) computed from the removed source counts and converted to physical units with the same spectral model, we obtain for XBOOTES a total extragalactic CXB surface brightness of $9.0\pm0.1\,\times10^{-12}\,\mathrm{erg\,cm^{-2}\,s^{-1}\,deg^{-2}}$.
			The values are summarized in \Tref{tab:CXB_all}.
			
			We can see in \Fref{fig:CXB_spec_full} that the extragalactic emission dominates the energy spectrum above $\sim1\,\mathrm{keV}$.
			For the \eExgalB{} it has a surface brightness of \Ac{$2.6\pm0.1\,\times10^{-12}\,\mathrm{erg\,cm^{-2}\,s^{-1}\,deg^{-2}}$},
			while the Galactic emission has less than $0.1\times10^{-12}\,\mathrm{erg\,cm^{-2}\,s^{-1}\,deg^{-2}}$, based on our spectral model.
			Therefore, the contribution by any known Galactic sources can be neglected, which makes the \eExgalB{} very suitable to compare our extragalactic CXB measurement with other studies.
			In \Fref{fig:CXB_comp} we show a summary of previous CXB measurement taken from \HickI{} (Table~5 and Table~6, last column) together with our measurement.
			A comparison reveals that our measurement is the highest but still in good agreement with almost all of these studies.
			Except for the measurement of \citet{Gendreau1995} with ASCA, the differences remains below $14\,\%$.
			We are consistent within one standard deviation with the CDF-North (CDF-N) ($4.8\pm0.4\,\times10^{-12}\,\mathrm{erg\,cm^{-2}\,s^{-1}\,deg^{-2}}$)
			and within two standard deviation with the CDF-South (CDF-S) ($4.4\pm0.4\,\times10^{-12}\,\mathrm{erg\,cm^{-2}\,s^{-1}\,deg^{-2}}$).
			Note that since both CDFs are deep pencil-beam surveys with a sky area of $\sim0.02$~deg$^2$ each, which is about $400$ times smaller than for XBOOTES,
			cosmic variance needs to be considered.
			\HickI{} estimate that this adds an additional uncertainty of about $\sim(10-20)\,\%$ to the measurement of the CDFs.
			Furthermore, the much bigger sky area used in our analysis (by a factor of $\sim100$) is the main reason of our much smaller statistical uncertainty in comparison to the CDFs.

			
				\begin{table}
					\caption{Extragalactic emission of XBOOTES \newline [$10^{-12}\,\mathrm{erg\,cm^{-2}\,s^{-1}\,deg^{-2}}$].}
					\label{tab:CXB_all}
					\begin{center}     
						\begin{tabular}{l c c}	
							\hline
							\hline
											&   \eSoft{}	& \eExgal{} \\
							\hline
							Unresolved$^{(\mathrm{a})}$	& $4.6\pm0.1$	& $2.58\pm0.05$ \\ 
							Resolved$^{(\mathrm{b})}$	& $4.4\pm0.1$	& $2.48\pm0.03$ \\ 
							\hline
							Total				& $9.0\pm0.1$	& $5.1\pm0.1$  \\
							\hline
						\end{tabular}
					\end{center}
					(a) From the spectral fit (\Tref{tab:CXB_spec});
					(b) From the removed source counts of the resolved sources in the background-subtracted maps (\Sref{ss:ResoScr}-\ref{ss:QuiBKG}).
				\end{table}

			\subsection{Flux budget} \label{sss:Unr_Exgal}
			
			The unresolved extragalactic emission  is a superposition of contributions of various types of sources of which most important are expected  to be:
			AGN, normal galaxies (no indication of AGN activity), and \GCG{},
			which we refer to in the following as the major X-ray source populations
			\citep[e.g.][]{Lumb2002,DeLuca2004,Hickox2006,Hickox2007,Kim2007,Georgakakis2008,Lehmer2012,Rosati2002,Finoguenov2007,Finoguenov2010,Finoguenov2015}.
			Here, we estimate the absolute and fractional contribution of each population to the unresolved extragalactic emission, which will be relevant for our fluctuation analysis (\Sref{sec:Ana} and \ref{sec:Excess}).
			We compute the surface brightness of the unresolved emission of an X-ray population as follows:
				\begin{align} \label{eq:SurfBright}
						B & = \int\limits_{S_\mathrm{Min}} \d S \; S \, \frac{\d N}{\d S} \; \left(1 - f(S)\right)
					\text{ .}	
				\end{align}
			Hereby, we use differential \LogNLogS{} from the literature and the normalized XBOOTES survey sensitivity curve  $f(S)$ for each source population.
			
			\citet[hereafter \Leh{}]{Lehmer2012} present the currently best measurement of the \LogNLogS{} of AGN and normal galaxies, which is based on source samples down to a flux limit of $S_{\eSoft}\sim10^{-17}\,\mathrm{erg\,cm^{-2}\,s^{-1}}$.
			They demonstrated that above this flux limit AGN dominate the source counts, while below it normal galaxies are becoming the dominant  source population  (see Fig.~6 of \Leh{}).
			We find a very good agreement between the \LogNLogS{} of \Leh{} and of XBOOTES (\Ken{}) down to the flux-limit of XBOOTES ($\sim2\times10^{-15}\,\mathrm{erg\,cm^{-2}\,s^{-1}}$).
			Based on the \LogNLogS{} of \Leh{}, we estimate a surface brightness of $B_\mathrm{AGN}\sim1.4\times10^{-12}\,\mathrm{erg\,cm^{-2}\,s^{-1}\,deg^{-2}}$ and $B_\mathrm{Nor.Gal.}\sim2.2\times10^{-13}\,\mathrm{erg\,cm^{-2}\,s^{-1}\,deg^{-2}}$ for the unresolved AGN and normal galaxy population in XBOOTES, respectively.
			For this we used the \Eqref{eq:SurfBright}, the point-source sensitivity distribution of XBOOTES (Fig.~12 of \Ken, see also \Sref{ss:ResoScr}), and a flux limit of \Leh{} of $S_\mathrm{Min}=10^{-17}\,\mathrm{erg\,cm^{-2}\,s^{-1}}$.
			If we extrapolate the \LogNLogS{} of \Leh{} by two orders of magnitude down to  $S_\mathrm{Min}=10^{-19}\,\mathrm{erg\,cm^{-2}\,s^{-1}}$, the surface brightness of AGN increases by $\sim6\,\%$ ($B_\mathrm{AGN}\sim1.5\times10^{-12}\,\mathrm{erg\,cm^{-2}\,s^{-1}\,deg^{-2}}$)
			and of normal galaxies by a factor of $\sim3.5$ ($B_\mathrm{Nor.Gal.}\sim7.8\times10^{-13}\,\mathrm{erg\,cm^{-2}\,s^{-1}\,deg^{-2}}$).

				\begin{figure} 
					\centering
					\resizebox{\hsize}{!}{\includegraphics{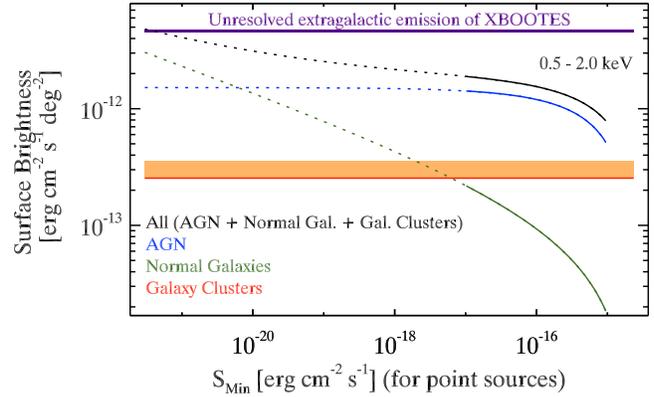}}   
					\caption{\label{fig:Exgal_Brght}%
						The predicted surface brightnesses of unresolved X-ray source populations in XBOOTES (\eSoft) as a function of the lower flux limit $S_\mathrm{Min}$ of \Eqref{eq:SurfBright} for point sources.
						The dotted part of a curve illustrates, where we extrapolate the used \LogNLogS{}.
						Note that these parts of the curves should be interpreted with some caution as they represent extrapolations over several orders of magnitude.
						\Ac{The orange bar shows the contribution by \GCG{}.
						The lower limit of the bar (red line) illustrates the default estimate of \Tref{tab:CXB_Contrib},
						while for the upper limit the XLF slope of \citet{Ebeling1997} is increased by $\sim10\,\%$.}
						Purple horizontal line:
							The unresolved extragalactic emission of XBOOTES (\Tref{tab:CXB_all}).
							The thickness of the line represents one standard deviation.
					}	
				\end{figure}

			This calculation  shows that the total flux of AGN is relatively independent of  $S_\mathrm{Min}$, while for normal galaxies it is very sensitive to $S_\mathrm{Min}$, as it is further illustrated in \Fref{fig:Exgal_Brght}.
			Such a behavior is caused by the steeper slope of the \LogNLogS{} of normal galaxies in comparison to AGN ($\beta\approx 2.2$ for normal galaxies, $\beta\approx 1.5$ for AGN) and makes it difficult to estimate accurately the contribution of normal galaxies to the unresolved CXB emission.
			
				\begin{table}
					\caption{Contributions of major X-ray source populations to the unresolved extragalactic emission of XBOOTES.}
					\label{tab:CXB_Contrib}
					\begin{center}     
						\begin{tabular}{l c c c}	
							\hline
							\hline
							Contribution				&  \multicolumn{2}{c}{Absolute$^{(\mathrm{a})}$}	& Fractional$^{(\mathrm{b})}$ \\
							Energy-Band$^{(\mathrm{c})}$ [keV]	&   $0.5-2.0$	& $1.0-2.0$ &   $0.5-2.0$ \\ 
							\hline
							AGN$^{(\mathrm{d})}$			& $\sim1.51$	& $\sim0.85$	& $\sim33\,\%$ \\ 
							Normal Galaxies$^{(\mathrm{d})}$	& $\sim1.35$	& $\sim0.76$	& $\sim29\,\%$ \\ 
							Galaxy Clusters$^{(\mathrm{e})}$	& $\sim0.25$	& $\sim0.15$    & $\sim6\,\%$ \\ 
							\hline
							Sum		& $\sim3.11$	& $\sim1.77$    & $\sim68\,\%$  \\
							\hline
						\end{tabular}
					\end{center}
					(a) In units of $10^{-12}\,\mathrm{erg\,cm^{-2}\,s^{-1}\,deg^{-2}}$.
					(b) In respect to the unresolved extragalactic emission of XBOOTES (\Tref{tab:CXB_all}).
					(c) Spectral models for energy band conversion are described in \Sref{sss:Exc_Spec}.
					(d) Lower flux limit: $S_\mathrm{Min}=10^{-20}\,\mathrm{erg\,cm^{-2}\,s^{-1}}$ (\eSoft).
					(e) Lower temperature limit: $T=0.2\,\mathrm{keV}$
					($L_\mathrm{\eSoft}\gtrsim10^{40.2}\,\mathrm{erg\,s^{-1}}$, $M_{500}\gtrsim10^{12.8}\,\mathrm{M_{\sun}\,h^{-1}}$ for $z=0$).
				\end{table}
			
			To estimate the contribution of \GCG{} to the unresolved emission of XBOOTES, we are using the X-ray luminosity function (XLF) of \citet{Ebeling1997}.
			It is based on the ROSAT Brightest Cluster sample \citep[BCS,][]{Ebeling1996}, which
			includes 199 sources with a flux above $\sim5\times10^{-12}\,\mathrm{erg\,cm^{-2}\,s^{-1}}$, 
			and covers redshifts up to $z\sim0.3$ and luminosities down to $5\times\sim10^{42}\,\mathrm{erg\;s^{-1}}$ ($0.1-2.4\,\mathrm{keV}$).
			Despite the small redshift range of the XLF, it predicts consistent number densities of galaxy clusters down to the currently deepest studies  with \xmm{} having the sensitivity limit for extended sources of $S_{\eSoft}\sim10^{-16}\,\mathrm{erg\,cm^{-2}\,s^{-1}}$ \citep[e.g.][]{Rosati2002,Finoguenov2007,Finoguenov2010,Finoguenov2015}.
			Further, the XLF does not incorporate any redshift evolution, which is consistent with more recent studies over a larger redshift range up to $z\sim1$ \citep[e.g.][]{Boehringer2014,Pacaud2016}.
			To compute the \LogNLogS{} from this XLF, we integrate the XLF over redshift and luminosity.
			For the K-correction in this integration, we are assuming an \APEC{} model, where we couple its temperature to the luminosity with the luminosity-temperature scaling relation of \citet[Table~2, bold font]{Giles2015} and assume a typical metallicity of $0.3$ of the solar value \citep{Anders1989}.
			As a lower temperature limit we use $T=0.2\,\mathrm{keV}$ ($\approx2.3\times10^6$~K). 
			This limits our integration to luminosities above $\sim10^{40.2}\,\mathrm{erg\,s^{-1}}$ (\eSoft) and DMH masses above $M_{500}\sim10^{12.8},\mathrm{M_{\sun}\,h^{-1}}$ ($z=0$), using the luminosity-mass relation of \citet{Anderson2015}. 
			It also leads to a decline of the differential \LogNLogS{} at fluxes below $\sim10^{-18}\,\mathrm{erg\,cm^{-2}\,s^{-1}}$.

			We rescale the \LogNLogS{} computed with XLF of \citet{Ebeling1997} by a factor of $0.60$ to match the observed $\log N - \log S$ of extended sources in XBOOTES (\Ken{}, Table 1).
			We note that the scaled $\log N - \log S$ is still consistent within one standard deviation with the measurements of \citet{Finoguenov2010,Finoguenov2015}.
			Based on the shape comparison of the \LogNLogS{} of \citet{Ebeling1997} and the XBOOTES (\Ken{}, Table 1), we estimate that the sensitivity limit for extended sources for XBOOTES is around $S_{\eSoft}=3\times10^{-14}\,\mathrm{erg\,cm^{-2}\,s^{-1}}$.
			With this we estimate a surface brightness of $\sim2.5\times10^{-13}\,\mathrm{erg\,cm^{-2}\,s^{-1}\,deg^{-2}}$ for the unresolved \GCG{} in XBOOTES.

			The computed surface brightness of unresolved \GCG{} depends only mildly on the assumed flux  limit of XBOOTES for  extended sources.
			For instance, the surface brightness only changes by $\sim10\,\%$, if we change the flux  limit by $-30\,\%$ or $+50\,\%$ ($(2.1-4.5)\times10^{-14}\,\mathrm{erg\,cm^{-2}\,s^{-1}}$).
			Also the exact choice of the lower temperature limit in the XLF integration is not very critical.
			The surface brightness increases only by $\sim2\,\%$, if we decrease the limit to $T=0.1\,\mathrm{keV}$ ($-50\,\%$)
			and it decreases only by $\sim10\,\%$, if we increase the limit to $T=0.3\,\mathrm{keV}$ ($+50\,\%$).
			However, the estimate of the surface  brightness of \GCG{} is rather sensitive to the assumed shape of their XLF.
			For example, increasing its slope by  $\sim10\,\%$, from $\alpha=1.85$ to $2.03$, would increase the flux from unresolved \GCG{} by $\sim50\,\%$.
			This makes the estimate of their contribution to the unresolved part of the CXB  less certain than that of AGN (but still more accurate than the estimate for normal galaxies).

			The  total flux budget for the unresolved CXB is summarised in Table \ref{tab:CXB_Contrib}. In computing these numbers we assumed the low flux limit for the \LogNLogS{} integration of AGN and normal galaxies to be equal to   $S_\mathrm{Min}=10^{-20}\,\mathrm{erg\,cm^{-2}\,s^{-1}}$ (note that as the contribution of \GCG{} is computed via their XLF, no explicit low flux limit is needed in this case). With this  AGN account for $\sim33\,\%$ of the unresolved CXB in XBOOTES, normal galaxies for $\sim29\,\%$ and  \GCG{} for  $\approx6\,\%$.  All together they account for $\approx68\,\%$ of the unresolved  CXB.
			About $\approx32\,\%$ of the unresolved emission remains unaccounted for in this calculation. This is not worrisome however, because of the rather large uncertainty for the  contribution of \GCG{} and normal galaxies.
			
			We illustrate the dependence of the surface brightness of unresolved AGN and normal galaxies on $S_\mathrm{Min}$ in \Fref{fig:Exgal_Brght}.
			As one can see, the contribution by normal galaxies strongly increases towards small $S_\mathrm{Min}$ and at $S_\mathrm{Min}\sim 10^{-21}\,\mathrm{erg\,cm^{-2}\,s^{-1}}$ can easily explain the remaining part of the unresolved CXB.  Although this conclusion  is based on a very significant extrapolation of the observed \LogNLogS{} and  should be interpreted with caution, it is   clear that normal galaxies are an important contributor to the unresolved CXB in XBOOTES. 	
			This inference  is further supported by the results of \citet{Hickox2007} who showed that faint optical/IR point sources can be associated with $\sim50\,\%$ of the extragalactic emission below $S_{\eSoft}\sim10^{-17}\,\mathrm{erg\,cm^{-2}\,s^{-1}}$. 
			Combining this result with the \LogNLogS{} of resolved normal galaxies at higher fluxes, we estimate that   normal galaxies account for at least $\sim 34\,\%$ of the unresolved extragalactic CXB in XBOOTES, which is close to the value derived above (\Tref{tab:CXB_Contrib}).

		\subsection{Redshift and luminosity distributions of unresolved populations}  \label{ss:UnresoAGN} 
		
				\begin{figure*}
					\centering
					\includegraphics[width=0.47\textwidth]{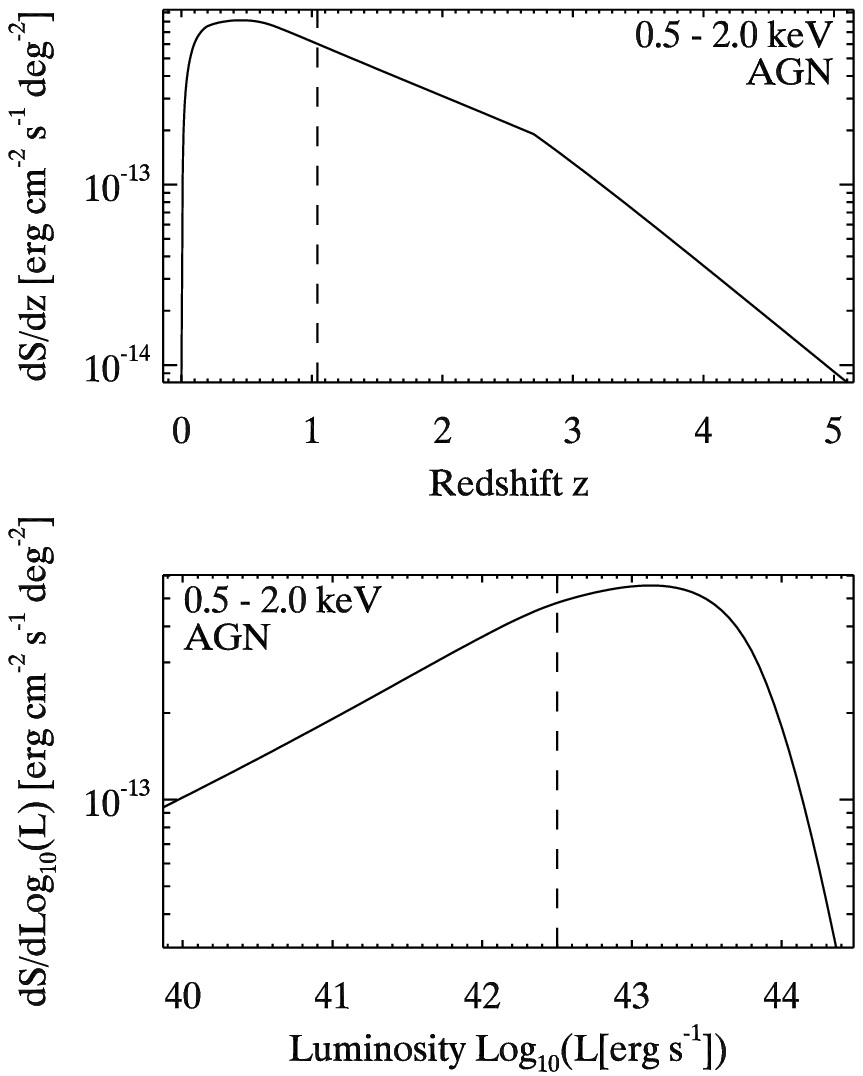}
					\hfill
					\includegraphics[width=0.47\textwidth]{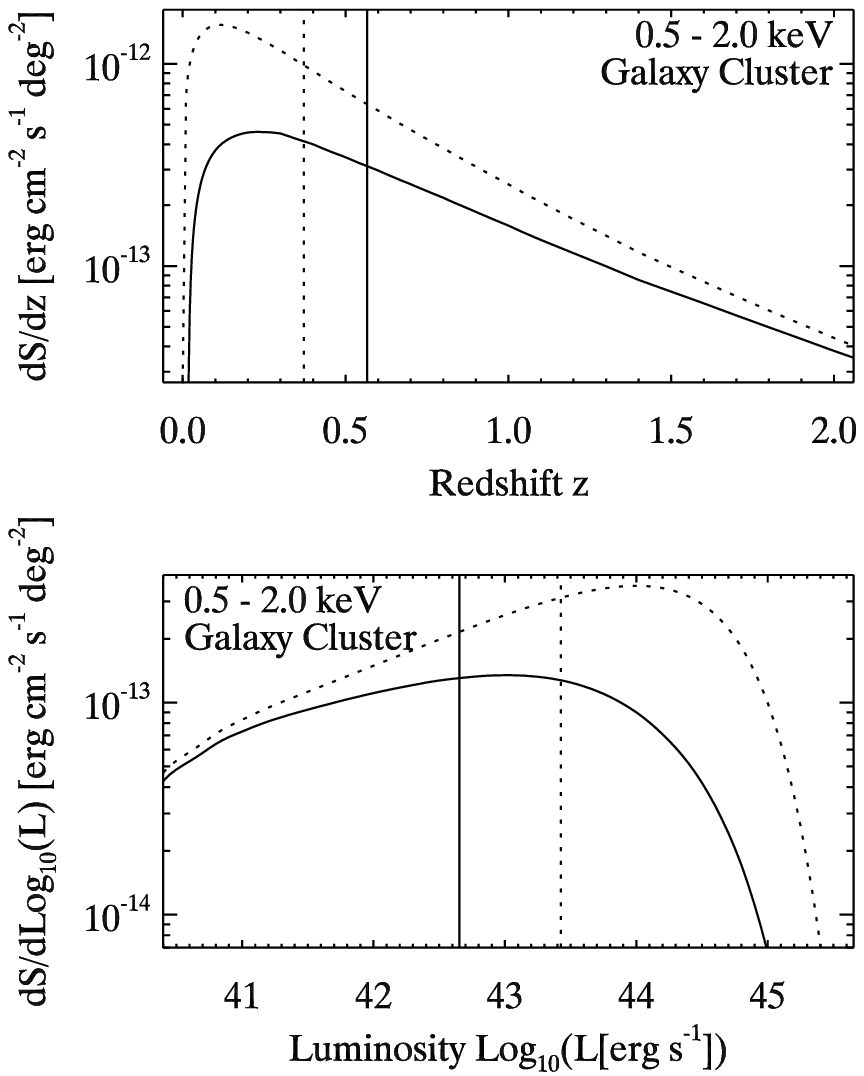}
					\caption{\label{fig:dSdz}%
					Flux production rate in the \eSoftB{} for different populations in XBOOTES survey versus redshift and luminosity:
					unresolved AGN (left)  and \GCG{} (right). 
					For the latter we show the flux production rate for both unresolved (solid curves) and total (dotted curve) populations.
					The vertical lines show corresponding median values.
					\Mn{Alex, we need to homogenize the left and right plots, e.g. add AGN and the band to the left plots or remove text from the right ones.}
					}
				\end{figure*}		

			In order to characterize the unresolved  populations, we calculate their \emph{flux production rate per solid angle} [erg cm$^{-2}$ s$^{-1}$ deg$^{-2}$] as a function of redshift and luminosity:
				\begin{align} \
					\frac{\d S (z)}{\d z} = &
						\int\!\d S \; \left\{ \dfrac{\Phi\!\left(\log_{10}(L_r[S,z]),z\right)}{L_r[S,z]  \; \log_\mathrm{e}(10)} \right\}
							\; \frac{\d^2 V(z)}{\d z \d\Omega}  \notag \\ & \times
							\; \frac{4\pi d^2_\mathrm{L}(z) }{K(z)}  \; S \; \left[1 - f(S)\right] 
						\label{eq:dSdz} \text{,} \\
					\frac{\d S (L)}{\d \log_{10}(L)} = &
						\int\!\d z \;  \Phi\!\left(\log_{10}(L_r),z\right) \; \frac{\d^2 V(z)}{\d z \d\Omega}  \notag \\ & \times
							\; S[L,z] \; \left[1 - f(S[L,z])\right]
						\label{eq:dSdL} \text{.} 
				\end{align}
			where $\Phi(\log_{10}(L_r),z)$ [h$^3$ Mpc$^{-3}$] is the luminosity function of sources, $f(S)$ is the normalized survey selection function of  for the given type of objects (point-like or extended), $K(z)$ is the K-correction,
			$L_r$ [erg s$^{-1}$] is the rest-frame luminosity ($L_r = L/ K(z)$),
			$\d^2 V(z)/\d z/\d\Omega$ [Mpc$^{3}$ h$^{-3}$ deg$^{-2}$] is the co-moving volume element, and
			$d_\mathrm{L}(z)$ [cm] is the luminosity distance \citep[e.g.][]{Hogg1999}.
	
			\subsubsection{AGN}	
				
			For the AGN luminosity function  $\Phi(\log_{10}(L_r),z)$ [h$^3$ Mpc$^{-3}$] we used results of  \citet{Hasinger2005} for the \eSoftB{} and for the $f(S)$ we use is the normalized survey selection function  for point sources (\Ken{}, Fig.~12). 	For the XLF we include the exponential redshift cutoff for $z>2.7$: $\Phi(z) = \Phi(z\!=\!2.7)\times10^{0.43\,(2.7 - z)}$, which was proposed by \citet{Brusa2009}.
			For computing the K-correction we assume a powerlaw with the photon-index of $\Gamma=1.7$, which simplifies the quantity to $K(z)=(1+z)^{2-\Gamma}$.
			The survey selection function is given in the \eFull{} band and we convert it to the \eSoftB{} with an absorbed powerlaw with a photon-index of $\Gamma=1.7$.
			Note, that the XLF of \citet{Hasinger2005} includes  type 1 AGN only, and is defined for the  minimum luminosity of $L_{\eSoft} \approx 10^{42}\,\mathrm{erg\;s^{-1}}$ and maximum redshift of $z\approx3-5$.
			Therefore  we have to correct the amplitude of the differential flux distributions to match the \LogNLogS{} of \Leh{} (factor of $\sim1.44$ increase).
			Secondly, one should be aware that the derived flux production rates   may be subject to large uncertainties at low luminosity ($<10^{42}\,\mathrm{erg\;s^{-1}}$) and large redshift ($z>3$)
			\citep[see also the discussion of the uncertainties of different XLF of AGN in Sect.~5 of][]{Kolodzig2012}.

			In \Fref{fig:dSdz} (left panels) we show the differential flux distributions of the unresolved AGN  for the \eSoftB{}.
			The redshift distribution peaks around $z\sim0.5$ with the median value of  $z\sim1.0$.
			About two thirds of unresolved AGN are located between redshift $\sim0.4$ and $\sim2.3$.
			The luminosity distribution peaks $\sim 10^{43.3}\,\mathrm{erg\;s^{-1}}$ and has the  median value of  $L \sim 10^{42.6}\,\mathrm{erg\;s^{-1}}$.
			About two thirds of unresolved AGN have the luminosity between $\sim 10^{41.0}$ and $\sim 10^{43.6}\,\mathrm{erg\;s^{-1}}$. 
			Thus, that with the unresolved CXB of XBOOTES  one can  study the clustering signal of relatively low-luminosity AGN located around redshift $z\sim1$. These objects  are largely   inaccessible for the conventional  clustering studies using resolved AGN \citep[e.g.][]{Cappelluti2012,Krumpe2013}.

			\subsubsection{Galaxy clusters \& groups} %
			As described in \Sref{sss:Unr_Exgal}, for \GCG{}  we use the XLF of \citet{Ebeling1997}, setting the lower integration limit corresponding to the gas temperature of  $T=0.2\,\mathrm{keV}$ (\Sref{sss:Unr_Exgal}).		
			The selection function $f(S)$ is set to be a step-function with the step at the flux corresponding to the average survey sensitivity for extended source  of $3\times10^{-14}\,\mathrm{erg\,cm^{-2}\,s^{-1}}$ (\Sref{sss:Unr_Exgal}). We also compute the flux production rate distributions for the entire population of \GCG{}, as it will be relevant for the  discussion in \Sref{sec:Excess}.
			In this case we assumed $f(S)=0$ for all fluxes.
			As with the AGN distributions, result of this calculation is subject to some uncertainty at low luminosities and large redshifts, where the XLF of the objects of interests is poorly constrained.

			The obtained distributions are shown in  \Fref{fig:dSdz}.
			For the unresolved population,
			the redshift distribution has a median redshift of $z\sim0.6$ and peaks around $z\sim0.2$. 
			\Ac{The flux-weighted mean redshift equals $\left<z\right>\sim0.3$.}
			About two third of the population are located between redshift $\sim0.2$ and $\sim1.3$.
			Their median and peak luminosity is around $L \sim 10^{42.7}$ and $\sim 10^{43.1}\,\mathrm{erg\;s^{-1}}$, respectively,
			and about two third have a luminosity between $\sim 10^{41.4}$ and $\sim 10^{43.8}\,\mathrm{erg\;s^{-1}}$.
			Thus, the unresolved population of \GCG{}  consists mainly of relatively low-luminosity and nearby objects, located around redshift $z\sim0.6$.
			They are more local and distributed over a more narrow redshift (and luminosity) range than unresolved AGN population.
			
			The total (resolved and unresolved) population of \GCG{} is located on average closer than its unresolved part, with a median redshift of $z\sim0.4$ (peak at $z\sim0.1$, flux-weighted mean of $\left<z\right>\sim 0.2$) and two third are located between redshift $\sim0.1$ and $\sim1.0$.
			The total population of \GCG{} is in average more luminous than its unresolved part, with the median luminosity of $\sim 10^{43.4}\,\mathrm{erg\;s^{-1}}$ (peak at $\sim10^{44.0}\,\mathrm{erg\;s^{-1}}$) and two thirds having the  luminosity between $\sim 10^{42.0}$ and $\sim 10^{44.4}\,\mathrm{erg\;s^{-1}}$.
			Both results are expect since the resolved fraction of the total population consist of \GCG{}, which contribute about $50\,\%$ to the total surface brightness of \GCG{} and are rather close by and more luminous \citep{Vajgel2014}.
			
			Based on the scaling relations  at the median redshift and luminosity,  we  estimate that the unresolved and total population of \GCG{} have in average an ICM  temperature of $T\sim1.4\,\mathrm{keV}$ ($\sim1.6\times10^7$~K) and $\sim2.9\,\mathrm{keV}$ ($\sim3.4\times10^7$~K), and a DMH mass of $M_{500}\sim10^{13.5}$ and $\sim10^{14.0}\,\mathrm{M_{\sun}\,h^{-1}}$, respectively.

	\section{Brightness fluctuations of the unresolved CXB} \label{sec:Ana}
		
		
		In the following we study the surface brightness fluctuations of the unresolved CXB by analyzing their \PoSp{}.
		In this study we focus on the angular scale range between $\sim3\arcsec$ and $\sim17\arcmin$ (angular frequencies of $0.001 - 0.300\InvArcSec$). 
		This range covers the spatial co-moving scales%
			\footnote{
			$x_\mathrm{spatial}(z)[\mathrm{Mpc\,h^{-1}}] = r[\mathrm{rad}] \; d(z)[\mathrm{Mpc\,h^{-1}}] $, using the co-moving distance $d(z)$ \citep[e.g.][Eq.~16]{Hogg1999}.
			}
		between $\sim0.03\,\mathrm{Mpc\,h^{-1}}$ and $[\sim1.5,\sim6.5,\sim11.4]\,\mathrm{Mpc\,h^{-1}}$ for the redshifts of $z=[0.1,0.5,1.0]$.
		Therefore, our measurement is sensitive to the small-scale ($\lesssim1\,\mathrm{Mpc\,h^{-1}}$) clustering regime, where the spatial correlation of galaxies and the ICM within the same DMH \citep[alias the \OHT{}, e.g.][]{Cooray2002} dominates the clustering signal.

		For our analysis we combine the \PoSpa{} of all considered XBOOTES observations.
		Since we compute the \PoSp{} for each observation individually, our maximum angular scale is defined by side size of the \acisi{} FOV\footnote{\url{http://cxc.harvard.edu/proposer/POG/html/chap6.html}}, which is about $17\arcmin$.
		We ignore in our analysis angular scales smaller than $\sim3\arcsec$ ($> 0.3\InvArcSec$) as, 
		at these angular frequencies, the source \PoSp{} is suppressed by more than a factor of $10$ due the PSF-smearing (see \Fref{fig:PS_PSF_smearing}).
	We demonstrate in \Aref{app:PSFsmear} -- \ref{app:Noise} and in \Fref{fig:PS_Signal_eSoft} that our PSF-smearing models can adequately describe the measured  \PoSp{} up to angular frequencies of $\sim 0.9\InvArcSec$ ($\gtrsim1.1\arcsec$, $\gtrsim2$ image pixels).

		\subsection{Formalism} \label{ss:Def}
		
			We study the surface brightness fluctuations $\delta F$ of the unresolved CXB via Fourier analysis.
			The Fourier transform of a density field $\delta F(\VEC{r})$ is defined in our study as:
			\begin{align} \label{eq:FrTrans}
				\widehat{\delta F}(\VEC{k}) & = \int\mathrm{d}^2r \,\delta F(\VEC{r}) \,\exp(- 2\pi \,\mathrm{i}\; \VEC{r} \cdot \VEC{k}  )
				\text{ .}	
			\end{align}
			We note that due to our choice of having a $2\pi$ in the exponent,  the angular scale is related to the angular frequency as $r=k^{-1}$.
 			Due to our measurement process, the field is transformed from a continuous into a discrete one. 
 			This changes \Eqref{eq:FrTrans} to a 2D discrete Fourier transform: %
			\begin{align} \label{eq:FrTransDis}
				\widehat{\delta F}(\VEC{k}) & =  \dfrac{1}{\sqrt{\Omega}} \; \sum_{i,j}  \,\delta F(\VEC{r}_{i,j}) \,\exp\left( -2\pi\,\mathrm{i}\; \VEC{r}_{i,j} \cdot \VEC{k}  \right)
				\text{ ,}	
			\end{align}
			where $\VEC{r}_{i,j}$ is the position of an image-pixel. 
			The normalization $\sqrt{\Omega}$ of the Fourier transform was chosen so that  the units of the resulting \PoSp{} (\Eref{eq:MeanPS}) are per deg$^2$.
			To compute the Fourier transform we use the FFTW library \citep[v.3.3.3,][\url{http://www.fftw.org}]{FFTW}.
			With the assumption of isotropy, we reduce our 2D Fourier transform to an one-dimensional \emph{\PoSp{}} as follows:
				\begin{align} \label{eq:MeanPS}
					\langle |\widehat{\delta F}(k)|^2 \rangle & =  \dfrac{2}{n(k)} \; \sum_l^{n(k)/2} |\widehat{\delta F}(\VEC{k}_l)|^2
					\text{ .}	
				\end{align}
			Hereby, the ensemble average $\langle \; \rangle$ is replaced with the average over all independent Fourier modes $\widehat{\delta F}(\VEC{k})$ per angular frequency $k$.
			There are $n(k)$ Fourier modes within the interval $[k-\Delta k/2,k+\Delta k/2[$ of the 2D Fourier transform, where $\Delta k = L^{-1}$ is defined by the angular size $L$ of the fluctuation map $\delta\mathbf{F}$.
			This size is defined to be equal for both dimensions ($L_x = L_y$) of the map and to be large enough to embed the entire FOV of an observation (\Sref{ss:MSK}).
			One can analytically approximate the value of $n(k)$ with $2\pi k/\Delta k$ 
			but we directly count the number of modes for each annulus $[k-\Delta k/2,k+\Delta k/2[$ of the 2D Fourier transform.
			Since the fluctuations map $\delta\mathbf{F}$ is a real quantity, half of the 2D Fourier transform is redundant ($\widehat{\delta F}(\VEC{k}) =  \widehat{\delta F}^\ast(-\VEC{k})$,  where $^\ast$ indicates the complex conjugate.
			Therefore, we only have to average in \Eqref{eq:MeanPS} over $n(k)/2$ independent Fourier modes.
			The number and range of independent Fourier frequencies $k$ is limited by the pixel size $\Delta p$, which sets the maximum angular frequency (or minimum angular scale), known as the Nyquist-Frequency, to $k_\mathrm{Ny}=(2\Delta p)^{-1}$, and the angular size $L$ of the fluctuation map,
			which sets the minimum angular frequency (or maximum angular scale) to $k_\mathrm{min}=\Delta k= L^{-1}$.
			In order to obtain the \emph{photon-shot-noise-subtracted \PoSp{}} $P(k)$, we subtract from the \PoSp{} $\langle |\widehat{\delta F}(k)|^2 \rangle$ the photon shot-noise estimate $P_\mathrm{Phot.SN}$, which is explained and discussed in detail in \Aref{app:Noise}, as following:
			\begin{align} \label{eq:DefPS}
				P(k)  & =  \langle |\widehat{\delta F}(k)|^2 \rangle - P_\mathrm{Phot.SN}
				\text{ .}	
			\end{align}
			\Mc{In the following, we will} refer to the resulting \PoSp{} $P(k)$ as the \emph{measured \PoSp} and we will only show it in instrumental units of [$\mathrm{(cts\,s^{-1})^2\,deg^{-2}}$].
			We do not convert it into physical units.
			Instead, we convert our clustering models into instrumental units. 
			
			Based on the assumption that our fluctuations are Gaussian distributed and superimposed by the photon shot noise, we can estimate the statistical uncertainty of $P(k)$ as follows: 
				\begin{align} \label{eq:PSErr}
					\sigma_P(k)  & =  \sqrt{\dfrac{2}{n(k)}} \; \langle |\widehat{\delta F}(k)|^2 \rangle
					\text{ .}	
				\end{align}
			Here, one uses the fact that for a given angular frequency the \PoSp{} $\langle |\widehat{\delta F}(k)|^2 \rangle$ follows a $\chi^2$ distribution \citep[e.g.][]{Klis1989}.
			For large angular frequencies (small angular scales) the number of modes per frequency-bin $[k-\Delta k/2,k+\Delta k/2[$ becomes large enough ($n(k) \gtrsim 100$) that one can assume a Gaussian distribution for the \PoSp{} thanks to the central-limit theorem.
			Due to the fact that we use an averaged \PoSp{} over more than $100$ \PoSpa{}, we can   assume a Gaussian distribution also for the smallest angular frequencies.
			This simplifies the error propagation.

			In order to directly compare \PoSpa{} of different energy bands (as in \Fref{fig:PS_eSoft_Extra03}) we use the \emph{flux-normalized \PoSp{}}, which we define as:
				\begin{align} \label{eq:PS_FluxNorm}
					Q(k)  & = \frac{P(k)}{ \left( \Omega^{-1} \, \sum_{i,j} F(\VEC{r}_{i,j}) \right)^2 }
					\text{ .}	
				\end{align}
			This characterizes the squared fractional amplitude of the fluctuations per unit frequency interval.
			Since the flux map $\mathbf{F}$ has units of [$\mathrm{cts\,s^{-1}}$], the flux-normalized \PoSp{} $Q(k)$ has units of [deg$^{2}$].

		\subsection{The measured \PoSp{}} \label{ss:StackPS}
			
				\begin{figure}
					\begin{center}
						\resizebox{\hsize}{!}{\includegraphics{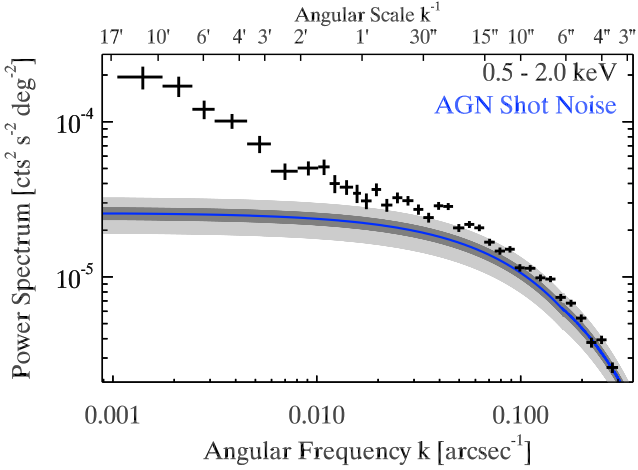}}
						\caption{\label{fig:PS_eSoft}%
							The measured \PoSp{} $P(k)$ of the brightness fluctuations of the unresolved CXB in XBOOTES in
							the \eSoftB{}.
							Blue curve shows the  AGN shot noise (\Sref{ss:shot_noise})  corrected for the PSF-smearing (\Aref{app:PSFsmear}).
							 Grey areas illustrate its dependence on the photon index assumed for the flux conversion:
							 the dark gray region shows the effect of the photon index varying from $1.6-1.8$, light grey from $1.4-2.0$.
							The larger  photon-index values result in  the larger  amplitude of the AGN shot noise.
						}	
					\end{center}
				\end{figure}

			\Mc{In \Fref{fig:PS_eSoft} we plot the measured (i.e. the photon shot-noise subtracted)  \PoSp{} in the \eSoftB{}.
			\Ac{In our current study we are interested primarily on the extragalactic part of the CXB.}
			Our spectral analysis  in \Sref{sec:Espec} suggests that \Ac{fluctuations} below $\sim1$~keV can potentially be contaminated by the  emission of the Galaxy  \Ac{since it} contributes about $\sim 40\,\%$ to the unresolved CXB flux in the \eSoftB{} (\Tref{tab:CXB_spec}).
			However, as we demonstrate in \Sref{sss:Exc_Spec}, the energy spectrum of CXB fluctuations is much harder than that of Galactic emission.
			This allows us to place an upper limit of about \Ac{$\approx 7\,\%$} on the  contribution of fluctuations of the Galactic emission to the \Ac{average} power in the \eSoftB{} \Ac{at large angular scales ($\sim3\arcmin - 8\arcmin$,} \Sref{sss:Exc_Spec}).
			Furthermore, we compared our results with the those obtained in the \eExgalB{} and which is virtually   uncontaminated by the  emission of the Galaxy and found fully consistent results.
			We will conduct our analysis in the \eSoftB{} for consistency and ease of comparison with previous work.}

		\subsection{Point-source shot noise} \label{ss:shot_noise} \label{ss:AGNmodel}
			
		Due to their discreteness, point-like X-ray sources (AGN and normal galaxies), give rise to a shot-noise component in the \PoSp{}%
			\footnote{
				For extended sources, the analog of the shot noise  has a more complex shape of the \PoSp{}, carrying information about the  spatial structure of their DMHs, and it is usually accounted through the \OHT{}, as discussed in the following section.
			} \citep[e.g.][]{Cooray2002}.
		The shot noise  is caused by  fluctuations of the  number of sources  per beam
		and is generally uncorrelated with the LSS signal itself,
		i.e. it is added linearly to the \PoSp{}.
		It  is an analog of the photon shot noise (\Aref{app:Noise}) and for the definition of the \PoSp{} used in this paper, is independent of the angular frequency.

		The shot noise of unresolved point sources can be computed as:
			\begin{align} \label{eq:PS_SN}
				P_\mathrm{SN} = \int\limits_{S_\mathrm{Min}} \d S \; S^2 \; \frac{\d N}{\d S} \; \left(1 - f(S)\right)
				\text{ .}	
			\end{align}
		where  $\d N/\d S$ is the  differential \LogNLogS{} distribution and $f(S)$ is the normalized selection function  for  point sources  (Fig.~12 of \Ken{}).
		Using the \LogNLogS{} distributions from \Leh{}, we obtain for the shot noise of AGN a value of $P_\mathrm{SN}^\mathrm{(AGN)} \approx 1.25 \times 10^{-27}\,\mathrm{(erg\,cm^{-2} s^{-1})^2\,deg^{-2}}$ in the \eSoftB.
		For this calculation we set  the lower flux limit in \Eqref{eq:PS_SN} to $S_\mathrm{Min}=10^{-17}\,\mathrm{erg\,cm^{-2}\,s^{-1}}$, however the result is nearly insensitive to the value of $S_\mathrm{Min}$. 
		Although normal galaxies make a comparable contribution to the unresolved CXB flux (\Sref{sss:Unr_Exgal}),
		due to their steeper \LogNLogS{} in combination with the $S^2$ term in \Eqref{eq:PS_SN} their shot noise is about $\sim 20$ times lower than the AGN shot noise and in the following it will be neglected.

		In instrumental units the AGN shot noise is  $\approx 2.49\times 10^{-5}\,\mathrm{(cts\,s^{-1})^2\, deg^{-2}} $, using an absorbed power law with a photon-index of $\Gamma=1.7$ (\Sref{ss:Espec_Exgal}) for the conversion from physical to instrumental units.
		The shot noise level is \Mc{moderately}  sensitive to the choice of the photon-index.
		Varying the latter between $\Gamma=1.4$ and $2.0$ (between $1.6$ and $1.8$) results in a variation of  the shot noise level by $\sim\pm25\,\%$ ($\sim\pm9\,\%$).

		The predicted shot noise of unresolved AGN, corrected for the PSF-smearing (\Aref{app:PSFsmear}), is shown in \Fref{fig:PS_eSoft}, along with the uncertainty due to variations of the  \Mc{ photon index $\Gamma$ used for the units conversion}. 
		From \Fref{fig:PS_eSoft}  we can see that at small angular scales, below $\sim15\arcsec$,
		the measured \PoSp{} agrees with the theoretical prediction for AGN shot noise \Mc{quite well,  within $\sim30\,\%$}.
		However, at larger angular scales, there is a clear LSS signal detected above the shot noise of unresolved AGN.
		Its origin will be investigated in the next section.

	\section{The origin of the LSS signal} \label{sec:Excess}

		\subsection{The (point-source shot-noise subtracted) \PoSp{}} \label{ss:lss_power_spectrum}
		
		\begin{figure}
			\begin{center} 
				\resizebox{\hsize}{!}{\includegraphics{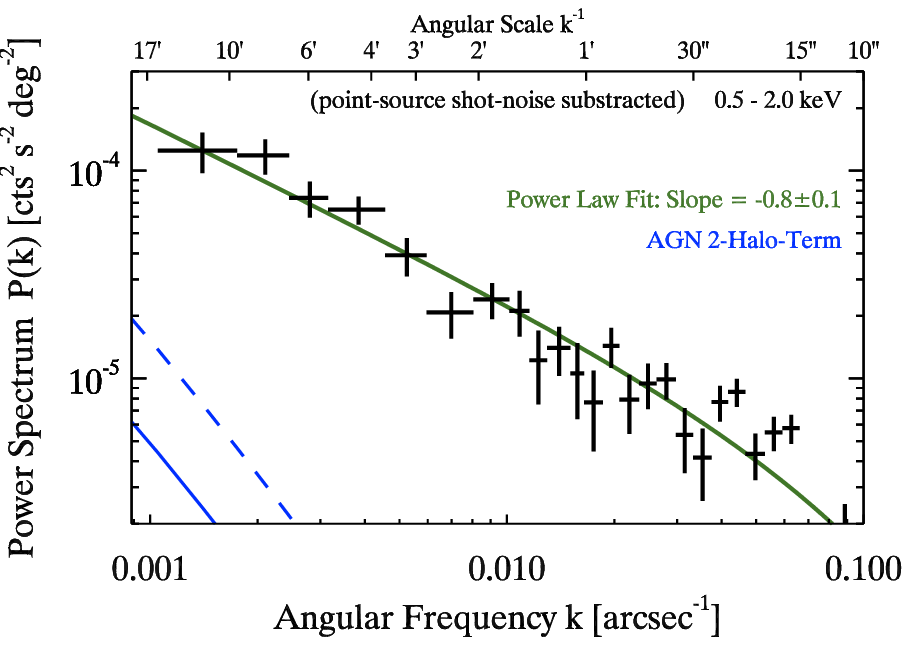}} 
				\caption{\label{fig:excess} \label{fig:agn_tht} 
					The \PoSp{} of the detected LSS signal obtained by subtracting the point-source shot-noise from the measured \PoSp{} shown in \Fref{fig:PS_eSoft}.
							The green solid line shows the power law with the slope of $-0.8$, best-fit to the power spectrum in the  $30\arcsec - 10\arcmin$ range of angular scales. 
					The blue curves show the \THT{} of unresolved AGN assuming $M_\mathrm{eff}=2\times10^{13}\,\mathrm{M_{\sun}\,h^{-1}}$ (solid line) and $M_\mathrm{eff}=1\times10^{14}\,\mathrm{M_{\sun}\,h^{-1}}$ (dashed line).
					Note that their  average slope is $\approx -2.3$.
					See \Sref{sss:Int_AGN} for details.
					}
			\end{center}
		\end{figure}

		\Mc{In order to characterize the  LSS signal  we subtract the point-source shot noise from the measured \PoSp{} in order to obtain the \emph{point-source shot-noise subtracted} \PoSp{}.
		Before the subtraction we renormalize the theoretically computed   AGN shot noise from \Sref{ss:shot_noise} to match the observed power in the  $[0.1,0.3[\InvArcSec$ frequency range ($\sim3\arcsec - 10\arcsec$).
		In this range, we can expect that the \PoSp{} is dominated by the point-source shot noise. 
		The renormalization of the point-source shot noise is needed because the theoretical calculation in  \Sref{ss:shot_noise} is subject to a number of uncertainties, the main of which are: 
			(i) the conversion from physical to instrumental units; 
			(ii) the accuracy of the AGN \LogNLogS{}; 
			(iii) conversion of the survey selection function ($f(S)$) from \eFull{} band to the \eSoftB{};
			(iv) neglected contribution from normal galaxies. 
		The shape of the  theoretical AGN shot noise is determined by the PSF-smearing model,
		which has an accuracy of $\sim5\,\%$ for $\gtrsim 2\arcsec$ (\Aref{app:PSFsmear}).
		The so computed correction factor is $1.17$ i.e. it is sufficiently close to unity.}

		\Mc{The resulting \PoSp{} of the LSS signal is shown in \Fref{fig:excess}.
		A power law fit in the $30\arcsec - 10\arcmin$ range of angular scales gives the value of the best fit slope of $-0.8\pm0.1$.
		Note that the power spectrum slope depends on the flux cut.
		This will be further discussed in  \Sref{sss:Ex_GC_limit}.}

		\subsection{Root-mean-square variation}  \label{sss:RMS}
	
			In order to compute the root-mean-square (RMS) variation, we first compute  for each observation  the variance of the flux map $\mathbf{F}$ in the spatial frequency range of interest:
				\begin{align} \label{eq:VAR}
						\langle(\delta\mathbf{F})^2\rangle_{(k_1,k_2)} & = \int\limits_{k_1}^{k_2}\mathrm{d}^2k \, P(k)
							 = (\Delta k)^2 \sum_{k_1}^{k_2} n(k) \, P(k)
						\text{ .}	
				\end{align} 
			As before, $P(k)$ is the measured \PoSp{} (\Eref{eq:DefPS}).
			Note that we used \Eqref{eq:MeanPS} in order to simplify the summation and that the leading coefficient in the above formula  depends on the definition of the Fourier transform (\Eref{eq:FrTrans}).

			With $\langle(\delta\mathbf{F})^2\rangle$ we can  compute the fractional variance $\langle(\delta\mathbf{F})^2\rangle/\langle F \rangle^2$ where $\langle F \rangle =\Sigma_{i,j} F_{i,j} / \Sigma_{i,j} M_{i,j}$, and its average over all considered observations.
			The square root of this value gives the average fractional RMS variation in the spatial  frequency range $[k_1,k_2]$.
			In order to estimate the uncertainties, we compute the standard deviation of the  fractional variance for individual observations and then use error propagation. 
			
			For the entire range of the spatial frequencies  $[k_\mathrm{Min},k_\mathrm{Ny}]$, we obtain the fractional RMS variation of $8\pm1$ for the \eSoftB.
			The  AGN shot noise model corrected for the  PSF-smearing predicts $\approx7$, 
			which  is fully   consistent with the measured value.
			However, the RMS variation in the full frequency range is dominated by the power at small angular scales (around $\sim10\arcsec$), where the product of $P(k)$ and the number of modes $n(k)$ is the largest, i.e. it is determined by the AGN shot noise and does not characterize the power at large angular scales. 
			
			In order to characterize the latter, we compute the RMS variation for the frequency interval $[0.002,0.006[\InvArcSec$ corresponding to the angular scales from $\sim3\arcmin - 8\arcmin$, where the detected LSS signal dominates the \PoSp{}.
			Note that the same frequency interval will be used for our spectral analysis in \Sref{sss:Exc_Spec}.
			We obtain a RMS variation  of {$(42\pm2)\,\%$} 
			while  our AGN model predicts {$\approx22\,\%$}. 
			%
			Subtracting quadratically the latter from the former we obtain the RMS variation of {$(36\pm2)\,\%$} for the LSS signal \Mc{on the angular scales  of arcminutes}.

		\subsection{Potential sources of contamination}  \label{ss:Exc_NoIB}
		
		The angular scales where the detected LSS signal is particularly pronounced ($\sim 1\arcmin -15\arcmin$) are comparable to the size of the \chandra{}  FOV. 
		There are three main potential sources of contamination at these angular scales:
		(i) the instrumental background,
		(ii) large scale spatial non-uniformity of the detector efficiency, and
		(iii) residual counts in the wings of the PSF from the removed resolved sources. 

		To investigate  significance of the first two factors, we use the instrumental background.
		In particular, we use both  the stowed background data and the XBOOTES data in the \ePartB{} which is entirely dominated by the instrumental background signal.
		In \Aref{app:ss:Bkg} we compute their \PoSpa{} and compare them with each other and with  the \PoSp{} of the unresolved CXB in XBOOTES.
		We do the comparison both in the units of flux [(cts s$^{-1}$)$^2$ deg$^{-2}$] 
		and of the squared fractional RMS [deg$^2$] (\Fref{fig:PS_eSoft_Extra03}).
		The former characterizes the significance of the additive contamination (the instrumental background), while the latter characterizes the role of the multiplicative factor (the non-uniformity of the detector efficiency).
		In both cases, the \PoSp{} of the instrumental background is  by more than an order of magnitude smaller then the \PoSp{} of the unresolved CXB at any spatial frequency considered here.
		This excludes a possibility of any significant contamination of the \PoSp{} of unresolved CXB due to spatial non-uniformity of the instrumental background or detector efficiency.

		Residual counts from the resolved sources on average amount to $\approx 2.6$ counts per image in the \eSoftB{} (\Sref{ss:PointScr}).
		This should be compared to the total of $\sim 600$ counts per image of unresolved emission ($\sim 200$ from unresolved CXB and $\sim 400$ from the instrumental background).
		This is obviously too small to produce fractional RMS of $\approx 40\,\%$ on the $\sim$arcminutes angular scales (\Sref{sss:RMS}).
		We also repeated the  entire analysis with the point source exclusion radius of $r=20\arcsec$ and did not find any significant changes in the \PoSp{}.

		Furthermore, the energy spectrum of the  LSS signal is much steeper than the energy spectra of the instrumental background and of the resolved sources  (\Sref{sss:Exc_Spec}).
		In the case of the resolved sources the energy spectrum of the residual counts in the wing of the PSF is yet harder than the intrinsic spectrum of  sources, due to the energy dependence of the PSF width\footnote{\url{http://cxc.harvard.edu/proposer/POG/html/chap4.html#fg:hrma_ee_pointsource}}.
		This adds further confidence in excluding  the  contamination by the instrumental effects. 
				
		Several other less significant systematic effects, such as the mask effect, are investigated in \Aref{app:Sys}.

		\subsection{Observational evidences}
				
		\Mc{If the detected LSS signal  is caused by one of the known X-ray source populations,  it should be present in the resolved part of the CXB as well.}
		Hence, one should expect that the  LSS signal is enhanced if (some fraction of) resolved sources are not removed from the analyzed images.

		In the course of our data preparation procedure we remove two types of resolved sources: point and extended sources.
		The point sources are associated predominantly with AGN and a some small contribution from normal galaxies.
		As the latter can not be always separated from the former, we investigate their effect together, noting that any LSS signal will be by far dominated by AGN. The extended resolved sources are \GCG{}.
		Below we will investigate  possible contribution of each source type to the large scale LSS signal.

			\begin{figure}
				\begin{center}
					\resizebox{\hsize}{!}{\includegraphics{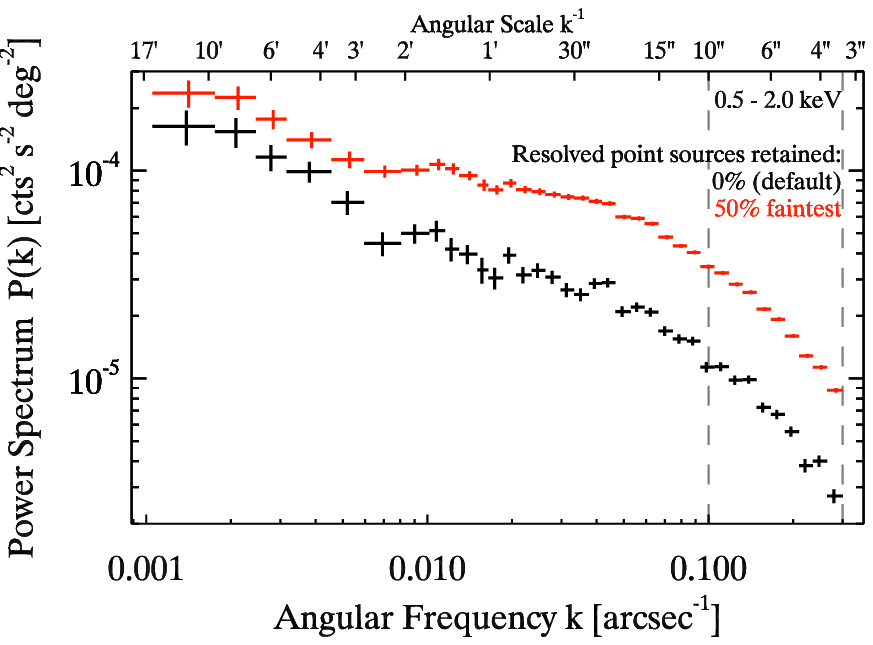}}
					\resizebox{\hsize}{!}{\includegraphics{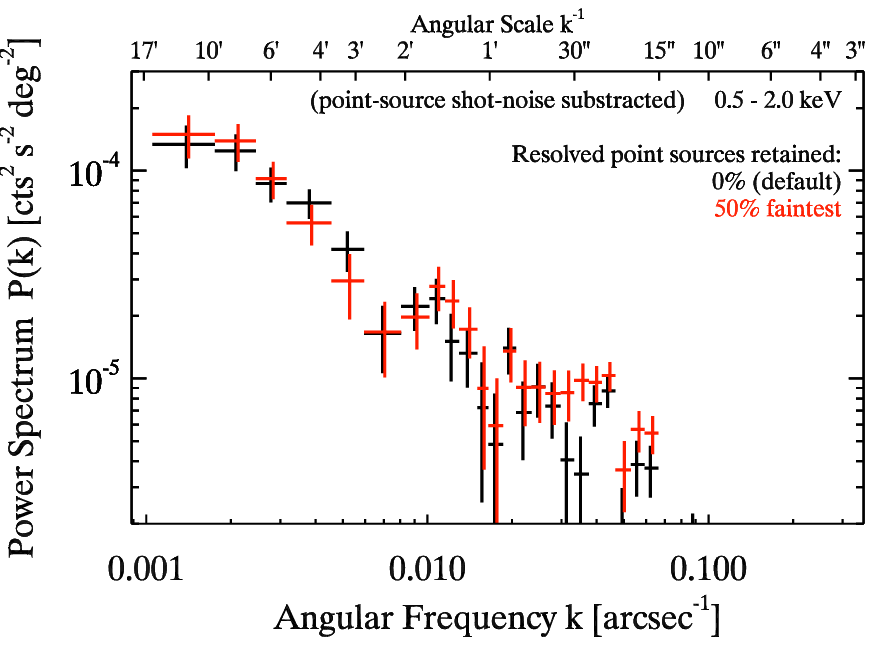}}
					\caption{\label{fig:Ex_PntScr_Limit}%
						\emph{Top}:
						The measured \PoSpa{} of the brightness fluctuations in the \eSoftB{} for different flux cuts for point source removal.
						\emph{Black} crosses: default case used throughout the paper, with all resolved sources removed.
						\emph{Red} crosses:
							the flux cut of $6.2\times10^{-15}\,\mathrm{erg\,cm^{-2}\,s^{-1}}$ for point sources was applied to the images, retaining $\approx50\,\%$ of faintest resolved point sources.
						\emph{Bottom}:
							Same as top panel but after subtracting the point-source shot noise (as described in \Sref{ss:lss_power_spectrum}).
							\Ac{The frequency range used to normalize the point-source shot-noise model is indicated with the two vertical dashed lines in the top panel.}
						The vertical bars are shifted slightly along the x-axis for visualization purposes. 
					}	
				\end{center}
			\end{figure}

		\subsubsection{AGN and normal galaxies}\label{ss:Ex_PntScr}

		In order to investigate the possible role of point sources (AGN and normal galaxies) we construct images keeping $50\,\%$ of faintest resolved point sources in the images.
		This corresponds to the flux cut of $6.2\times10^{-15}\,\mathrm{erg\,cm^{-2}\,s^{-1}}$ for the \eSoftB{}.
		With this flux cut, we on average retain 14 resolved point sources per observation, thus approximately doubling the average surface brightness. 
		All other data preparation steps are same as for the default case (\Sref{sec:DataProc}).

		The \PoSpa{} for the two cases are shown in the top panel of \Fref{fig:Ex_PntScr_Limit}.
		We can see that for the higher flux cut (red crosses) the point-source shot noise noticeably increased, as it should be expected (\Eref{eq:PS_SN}).
		In order to study the amplitude of the LSS signal, we subtract the point-source shot noise from the both spectra, as described in \Sref{ss:lss_power_spectrum}.
		The result is shown in the bottom panel of \Fref{fig:Ex_PntScr_Limit}, where we can see that the two spectra are nearly identical. 
		The small difference at small angular scales $\la2\arcmin$ is likely related to the imperfect shot noise subtraction.
		
		We thus conclude that the LSS signal at large angular scales can not be produced by AGN.

		\subsubsection{Galaxy clusters \& groups}\label{sss:Ex_GC_limit}

		We perform similar analysis for resolved extended sources.
		Due to their relatively small number in the survey (43  sources, see \Ken), we compute the \PoSp{} retaining  all resolved extended sources.
		However, the source area of almost every resolved extended source overlaps with an exclusion area of at least one resolved point source, leading to the loss of about $\sim 50\,\%$ of counts from  extended sources.
		To preserve the extended sources counts, in this analysis we reduced the circular exclusion area to the constant radius of $20\arcsec$ for all resolved point-source sources located within the exclusion area of resolved extended sources.
		 This does not contaminate the signal as resolved point sources do not contribute to the \PoSp{} in the frequency range of interest, apart from their shot-noise component (also see \Sref{ss:Ex_PntScr}). 		
			
		Since not all of 118 considered XBOOTES fields contain a \rES{},
		we have a possibility to   compute the \PoSp{} for three different field selections.
		This gives us a more detailed view of the dependence of the \PoSp{} on the presence or absence of resolved extended sources.  
		Selection A includes all 118 fields.
		Selection B includes only fields with resolved extended sources.
			To do the filtering we used the catalog  
			of extended sources from  \citet[Table~1]{Vajgel2014} which is a result of a more strict selection  than the catalog of \Ken{} (Table~1).
			This selection includes  26 fields containing  29 individual \rES{}s.
		Finally, selection C is composed of fields without resolved extended sources. The filtering is based on the extended source catalogs of both \citet[Table~1]{Vajgel2014} and \Ken{} (Table~1).
			Furthermore, we also excluded from this selection fields containing the exclusion area of an extended sources located in  an adjacent field.  
			This selection is composed of 76 fields.	 
		It can be considered as our control sample, and we expect its \PoSp{} to be consistent with the default one (from which all resolved sources are removed).

		\begin{figure}
		\begin{center} 
			\resizebox{\hsize}{!}{\includegraphics{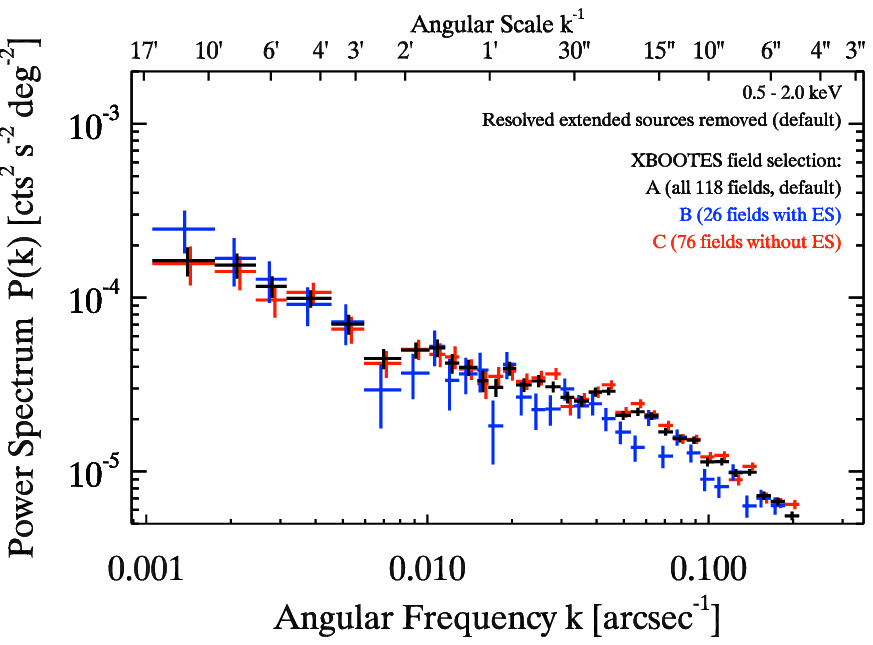}}
			\resizebox{\hsize}{!}{\includegraphics{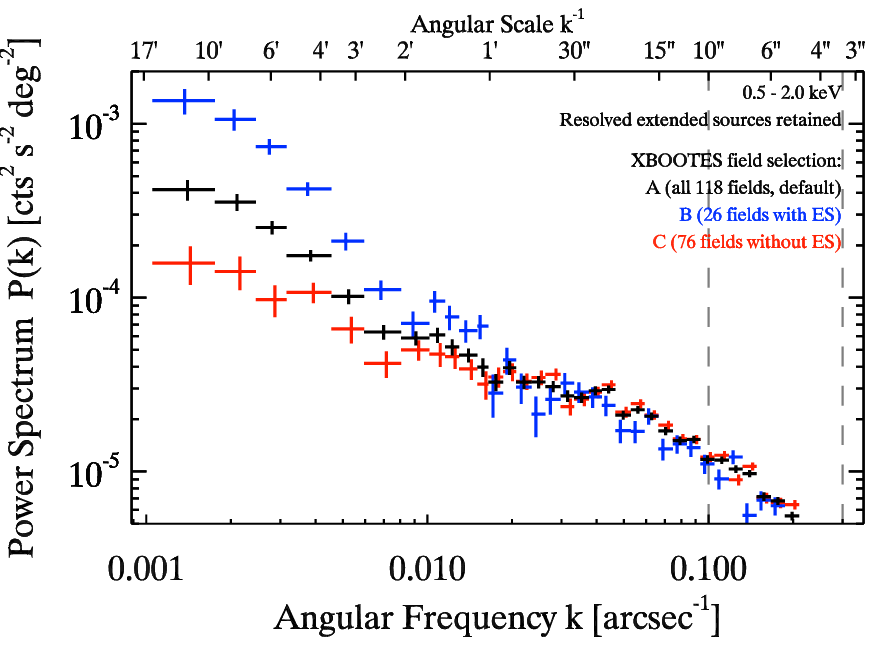}}
			\resizebox{\hsize}{!}{\includegraphics{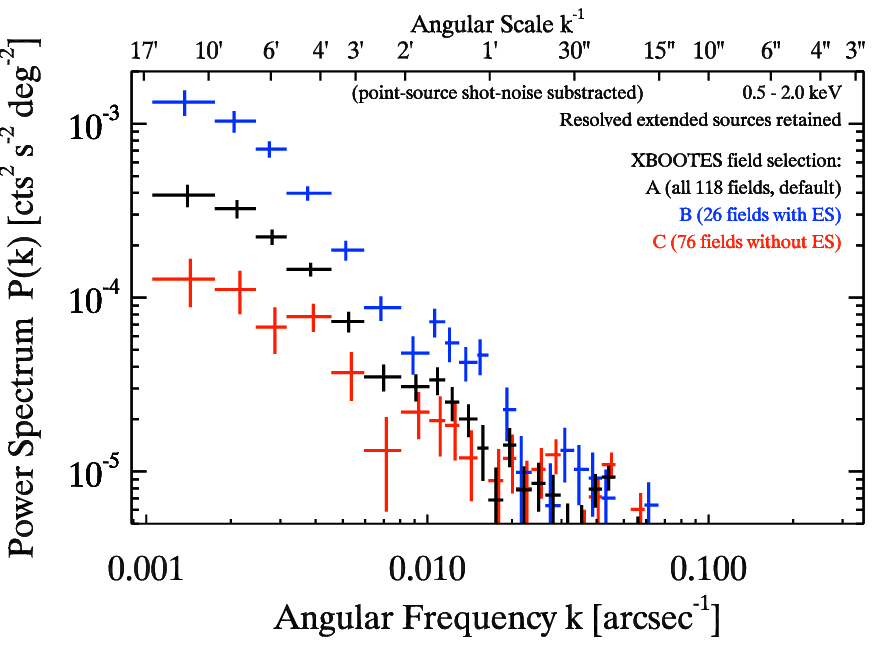}}
			\caption{\label{fig:Ex_ES_Limit}%
				The measured \PoSpa{} in the \eSoftB{} for different masks and different XBOOTES field selections.
				\emph{Top} panel:	default mask used throughout the paper, where all resolved sources (point-like and extended) are excluded.
				\emph{Middle} panel: special mask described in \Sref{sss:Ex_GC_limit}, retaining all resolved extended sources.
				\emph{Bottom}:
					Same as middle panel but after subtracting the point-source shot noise (as described in \Sref{ss:lss_power_spectrum}).
					\Ac{The frequency range used to normalize the point-source shot-noise model is  indicated with the two vertical dashed lines in the middle panel.}
				The vertical bars for selection B and C are shifted slightly along the x-axis for visualization purposes.
			}	
		\end{center}
		\end{figure}

		In \Fref{fig:Ex_ES_Limit} we show the \PoSp{} for the default mask (top panel, excluding all resolved sources) and the special mask retaining all resolved extended sources  (middle and bottom panels) for our three different field selections.
		From the top panel we can see a good agreement between different field selections%
			\footnote{
				The \PoSp{} of selection B (blue crosses in \Fref{fig:Ex_ES_Limit}) shows a  weaker point source shot-noise, indicated by the lower \PoSp{} at angular scales of $\lesssim2\arcmin$, in comparison with the two other field selections.
				This results from the fact that the average exposure per field in selection B is slightly higher than for the other selections, which leads to a smaller sensitivity limit for point sources.
			}
		On the other hand,  in the middle and bottom panels we can see  that the \PoSp{} increases very significantly for angular scales of $\gtrsim2\arcmin$ for field selection A and B, when we retain  all \rES{} in the images.
		This strongly suggests that the LSS signal at large angular scales is produced by extended sources (\GCG), resolved as well as unresolved.	
		
		It is interesting to note that  not only the amplitude but also the shape of the LSS power spectrum depends on the flux cut for extended sources. While the  power spectrum of unresolved CXB has a slope of $-0.8\pm0.1$ (see also \Fref{fig:excess}), after we retain resolved extended sources, the best fit slope of the power spectrum changes to $-1.2\pm0.1$ (see also \Fref{fig:Ex_rES_Cluster}).

			\begin{figure}
				\begin{center}
				\resizebox{\hsize}{!}{\includegraphics{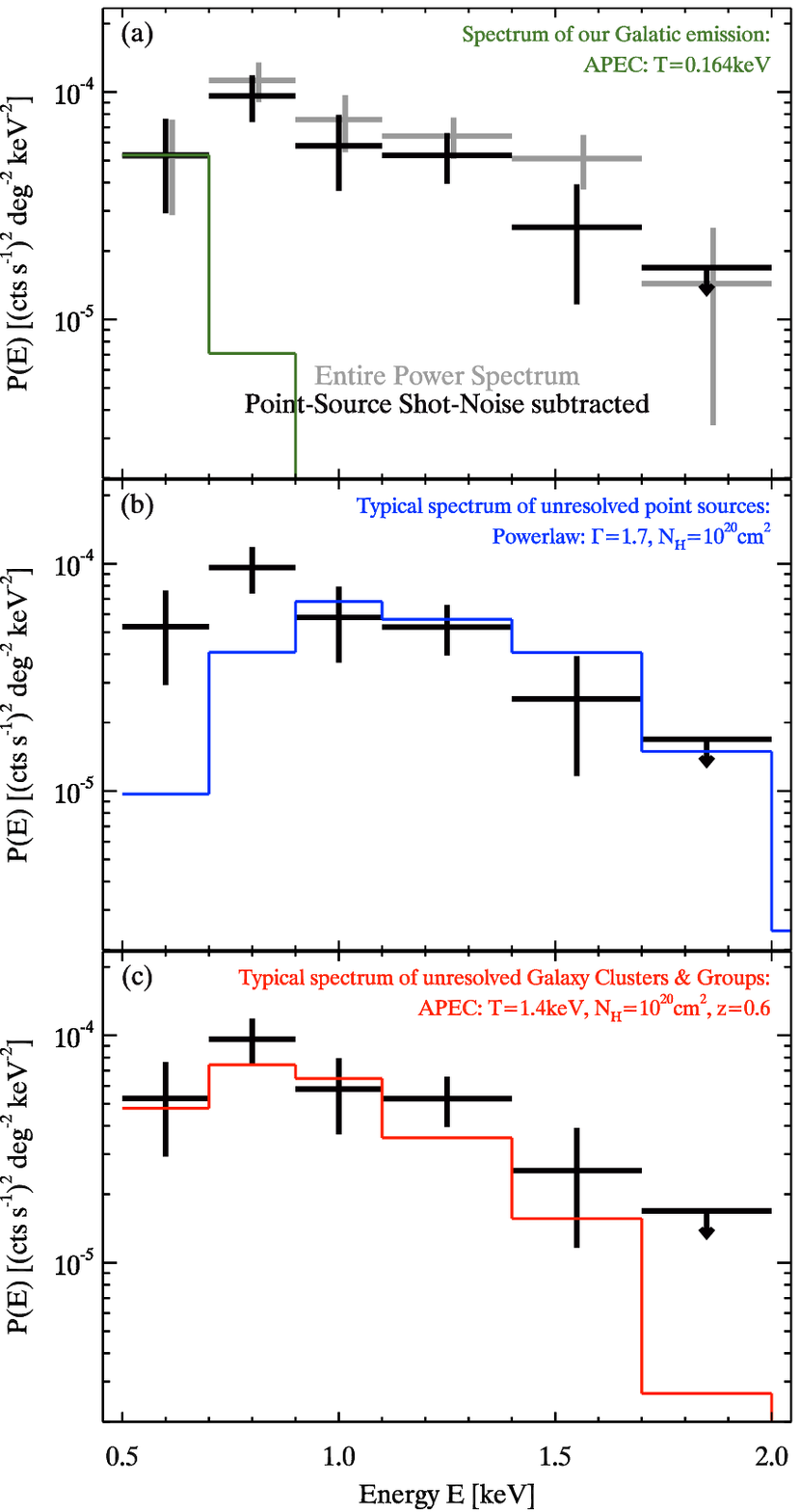}}
				\caption{\label{fig:PS_spec_Excess}%
					Average power of  CXB fluctuations versus energy along with predictions of different spectral models representing 
						the Galactic emission (green, panel \textbf{(a)}),
						AGN and normal galaxies (blue, panel \textbf{(b)}), and
						unresolved \GCG{} (red, panel \textbf{(c)}).
					The power is computed from the frequency interval $[0.002,0.006[\InvArcSec$ ($\sim3\arcmin - 10\arcmin$) of the measured \PoSp{} (\Eref{eq:PS_Excess}).
					The point-source shot noise is subtracted (as in \Sref{ss:lss_power_spectrum}), except for the spectrum shown in the upper panel in grey.
					The points, which are consistent with zero within one standard deviation, are replaced with $1\sigma$ upper limits.
				}	
				\end{center}
			\end{figure}

		\subsubsection{Energy dependence of the CXB fluctuations}  \label{sss:Exc_Spec}

		The conclusion regarding the association of the LSS signal with \GCG{} is further supported by the analysis of the energy spectrum of the CXB fluctuations. 
		To construct  the latter,
		we first compute a series of \PoSpa{} in contiguous energy bins in the \eSoftB{} and then for each energy bin compute  the average power in the range of angular frequencies  of interest, subtracting the point-source shot-noise as follows:
		\begin{align} \label{eq:PS_Excess}
			\langle P(E) \rangle = & 
				\frac{ \Sigma_{k_1}^{k_2} \; P(k)}{\Sigma_{k_1}^{k_2}}
				- \frac{ \Sigma_{k_3}^{k_4} \; P(k)}{\Sigma_{k_3}^{k_4}  P_\mathrm{PSF}(k)} 
				\; \frac{ \Sigma_{k_1}^{k_2} P_\mathrm{PSF}(k)}{\Sigma_{k_1}^{k_2}} 
			\text{ .}	
		\end{align}
		where  $[k_1,k_2]$ is the frequency interval for which we used  $[0.002,0.006[\InvArcSec$ ($\sim3\arcmin - 8\arcmin$), 
		$P_\mathrm{PSF}(k)$ is the  PSF-smearing model (\Aref{app:PSFsmear}), 	
		and for estimating the point-source shot-noise we use the frequency interval $[k_3,k_4]=[0.1,0.3[\InvArcSec$ ($\sim3\arcsec - 10\arcsec$). 
		This definition is equivalent to the \Mc{re-normalized} AGN shot noise model of \Sref{ss:lss_power_spectrum}.
				
		The so computed average power is plotted versus energy in \Fref{fig:PS_spec_Excess}.
		The panel \textbf{(a)} also shows the energy spectrum before subtracting the point-source shot-noise (gray crosses), demonstrating that for the chosen range of angular frequencies its effect is not very significant.
		Note that taking a square root in \Eqref{eq:PS_Excess} one could obtain a quantity similar to a normal energy spectrum.
		However, the rather complicated procedure involved in computing the energy spectrum makes the error distribution rather complex and non-Gaussian, so that  the conventional spectral fitting techniques may not be directly applied.
		We therefore chose to consider the energy dependence of the average power expressed in the units of square of the instrumental flux and compare it with predictions of various spectral models.
		The latter were computed by convolving a spectral model with the energy response of \chandra{} and squaring the result.
		The normalizations of the models shown below are arbitrary. 
		In computing the energy spectra of extragalactic sources we assumed the Galactic absorption of $\NH=10^{20}\,\mathrm{cm^{-2}}$ \citep[\Ken]{Kalberla2005}.

		To illustrate the amplitude of the possible contamination by the Galactic emission we plot with green lines in panel \textbf{(a)} the \APEC{} model with $T=0.164\,\mathrm{keV}$ (\Sref{ss:Espec_Scr}), normalizing it to the "flux"
		of the lowest energy bin (\eGal).
		This exercise shows that even in the extremely unlikely case that the lowest energy bin is entirely contaminated by the fluctuations of the surface brightness of the Galactic emission, their contribution to the adjacent $0.7-0.9\,\mathrm{keV}$ energy bin \Mc{will not exceed $\approx7\,\%$}
		and their contribution to the entire \eSoftB{} will not be larger than $\approx7\,\%$.
		This justifies using the \eSoftB{} for fluctuation studies of the extragalactic CXB.
						
		In panel \textbf{(b)} we show  a typical energy spectrum of AGN,
		representing it with a power law  with a photon-index of $\Gamma=1.7$ and modified by the Galactic absorption \citep[e.g.][]{Reynolds2014,Ueda2014,Yang2015}. 
		Note that the shape of the power law spectrum is not affected by the redshift. 
		We can see that the energy dependence of power predicted by this power law model is significantly harder that the observed dependence. 
		In order to describe the entire energy spectrum with a power law, one would  need to assume a very steep slope  of $\Gamma\approx3$, which is not feasible for AGN.
		
		In panel \textbf{(c)} we plot  a typical spectrum of \GCG{}, for which we used an \APEC{} model modified by the Galactic absorption. 
		For the temperature and redshift we assumed  $T=1.4\,\mathrm{keV}$ and $z=0.6$, which are the median values for the unresolved population of \GCG{} in XBOOTES (\Sref{ss:UnresoAGN}).
		We can see that the model describes reasonable well the observed energy dependence of the average power of the CXB fluctuations. 
			
		To conclude, results of this analysis support the conclusion made earlier in this section that the \Mc{LSS signal is associated with the emission from \GCG{}, but not from AGN}.

	\subsection{Comparison with theoretical predictions}  \label{sss:Ex_GC_Clus}
	
		\subsubsection{AGN} \label{sss:Int_AGN}
		The inability to explain the detected  LSS signal  by the clustering signal of AGN is not surprising, given our knowledge about their clustering properties.  To demonstrate this, we compute the \THT{} for unresolved AGN using  Limber's approximation (assuming small angles, $k^{-1} \ll 1\, \mathrm{rad} $) as follows:
			\begin{align} \label{eq:PS_2H_AGN}
				P^{(2H)}_\mathrm{AGN}(k) = & \int\d z
					\; \left( \frac{\d S}{\d z} \right)^2 
					\; \left( \frac{\d^2 V(z)}{\d z \d\Omega} \right)^{-1}  \notag \\ & \times
						\;  P_\mathrm{3D,AGN}\left(k_\mathrm{3D} = \tfrac{k}{\beta}\,\tfrac{\alpha}{d(z)},z\right)
				\text{ .}	
			\end{align}
		where 
		$\d S(z)/\d z$ [erg cm$^{-2}$ s$^{-1}$ deg$^{-2}$] is the AGN flux production rate  as function of redshift, defined by \Eqref{eq:dSdz} in \Sref{ss:UnresoAGN},
		$\d^2 V(z)/\d z/\d\Omega$ [Mpc$^{3}$ h$^{-3}$ deg$^{-2}$] is the co-moving volume element, and
		$d(z)$ [Mpc h$^{-1}$] is the co-moving distance to redshift $z$ \citep[e.g.][]{Hogg1999}.
		The $\alpha$ and $\beta$ are equal to 1, if there is a $2\pi$ in the exponent of the  Fourier transform, and they are equal to $2\pi$ otherwise.
		The AGN 3D \PoSp{} is computed as following \citep[e.g.][]{Cooray2002}:
			\begin{align} \label{eq:PS_3D_AGN}
				P_\mathrm{3D,AGN}(k_\mathrm{3D},z) = b(M_\mathrm{eff},z)^2 \, g(z)^2 \,  P_\mathrm{lin}(k_\mathrm{3D})
				\text{ .}	
			\end{align}
		Hereby,
		$P_\mathrm{lin}(k_\mathrm{3D})$ [Mpc$^{3}$ h$^{-3}$] is the 3D linear $\Lambda$CDM \PoSp{} at $z=0$, which we computed using the fitting formulae of \citet{Eisenstein1998}, 
		$g(z)$ is the linear growth function \citep[e.g.][]{Dodelson2003}, and 
		$b(M_\mathrm{eff},z)$ is the AGN linear clustering bias factor, computed with the analytical model of \citet{Sheth2001}.
		For the effective mass $M_\mathrm{eff}$ of the DMH, where the AGN reside, we use $2\times10^{13}\,\mathrm{M_{\sun}\,h^{-1}}$,
		which is consistent with recent observations up to $z\sim3$ \citep[e.g.][]{Allevato2011,Krumpe2012,Mountrichas2013}.
		
		The predicted \THT{} of unresolved AGN is compared with the  measured \PoSp{} in \Fref{fig:agn_tht},
		which demonstrates that the predicted signal is by nearly two orders of magnitude smaller that the observed one.
		In order to match the amplitude of the observed LSS signal, one would need to assume the effective DMH mass to be much larger than $M_\mathrm{eff}\sim 10^{14}\,\mathrm{M_{\sun}\,h^{-1}}$ \Ac{for the unresolved AGN population of XBOOTES}, which is physically unrealistic \citep[e.g.][]{Cappelluti2012,Krumpe2013}. 
		\Mc{Furthermore,  the expected power spectrum of the AGN two-halo term is significantly steeper than the observed power spectrum.
		In the $30\arcsec - 10\arcmin$ angular scales range, the former has an average slope of $-2.3$  while the measured power spectrum has a best-fit slope of $-0.8\pm0.1$ (\Fref{fig:excess}).  } 
		
		\Mc{At sub-arcminute angular scales, the shape of the power spectrum is sufficiently well described by the shot noise of unresolved point sources, modified by the PSF smearing effects (\Fref{fig:PS_eSoft}).
		The residuals seen at $\sim 15\arcsec-30\arcsec$ angular scales (\Fref{fig:Ex_rES_Cluster}) are likely due to the correlation between adjacent Fourier modes caused by the mask.
		Thus, we do not see any significant evidence for the signal in the power spectrum due to the \OHT{} of AGN.
		This is consistent with the  theoretical prediction that  low luminosity AGN reside alone in their DMHs \citep[e.g.][]{Leauthaud2015,Fanidakis2013}.
		We will defer any quantitative constrains of the \Ac{halo occupation distribution of AGN} based on these data to future work.}

		\subsubsection{Clusters and groups of galaxies}	\label{sss:Int_GC}	
		The flux production rate for unresolved \GCG{} at the depth of XBOOTES survey peaks at the redshift $z\approx 0.2$ with the median value of $z\approx 0.6$ (\Sref{ss:UnresoAGN}). 
		At these redshifts, the angular scales of $\lesssim 10\arcmin$ correspond to spatial scales \Mc{of $\lesssim\ $few Mpc, suggesting that the LSS signal is produced by the internal structure of ICM within the same dark matter halo  (i.e. their\  \OHT{}), rather than by the cross-correlation between different objects (\THT{}).
		This conclusion is supported by the results of analytical calculations of the X-ray power spectrum of groups and clusters of galaxies which demonstrated  that at angular scales below $\sim10\arcmin$ their two-halo term can be neglected \citep[e.g.][]{Komatsu1999,Cheng2004}.}

		\Mc{ The \OHT{} on the  angular scales of $\gtrsim 1\arcmin$  is dominated by the nearby objects, located at  redshift $z\lesssim0.1$ \citep{Cheng2004}. 
		At these redshifts, angular scales of $\sim 5\arcmin-10\arcmin$ correspond to spatial scales of  $\sim 0.5-1$~Mpc,
		i.e. of the order of the $R_{200}$ of a DMH with a mass of $M_{200}\sim10^{13}-10^{14}\,\mathrm{M_{\sun}}$.
		%
		%
		Thus, the \PoSp{} of CXB fluctuations carries information about ICM  in cluster  outskirts.
		The shape of the \PoSp{}  is determined by the intrinsic structure of the ICM  convolved with the redshift distribution of  unresolved clusters and groups of galaxies. The latter depends on the flux cut for the resolved objects, opening prospects for redshift-resolved studies, provided the data of sufficient depth. The normalisation of the power spectrum is proportional to the square of the volume density of clusters and groups of galaxies, making the powers spectrum  a sensitive diagnostics of the volume density of these objects.}
				
		At small angular frequencies corresponding to \Ac{large angular scales,}
		the \PoSp{} is expected to flatten \Ac{based on the model of \citet{Diego2003}, while the model of \citet{Cheng2004} does not show such a strong feature.}
		These angular scales are outside the frequency range studied here, but should become accessible when the full range of scales provided by the XBOOTES ($\sim3$~deg) and other wide angle surveys will be utilized. 

		Thus, the \PoSp{} of CXB fluctuations could potentially provide a new tool to probe the structure of ICM, out to the linear scales beyond $R_{500}$,  which are expensive to study with direct imaging observations.
		This is a valuable possibility. 
		Indeed, due to the complex nature of ICM, theoretical predictions suffer from large uncertainties \citep[e.g.][]{Rosati2002,Kravtsov2012}.
		Depending on the assumed characteristics of  gas cooling and heating, 
		the predictions of the structure of ICM and for the clustering strength in particular for the \OHT{} can vary dramatically,
		as analytical studies \citep[e.g.][]{Cheng2004} 
		and cosmological hydrodynamical simulations \citep[e.g.][]{Roncarelli2012} 
		show.
		For the same reason it is also difficult to accurately predict  the total contribution of \GCG{} to the CXB 
		(e.g.\ \citealt{Roncarelli2006}).
		The latter is also difficult to achieve  observationally, given the insufficient depth of the current measurements of the $\log N - \log S$ of \GCG{}, reaching the flux limit of $\sim10^{-16}\,\mathrm{erg\,cm^{-2}\,s^{-1}}$ at the best \citep{Finoguenov2015}.
		Due to the steep slope of the $\log N - \log S$ a significant fraction of the total emission from \GCG{} is produced by below this flux limit.

		
			\begin{figure}
				\begin{center} 
					\resizebox{\hsize}{!}{\includegraphics{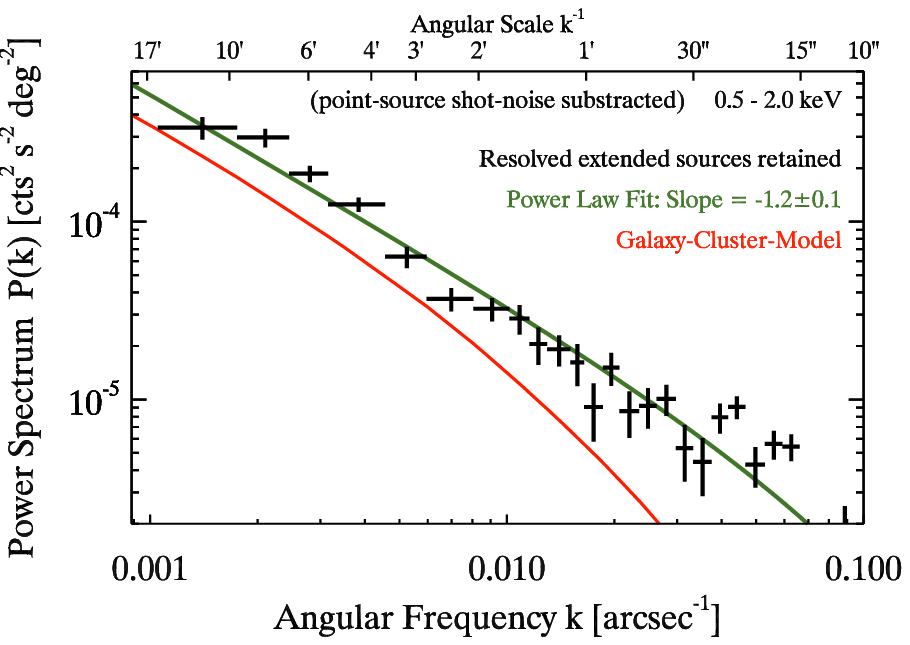}}
					\caption{\label{fig:Ex_rES_Cluster}%
						The \PoSp{} of the brightness fluctuations of the unresolved CXB of XBOOTES in the \eSoftB{}, when all \rES{} are included in the flux map (using our special mask, \Sref{sss:Ex_GC_limit}).
						The point-source shot noise is subtracted from the measured \PoSp{} (\Sref{ss:lss_power_spectrum}).
						The green solid line shows a power law with the slope of $-1.2$, which gives best fit to the observed power spectrum in the angular scale range of $30\arcsec - 10\arcmin$.
						The red curve shows the  model for \GCG{} based on results of \citet{Cheng2004},					
					}	
				\end{center}
			\end{figure}

		Keeping these uncertainties in mind, we compare our measured \PoSp{}  with theoretical predictions by \citet{Cheng2004}, whose calculations are based on a simple model assuming that ICM is in hydrostatic equilibrium.
		This model is in a good agreement with hydrodynamical simulations of \citet{Roncarelli2006}, including more complex physical processes, such as star-formation and supernova feedback. 
		Since \citet{Cheng2004} computed \PoSp{} of all \GCG{} without any flux cut, we will be comparing their results with the corresponding \PoSp{} from \Sref{sss:Ex_GC_limit}, including all extended sources.
		This \PoSp{} is shown with  black symbols in the bottom panel of \Fref{fig:Ex_ES_Limit}.
		
		For  comparison with our measurement, the \PoSp{} from \citet{Cheng2004} needs to be converted to flux units, the conversion being achieved by multiplying their \PoSp{} with the square of the surface brightness of \GCG{}.
		To estimate the latter, we compute the combined surface brightness of resolved extended objects from the catalog of \Ken{} above the flux cut of $S=3\times10^{-14}\,\mathrm{erg\,cm^{-2}\,s^{-1}}$ (\eSoftB{}), obtaining $\sim1.0\times10^{-13}\,\mathrm{erg\,cm^{-2}\,s^{-1}\,deg^{-2}}$.
		We estimate the surface brightness of unresolved \GCG{} using the results of   \Sref{sss:Unr_Exgal} and obtain $\sim2.5\times10^{-13}\,\mathrm{erg\,cm^{-2}\,s^{-1}\,deg^{-2}}$.
		Hence, the total surface brightness of \GCG{}  is $\sim3.5\times10^{-13}\,\mathrm{erg\,cm^{-2}\,s^{-1}\,deg^{-2}}$.
		We convert this to instrumental units with a \APEC{} model modified by the Galactic absorption. 
		For the temperature and redshift we assume  $T=2.9\,\mathrm{keV}$ and $z=0.4$, which are the median values for the total population of \GCG{} (\Sref{ss:UnresoAGN}).
		The conversion depends weakly ($\lesssim 10\,\%$) on the assumed temperature for $T\gtrsim1.5\,\mathrm{keV}$.

		Comparison  with our measurement in  \Fref{fig:Ex_rES_Cluster} shows  that the model of \citet{Cheng2004}  predicts the shape of the \PoSp{}
		remarkably well.
		However the model \Mc{normalisation} is by a factor of $\approx 2$ smaller.
		The discrepancy in \Mc{normalisation} is not too dramatic given the number of uncertainties and simplifications involved in the model of \citet{Cheng2004}. 
		The easiest explanation for it could be the method for computing the model normalisation we used above. 
		Indeed, the estimate of the surface brightness of unresolved \GCG{} depends on the assumed slope of their $\log N - \log S$ distribution, for which we used \citet{Ebeling1997} results. 
		As explained  in \Sref{sss:Unr_Exgal}, if we increased the slope of the XLF of \citet{Ebeling1997} by $\sim10\,\%$, the flux from unresolved \GCG{} would increase by $\sim50\,\%$ and the \citet{Cheng2004} model would match our measured LSS signal within $\sim 30\,\%$ accuracy.

		The \PoSp{} of \GCG{} can be modified significantly by non-gravitational effects \citep[e.g.\  Fig.~3 in ][]{Cheng2004}. 
		Cooling and heating of the ICM can significantly affect its surface brightness distribution which can change the \PoSp{} by as much as an  order of magnitude. Therefore the agreement between a simple semi-analytical model and our data is remarkable and demonstrates the potential of the \PoSp{} of CXB fluctuations in constraining theoretical models.

		\begin{figure}
			\begin{center}
			\includegraphics[width=0.45\textwidth]{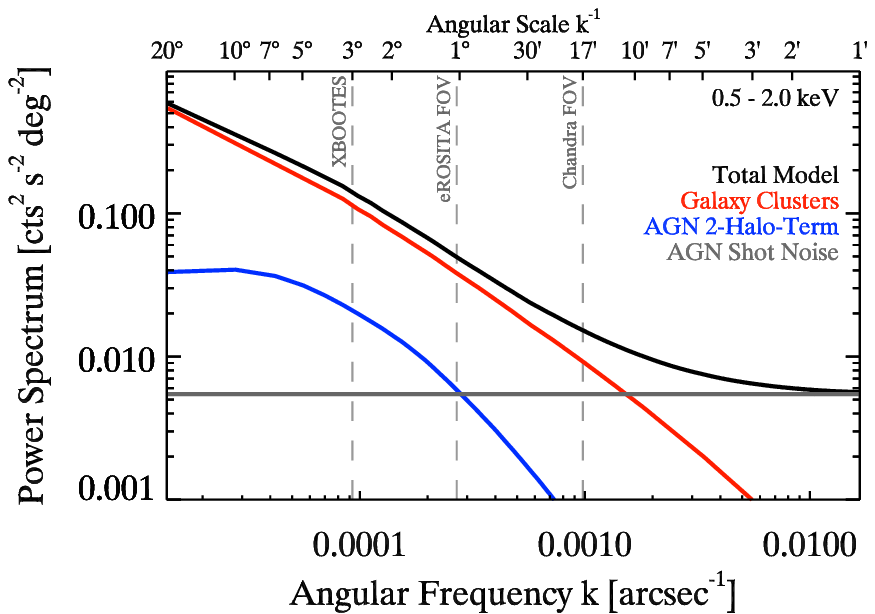} 
			\caption{\label{fig:PS_eRosita}%
			Predicted  \PoSp{} of the brightness fluctuations of the CXB in the SRG/eROSITA all-sky survey in the \eSoftB{} for the sky-averaged case.
			The curves are: 
				clustering signal of (resolved and unresolved) galaxy clusters (red curve), 
				\THT{} (blue curve), and shot-noise level (gray line) for AGN,
				and the sum of all components (black curve). 	
				The PSF-smearing is not taken into account but should only become important at angular scales below $\sim2\arcmin$.
			The  vertical dashed lines indicate the size of the instruments' FOVs and of the XBOOTES survey. 
			The \PoSp{} is shown in instrumental units and we used the response matrix of SRG/eROSITA (\texttt{erosita\_iv\_7telfov\_ff.rsp}) for the conversion.
			}	
			\end{center}
		\end{figure}
		
	\section{SRG/eROSITA forecast}
		The upcoming all-sky survey of SRG/eROSITA (eRASS, $\sim$2017-2021) will lead us in a new area of high precision LSS studies with $\sim 3$ million of resolved AGN  \citep{Kolodzig2012,Kolodzig2013,Huetsi2013} and $\sim 100\,000$ galaxy clusters \citep{eROSITA.SB}.  
		However, eRASS will also produce an all-sky map of the unresolved CXB with the usable area of the order of $\sim34\,000$~deg$^2$ (excluding the Galactic plane, $|b|>10^\circ$) and the point-source \Acc{sensitivity} of  $S_{\eSoft}\approx1.1\times10^{-14}$ $\mathrm{erg\,cm^{-2}\,s^{-1}}$ \Acc{(sky average)}.
		With  the  FOV-averaged PSF of $\sim30\arcsec$ \citep{eROSITA.SB}, it will cover 
		a very large range of angular scales,  from $\sim$arcminute to the  full sky.
		Given the PSF size, a point-source exclusion region radius of  $140\arcsec$ would be sufficient, leaving unmasked about $\approx 60\,\%$ of the sky area ($\sim21\,000$~deg$^2$).
		
		\Acc{Due to the survey strategy of eRASS, it will also have deep survey regions at the northern and southern ecliptic poles \citep[NEP and SEP,][]{eROSITA.SB}.
		These regions are of particular interested since they will offer a comparable point-source sensitivity as the largest surveys of the current generation of X-ray telescopes, e.g.\ XBOOTES ($\sim9$~deg$^2$, \chandra{}, \Ken{}), Stripe 82 \citep[$\sim31$~deg$^2$, \xmm,][]{LaMassa2016}, and XXL \citep[$\sim25\times2$~deg$^2$, \xmm,][]{XXL} but with a much larger sky area of $\sim100$~deg$^2$ or even up to $\sim1000$~deg$^2$.
		Further, the angular resolution within these regions can be improved by a factor of $\sim$two by only using the central part of the eROSITA-FOV, making it comparable to the the angular resolution of \xmm{}\footnote{\url{http://heasarc.gsfc.nasa.gov/docs/xmm/uhb/onaxisxraypsf.html}}.}
		
		\Acc{Given these features of eRASS, it} will permit not only to dramatically broaden the range of the angular scales, but  also to  improve the accuracy of the measurement of the surface brightness fluctuations of the unresolved CXB, increasing in the S/N ratio by a factor of about $\sim 50$, as compared to XBOOTES data.
		This will  allow one to test clustering models of \Acc{unresolved} AGN and galaxy clusters to \Acc{an unprecedented} precision.
		Importantly, these studies do not need redshifts of objects i.e. are independent on the optical follow-up studies. 
		
		In \Fref{fig:PS_eRosita} we show the expected \PoSp{} of CXB fluctuations in eRASS in the \eSoftB{} for the sky-averaged case.
		The shot-noise level (gray line) and \THT{} (blue curve) of unresolved AGN were computed as described in \Sref{ss:shot_noise} and \ref{sss:Int_AGN}, respectively, using the sky-averaged point-source sensitivity of eRASS.
		The power spectrum of (resolved and unresolved) clusters of galaxies (red curve) is based on the theoretical prediction by \citet{Cheng2004}, which was normalized as described in \Sref{sss:Int_GC}.
		One can see from \Fref{fig:PS_eRosita} that the CXB brightness fluctuations will be dominated by the clustering signal of \GCG{} at angular scales larger than $\sim10\arcmin$.
		At smaller angular scales the dominant contribution will come from AGN.
		The unresolved AGN  population of eRASS will be located at a median redshift of $z\approx1.2$ and will have a moderate median luminosity of $L_{\eSoft} \sim 10^{43.4}\,\mathrm{erg\;s^{-1}}$.
		Given the angular resolution of SRG/eROSITA and the expected high S/N of the \PoSp{}, one should be able to measure or constrain the \OHT{} of AGN at the angular scales of $\sim1\arcmin-10\arcmin$ to a high precision.
		This may permit us to finally measure or constrain the mean occupation number of AGN per DMH, which is an important parameter for the  AGN triggering and fueling theories which  has not yet been reliably  constrained  \citep[e.g.][]{Cappelluti2012,Krumpe2013,Leauthaud2015}.

	\section{Summary}
		Surface brightness fluctuations of the unresolved CXB present  a great opportunity to study  faint source populations
		which are yet beyond the reach of more conventional studies of resolved sources.
		The renaissance of this field   was  facilitated by the fact that wide angle  X-rays surveys covering tens of deg$^2$ have been  undertaken by the modern X-ray telescopes aboard \chandra{} and XMM-Newton, featuring  superb angular resolution \citep[e.g.][]{Brandt2015}.
		In the work reported in this paper we  used the data of XBOOTES,  the presently largest continuous \chandra{} survey  covering the  area of $\sim9$~deg$^2$, to conduct   the most accurate measurement to date of the brightness fluctuations of unresolved CXB in the  angular scales ranging from $\sim3\arcsec$ to  $\sim 17\arcmin$. 
		
		The XBOOTES survey with its average exposure of $\sim5$~ksec per field  has an average point-source sensitivity of $\sim2\times10^{-15}\,\mathrm{erg\,cm^{-2}\,s^{-1}}$ in the \eSoftB{} (\Ken) and an average extended-source sensitivity of $\sim3\times10^{-14}\,\mathrm{erg\,cm^{-2}\,s^{-1}}$ (\Sref{sss:Unr_Exgal}).
		Given these flux limits, unresolved AGN and normal galaxies make the major contribution to the surface brightness of the unresolved CXB, accounting each for $\sim30\,\%$, with the galaxy clusters making a more modest contribution at the level of $\sim6-8\,\%$ (\Sref{sss:Unr_Exgal}).
		However,  estimates for normal galaxies and galaxy clusters are highly uncertain, thus explaining that about $\sim 1/3$ of the CXB flux remains unaccounted for in this calculation.
		At a point-source sensitivity level of $\sim10^{-15}\,\mathrm{erg\,cm^{-2}\,s^{-1}}$ normal galaxies are not expected to contribute significantly to the unresolved CXB fluctuations. 
		The unresolved AGN have a median redshift of  $z\sim1.0$ and median  luminosity of $L_{\eSoft} \sim 10^{42.6}\,\mathrm{erg\;s^{-1}}$ (\Sref{ss:UnresoAGN}).
		The unresolved galaxy clusters are about twice as close, with a median redshift of $z\sim0.6$ \Ac{(flux-weighted mean of $\left<z\right>\sim0.3$)} and have comparable median luminosity of $L_{\eSoft} \sim 10^{42.7}\,\mathrm{erg\;s^{-1}}$.
		With the standard scaling relations, these numbers correspond to an ICM temperature of $T\sim1.4\,\mathrm{keV}$, and a DMH mass of $M_{500}\sim10^{13.5}$ (\Sref{sss:Unr_Exgal}).

		After masking out resolved (point and extended) sources, we obtained the \PoSp{} of surface brightness fluctuations of unresolved CXB in the $3\arcsec-17\arcmin$ range of angular scales (\Fref{fig:PS_eSoft}).
		At sub-arcminute  angular scales, the obtained \PoSp{} is consistent with the predicted point source  shot noise of unresolved AGN, corrected for the PSF-smearing.
		However, at the  angular scales exceeding $\gtrsim1\arcmin$ we detect a clear and highly statistically significant LSS signal above the AGN shot noise.
		After subtracting the point-source shot noise, we obtain the \PoSp{} of the LSS signal, which follows an approximate power-law shape with the slope of $-0.8\pm0.1$ (\Fref{fig:excess})
		and has normalization corresponding to the fractional RMS variation of $(36\pm2)\,\%$. 

		The  detected LSS signal is by \Mc{almost two  orders} of magnitude stronger than that expected from the AGN \THT{} (\Fref{fig:agn_tht}).
		We present strong evidence that it is associated with the \Mc{ICM of unresolved galaxy clusters, namely with their one-halo term. }
		In particular, we show that the LSS signal is not present in the \PoSp{} of resolved AGN  (\Fref{fig:Ex_PntScr_Limit}),
		and it is much enhanced when resolved galaxy clusters are retained on the images (\Fref{fig:Ex_ES_Limit}). 
		The energy dependence of the mean power  of the CXB fluctuations at a $\sim$few arcmin angular scales is consistent with the energy spectrum of an optically thin plasma with temperature of $T=1.4$~keV redshifted to $z=0.6$, corresponding to the \Ac{average} ICM temperature and  \Ac{median} redshift of unresolved galaxy clusters in the XBOOTES survey (\Fref{fig:PS_spec_Excess}).
		The shape of the \PoSp{} can be remarkably well described by the model of \citet{Cheng2004},
		although the normalization of the theoretical spectrum is by about a factor of $\approx 2$ smaller than observed (\Fref{fig:Ex_rES_Cluster}).		
		
		\Mc{The power spectrum of fluctuations of unresolved CXB carries information about the ICM structure in the outskirts  (out to $\sim R_{200}$) of nearby ($z\lesssim0.1$) \GCG{}   \citep[e.g.][]{Cheng2004}.}
		These scales are difficult to reach by conventional studies based on the surface brightness distribution of individual and usually relatively nearby objects. 
			The shape of the observed \PoSp{} is determined by  the spatial structure  of ICM, and the redshift distribution of clusters and groups of galaxies, while its normalization is proportional to the square of their volume density.
		This underlines  the enormous  diagnostic potential  of the unresolved CXB fluctuation analysis.

	\section*{Acknowledgments}
		We have enjoined helpful discussions with M. Anderson, M. Krumpe, K. Helgason, N. Cappelluti, G. Hasinger, and N. Clerc.
		A. Kolodzig acknowledges support from and participation in the International Max-Planck Research School (IMPRS) on Astrophysics at the Ludwig-Maximilians University of Munich (LMU) and by China Postdoctoral Science Foundation, Grant No. 2016M590012.
		M. Gilfanov and R. Sunyaev acknowledge partial support by Russian Scientific Foundation (RNF), project 14-22-00271.
		The scientific results reported in this article are based on data obtained from the \chandra{} Data Archive.
		This research has made use of software provided by the \chandra{} X-ray Center (CXC) in the application package CIAO. 

	\appendix

	\section{Spectral model of the instrumental background} \label{app:SpecPB} 
		
\begin{figure}
					\resizebox{\hsize}{!}{\includegraphics[angle=270]{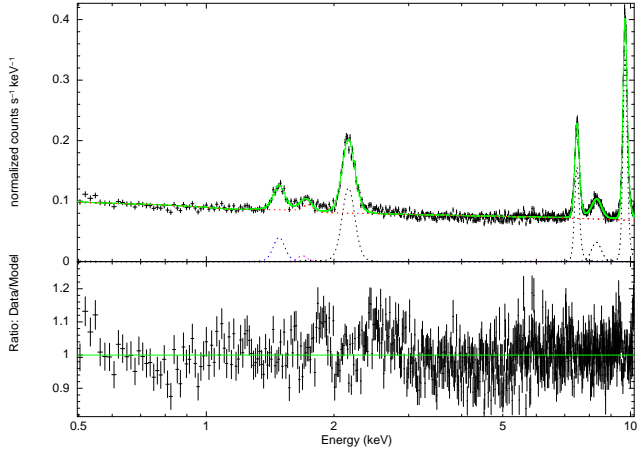}}   
					\caption{\label{fig:PB_spec}%
						Energy spectrum of the \acisi{} stowed background along with the instrumental background model.
						Black crosses: data points with one standard deviation as error-bars.
						Green Curve: Total background model.
						Dotted Curves: single components of the model (see \Tref{tab:PB_spec}).
					}
				\end{figure}

		In order to study the energy spectrum of the unresolved CXB, one needs a spectral model for the instrumental background.
		We create such a model using the \acisi{} stowed background data%
				\footnote{\url{http://cxc.harvard.edu/contrib/maxim/acisbg/}; ``\mbox{acis[0-3]D2000-12-01bgstow\_ctiN0004.fits}''
				}.
		The data is processed as recommended by the corresponding CIAO threads.
		We fit its energy spectrum between $0.5$ and $10.0\,\mathrm{keV}$ with a spectral model consisting of a power law with   six Gaussians to account for the instrumental emission lines of Al~K$_\alpha$, Si~K$_\alpha$, Au~K$_{\alpha,\beta}$, Ni~K$_\alpha$ and Au~L$_\alpha$, \citep[e.g. Fig.~3, left panel]{ACISMemo162} and for an additional feature around $8.3\,\mathrm{keV}$. 
		\Mc{The stowed background spectrum and the best-fit model is shown in \Fref{fig:PB_spec}.
		The model describes the energy spectrum of the instrumental background  with the accuracy sufficient for this study, which main purpose is decomposition of the CXB spectrum below 2 keV into Galactic and extragalactic components and measurement of their fluxes. We therefore use this model to fit the spectrum of the unresolved CXB in XBOOTES in  \Sref{ss:Espec_Scr}. In the spectral fit to the XBOOTES data we fixed the slope of the power law of the instrumental background and the parameters of the two low energy lines as indicated in  \Tref{tab:PB_spec} at the best-fit values to the stowed background data. Other parameters were left free. Their best-fit values are listed in \Tref{tab:PB_spec}.}

				\begin{table}
					\caption{Best-fit values of the instrumental background model (pink curve in \Fref{fig:CXB_spec_full}) obtained in the spectral fit to the unresolved CXB in XBOOTES (\Sref{sec:Espec})}
					\label{tab:PB_spec}
					\begin{center}     
						\begin{tabular}{l l c}	
							\hline
							\hline
							Component-Name  & Parameter & Value  \\
							\hline
							powerlaw	& Photon Index	& 0.107 (fixed) \\
							1. Gaussian	& Center	& 1.490 keV \\
							(Al K$_\alpha$)	& Width		& 19.4 eV (fixed) \\
							2. Gaussian	& Center	& 1.678 keV (fixed) \\
							(Si K$_\alpha$)	& Width		& 0.1 eV (fixed) \\
							3. Gaussian	& Center	& 2.162 keV \\
							(Au K$_{\alpha,\beta}$)	& Width	& 52.0 eV \\
							4. Gaussian	& Center	& 7.473 keV \\
							(Ni K$_\alpha$)	& Width		& 11.8 eV \\
							5. Gaussian	& Center	& 9.700 keV \\
							(Au~L$_\alpha$)	& Width		& 35.0 eV \\
							6. Gaussian	& Center	& 8.294 keV \\
								& Width		& 185.1 eV \\
							\hline
						\end{tabular}
					\end{center}
					The width is the standard deviation of the Gaussian.
					The fixed values are from the fit of the \acisi{} stowed background spectrum (\Fref{fig:PB_spec}).
				\end{table}

				\begin{table}
					\caption{
						Surface brightness of the instrumental background of XBOOTES in the \eSoftB{}
						\Mc{determined from the spectral fit and using the background map}.
					}
					\label{tab:PB_BkgSB}
					\begin{center}     
						\begin{tabular}{l c}	
							\hline
							\hline
							Estimated based on 	&  Value \\
								&  [cts s$^{-1}$ deg$^{-2}$] \\
							\hline
							Spectral fit (\Sref{ss:Espec_Scr})	& $1.548\pm0.004$ \\
							Background maps (\Sref{ss:QuiBKG})	& $1.55\pm0.01$  \\
							\hline
						\end{tabular}
					\end{center}
				\end{table}
	
		One can see in \Fref{fig:CXB_spec_full} that the instrumental background model has higher continuum level than the CXB, which makes the estimate of the CXB flux sensitive to the slope of the power law component of the background model.
		Therefore it is reassuring that the surface brightness of the instrumental background obtained from the spectral fit of the XBOOTES data is consistent with the average surface brightness of the background maps $\mathbf{C}^\mathrm{BKG}$ \Mc{obtained \Ac{with the help of} the stowed data} in \Sref{ss:QuiBKG}, see \Tref{tab:PB_BkgSB}.
		The latter uses the method of \HickI{} to estimate the  background spectrum normalisation  from the \ePartB{} (\Eref{eq:BKG}),
		whereas for the former we  fit the data below $10.0\,\mathrm{keV}$.

	\section{PSF-smearing model} \label{app:PSFsmear}
		
			\begin{figure}
				\centering
				\resizebox{\hsize}{!}{\includegraphics{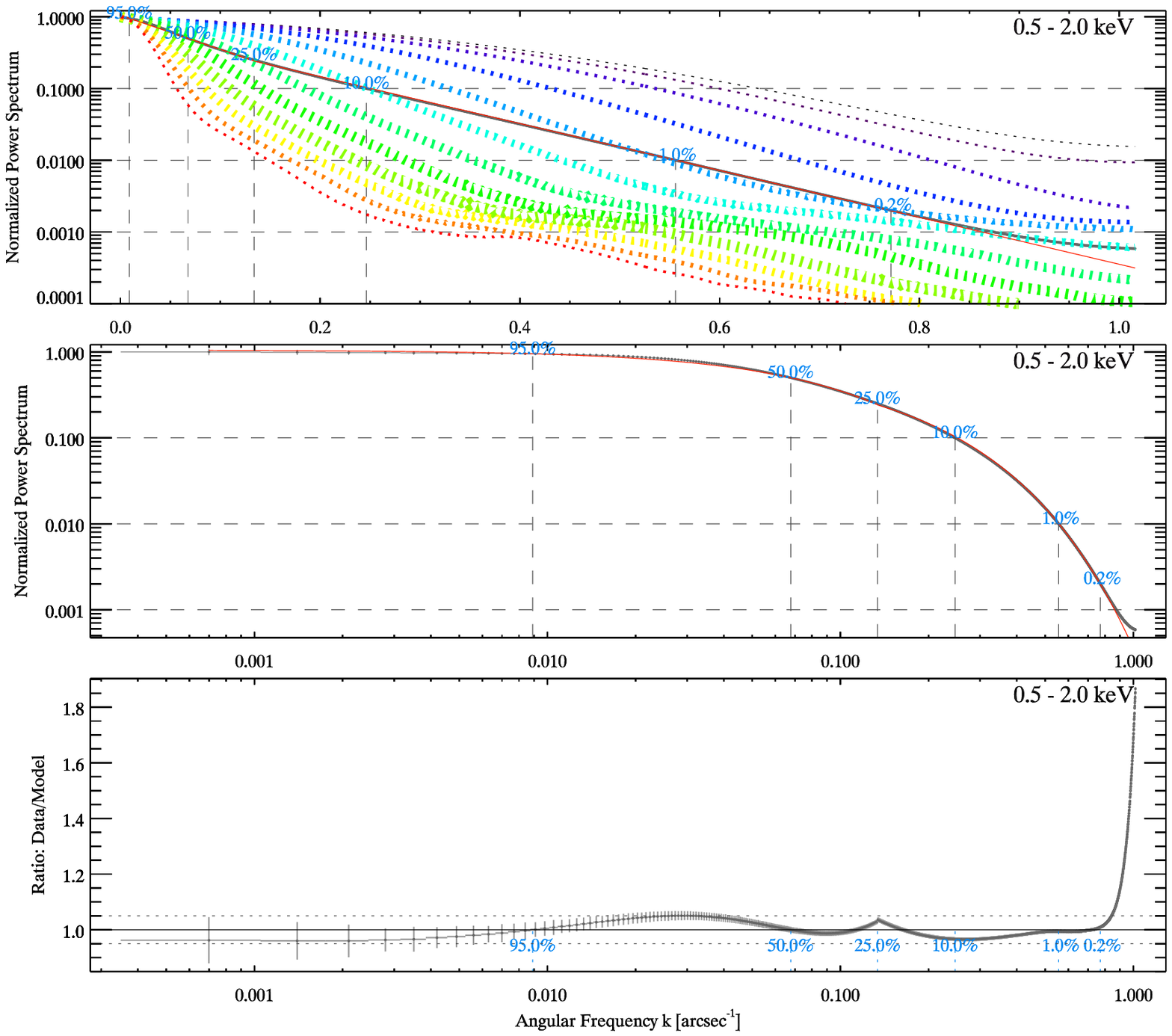}}
				\caption{\label{fig:PS_PSF_smearing}%
					\Mc{ 
					\emph{Top:} 
					The azimuthally averaged power spectra of Chandra  PSF  (dotted curves) for different offset angles from $0\arcmin$ (black) to $11\arcmin$ (red) with the step of $1\arcmin$ for the \eSoftB{}. 
					The thickness of the curves indicates the weight used in the  averaging over offset angles.
					\emph{Top and Middle:}
						Average \PoSp{} of the PSF of \chandra{} \acisi{} (gray crosses, average over all offset angles and CCDs) and the PSF-smearing model (red curve, \Eref{eq:PSF_model}).
					\emph{Bottom:}
						Ratio between the average PSF \PoSp{} and the PSF-smearing model.
					}
				}
			\end{figure}

			\begin{table}
				\caption{Parameters of the PSF-smearing model (\Eref{eq:PSF_model}) in the \eSoftB{}.}
				\label{tab:PSF_Model}
				\begin{center}     
					\begin{tabular}{l r c}	
						\hline
						\hline
						Symbol  & Value & Unit \\
						\hline
						$k_b$		&  0.134 & arcsec$^{-1}$ \\
						$\alpha_1$	& -4.740 & arcsec \\
						$\beta_1$	&  0.020 & - \\
						$\alpha_2$	& -3.270 & arcsec \\
						$\beta_2$	& -0.180 & - \\
						\hline
					\end{tabular}
				\end{center}
			\end{table}
	
		The PSF of \chandra{} will smooth out any fluctuation signal caused by source photons on the angular scales smaller than its size. 
		Note that  fluctuation caused by instrumental background  are not affected.
		This leads to a drop of the source \PoSp{} $P_{Scr}(k)$ amplitude at small scales (e.g.\  \Fref{fig:PS_eSoft}).
		Since the shape of the PSF is rather complex and changes with the offset angle ($\theta$) and azimuthal angle ($\phi$) from the focal point (see \Fref{fig:PSF_shape}), one can not derive a simple analytical expression to describe this effect.
		Therefore, we use an empirical approach and compute an average PSF \PoSp{} from the measured PSF \PoSpa{} of our PSF simulations in \Sref{ss:ResoScr} for the \eSoftB{}.
		Thereby, we first average over all CCDs and then compute a weighted average over all offset angles.
		For the weights we use the surface area of the annulus of each offset angle.
		We show the average PSF \PoSp{} in \Fref{fig:PS_PSF_smearing} (grey crosses) along with our simple empirical \emph{PSF-smearing model} (red curve),
		which is a broken exponential function in the base of ten:
			\begin{align} \label{eq:PSF_model}
				\log_{10}\left[P_\mathrm{PSF}(k)\right] =
					\begin{cases}
					\alpha_1 \, k + \beta_1 & \text{for $k \leq k_b$},\\
					\alpha_2 \, k + \beta_2 & \text{for $k \geq k_b$}.
					\end{cases}					
			\end{align}
		The parameters of the model are listed in \Tref{tab:PSF_Model}.
		Our model is able able to describe the PSF \PoSp{} up to a frequency of $k\approx 0.9\InvArcSec$ \Ac{($\gtrsim1.1\arcsec$, $\gtrsim2$ image pixels)} with an accuracy of $\approx5\%$, which is sufficient for the given S/N of our fluctuation measurements.
		We can see also in \Fref{fig:PS_Signal_eSoft} that our PSF-smearing model is able to describe well the measured \PoSp{} \Mc{ in the entire range of angular scales of interest.}
	
		\section{Photon shot noise} \label{app:Noise}
			When measuring the angular fluctuations of the unresolved CXB via Fourier analysis, we have to take into account that the source \PoSp{} $P_\mathrm{Scr}(k)$ is superimposed by the \emph{photon shot noise} $P_\mathrm{Phot.SN}$
			in our measured \PoSp{}:
				$\langle |\widehat{\delta F}(k)|^2 \rangle = P_\mathrm{Src}(k) + P_\mathrm{Phot.SN}$. 			
				In the ideal case the shot noise is flat and independent of the Fourier frequency. 
			Its mean amplitude is inversely  proportional to the total number of counts 
			and \Mc{it contributes to  the statistical noise in the  source \PoSp{}: $\sigma_{P_\mathrm{Scr}}(k) \propto P_\mathrm{Src}(k) + P_\mathrm{Phot.SN}$.
			}

			\subsection{Estimators} \label{app:Est}
			Below we discuss three different photon shot-noise estimators. 
						
				\subsubsection{Analytical estimate} \label{app:Anoise}
				Ignoring the effects of the mask (\Sref{app:ss:MCS}) and  vignetting, 
				one can use a simple analytical expression to estimate the shot-noise amplitude, which we derive here for completeness.
				For simplicity we use in the following a single index ($j$) for the summations over all pixels $N$ of a 2D map.
				Based on our definition in \Sref{ss:Def}, we can write the measured \PoSp{} as following:
					\begin{align}
						\left\langle|\widehat{\delta F}(\VEC{k})|^2\right\rangle
							= & \dfrac{1}{\Omega} \! \left\langle \!
							\left( \sum_j^{N}\delta F_j\,\mathrm{e}^{-2\pi\mathrm{i}\,\VEC{r}_j\VEC{k}} \right) \!\!
							\left( \sum_l^{N}\delta F_l\,\mathrm{e}^{+2\pi\mathrm{i}\,\VEC{r}_l\VEC{k}} \right) \!
							\right\rangle \notag \\
							= &  \dfrac{1}{\Omega} \!\left\langle  \sum_j^{N} \delta F_j^2 \right\rangle \notag \\ &
							  + \dfrac{1}{\Omega} \!\left\langle \sum_j^{N}  \sum_{l \neq j}^{N-1} \delta F_j \, \delta F_l \,\mathrm{e}^{-2\pi\mathrm{i}\,\VEC{k} ( \VEC{r}_j - \VEC{r}_l )} \right\rangle
								\notag
						\text{ .}	
					\end{align}
				The second term represent the actually source \PoSp{} $P_\mathrm{Scr}(k)$, whereas the first term is the shot noise.
				We can reduce the first term further to:
					\begin{align}
						\left\langle \sum_j^{N} \delta F_j^2 \right\rangle
							& = \sum_j^{N} \left\langle \delta F_j^2\right\rangle 
							= \sum_j^{N} \left(\left\langle F_j^2\right\rangle - \langle F_j\rangle^2 \right) \notag \\
							& = \sum_j^{N} \sigma^2(F_j) 
							= \sum_j^{N} \dfrac{\sigma^2(C_j)}{E_j^2} 
							\approx \sum_j^{N} \dfrac{C_j}{E_j^2} \notag
							\text{ .}
					\end{align}
				Here, we use the fact that $\delta F_j = F_j - \langle F_j\rangle$ and $F_j = C_j/E_j$
				and that the variance $\sigma^2(x)$ of a Poissonian-distributed quantity $x$ is equal to its mean.
				Since we do not know the actually mean of $C_j$ we approximate it with its own value.
				Based on this derivation, we define our shot noise estimate as following:
					\begin{align} \label{eq:AnaSNoise}
						P_\mathrm{Phot.SN} & = \dfrac{1}{\Omega} \sum_j^{N} \dfrac{C(\VEC{r}_j)}{E^2(\VEC{r}_j)}
						\text{ .}
					\end{align}
				When we are using the average exposure map (\Eref{eq:AveETM}) for computing the fluctuation map $\delta\mathbf{F}$, the definition changes to:
					\begin{align}
						P_\mathrm{Phot.SN} & = \dfrac{1}{\Omega} \dfrac{\sum_j^{N} C(\VEC{r}_j)}{\langle E\rangle ^2}
						\text{ .} \notag
					\end{align}
				We refer to both definitions as the \emph{analytical shot-noise estimate}, which is our default shot-noise estimate.
				
				Note that this definition is only valid for the total-count map $\mathbf{C}_X^\mathrm{Total}$ (\Sref{ss:QuiBKG}).
				For the background-subtracted map $\mathbf{C}_X^\mathrm{CXB}$ (\Eref{eq:NetCtsMap}) one  has to \Mc{explicitly add the  shot noise  due to the instrumental background events, as the flux of the the latter was subtracted from $C(\VEC{r}_j)$ in \Eref{eq:AnaSNoise}:}
					\begin{align}
						P_\mathrm{Phot.SN}^\mathrm{(CXB)} & = P_\mathrm{Phot.SN} + P_\mathrm{Phot.SN}^\mathrm{(BKG)} \text{ ,} \notag
					\end{align}
				with (using \Eref{eq:BKG})
					\begin{align}  
						P_\mathrm{Phot.SN}^\mathrm{(BKG)} & = \dfrac{ \sum_j^N C_j^\mathrm{Stow} }{ \Omega \, \langle E^\mathrm{Stow}\rangle^2 }
							\cdot \left(\dfrac{ \sum_j^N \, C_{\ePart}^\mathrm{Total} }{\sum_j^{N} \, C_{\ePart}^\mathrm{Stow}}\right)^2
						\text{ .} \notag
					\end{align}

				\subsubsection{High-frequency based estimate} \label{app:HFnoise}
				The \emph{high-frequency based shot-noise estimate} uses the fact that at very high frequencies the measured \PoSp{} converges eventually to the shot-noise amplitude (\Sref{app:PSFsmear}).
				Hence, we can estimate the shot noise by taking the average ($P^\mathrm{(HF)}_\mathrm{Phot.SN}$) of the \PoSp{} for some frequency interval $[k^\mathrm{(HF)}_\mathrm{min},k^\mathrm{(HF)}_\mathrm{max}]$. \Mc{It is natural to choose   the upper limit at the Nyquist-Frequency: $k^\mathrm{(HF)}_\mathrm{max}=k_\mathrm{Ny}$.
				The choice of the lower limit is somewhat arbitrary and depends on the amplitude and the slope of the power spectrum of the signal.}
				We use $k^\mathrm{(HF)}_\mathrm{min}=k_\mathrm{Ny} \times 0.80 \approx 0.81\InvArcSec$.
				With this choice, the interval encapsulates angular scales of the size $\approx2.5$ \acisi{} CCD pixels or smaller ($\leq1.2\arcsec$).
				At these scales the amplitude of the source \PoSp{} is suppressed by more than $500$ times due to PSF-smearing (\Fref{fig:PS_PSF_smearing}) and 
				the shot-noise-subtracted \PoSp{} in the \eSoftB{} is more than $\sim2000$ times smaller than the shot-noise amplitude itself (computed using the analytical shot-noise estimate).
				\Mc{Therefore, within this frequency interval  the source \PoSp{} can be neglected (cf. \Fref{fig:PS_Signal_eSoft}).}
				We also tested larger and smaller values for $k^\mathrm{(HF)}_\mathrm{min}$ and find that $k^\mathrm{(HF)}_\mathrm{min}=k_\mathrm{Ny} \times 0.80$ is a good compromise \Mc{. Furthermore,   using  $k^\mathrm{(HF)}_\mathrm{min}=k_\mathrm{Ny} \times 0.40 \approx 0.41\InvArcSec$ we obtain the the shot-noise-subtracted  power spectrum which is nearly identical  to the default one in the  frequency range of interest, at frequencies $\lesssim0.3\InvArcSec$.
				It should be noted that this \Ac{shot-noise estimator} requires the maximal angular resolution of Chandra, i.e. needs the  image pixel binning of one.}
								
				As we show in the \Sref{app:ss:test}, the high-frequency based estimate works as well as the analytical estimate and \Mc{produces results compatible with the latter.}
				\Mc{The caveat is that it may be subject to contamination by the signal, i.e. prone to overestimating the shot-noise level  if there is this still a significant source signal  at frequencies  $\ga k^\mathrm{(HF)}_\mathrm{min}$.
				Nevertheless, this estimate serves as a useful  independent test for the analytical estimate}.

				\subsubsection{Observation-splitting based estimate} \label{app:EOnoise}
				The \emph{observation-splitting shot-noise based estimate}, also called the $A-B$ technique, was introduce by \citet{Kashlinsky2005} for infrared data and was also used for X-ray data \citep[e.g.][]{Cappelluti2013,Helgason2014}.
				Hereby, one splits the observation into even and odd time-frames (or events) and creates a fluctuation map for each subset ($\delta\mathbf{F}_A$ for even frames and $\delta\mathbf{F}_B$ for odd frames).
				The difference between these two maps \Mc{should be free of  any source or instrumental  signal which is steady in time,} because both subsets were observed nearly simultaneously, and  only contain the random noise of the observation.
				Hence, the measured \PoSp{} $P^\mathrm{(OS)}_\mathrm{Phot.SN}$ of the difference $\delta\mathbf{F}_D = (\delta\mathbf{F}_A -\delta\mathbf{F}_B)/2$ should represent the shot noise for this observation.
				For the shot-noise-subtracted \PoSp{}, we take the statistical uncertainties of the measured \PoSp{} and of $P^\mathrm{(OS)}_\mathrm{Phot.SN}$ into account and use the normal error propagation.
				
				As we show in the \Sref{app:ss:test}, observation-splitting based estimate gives in average consistent results in comparison to the analytical estimate.
				However, a major concern with this estimate is that this estimate is itself subject to noise.
				Combined with correlations between different Fourier modes caused by the mask effect this leads to appearances of irregularities in the resulting shot-noise-subtracted \PoSp{}.
				This can been seen for instance in \Fref{fig:PS_Signal_eSoft} (lower panels, green crosses) for in the interval of $0.5 \lesssim k[\mathrm{arcsec^{-1}}] \lesssim 0.7$,
				where the observation-splitting based estimate leads to a significant underestimation of the \PoSp{}.
				This also makes the observation-splitting based estimate somewhat less reliable in comparison to the analytical or high-frequency based estimate.
				However, it serves as \Mc{another  independent probe of the photon shot-noise level.}

			\subsection{Comparison of  different estimators} \label{app:ss:test}
				In the following, we compare  different shot-noise estimates.
				Thereby, we focus mostly on the high frequency part of the \PoSp{}, where the source \PoSp{} is of the order of the shot-noise amplitude or smaller. 
				For a better visualization, we show the \PoSp{} in linear scale and use a linear binning, although all calculations were performed with the unbinned \PoSp{}.

			\subsubsection{The \protect\ePart{} band}  \label{app:sss:ePart}
					\begin{figure}
						\begin{center}
							\resizebox{\hsize}{!}{\includegraphics{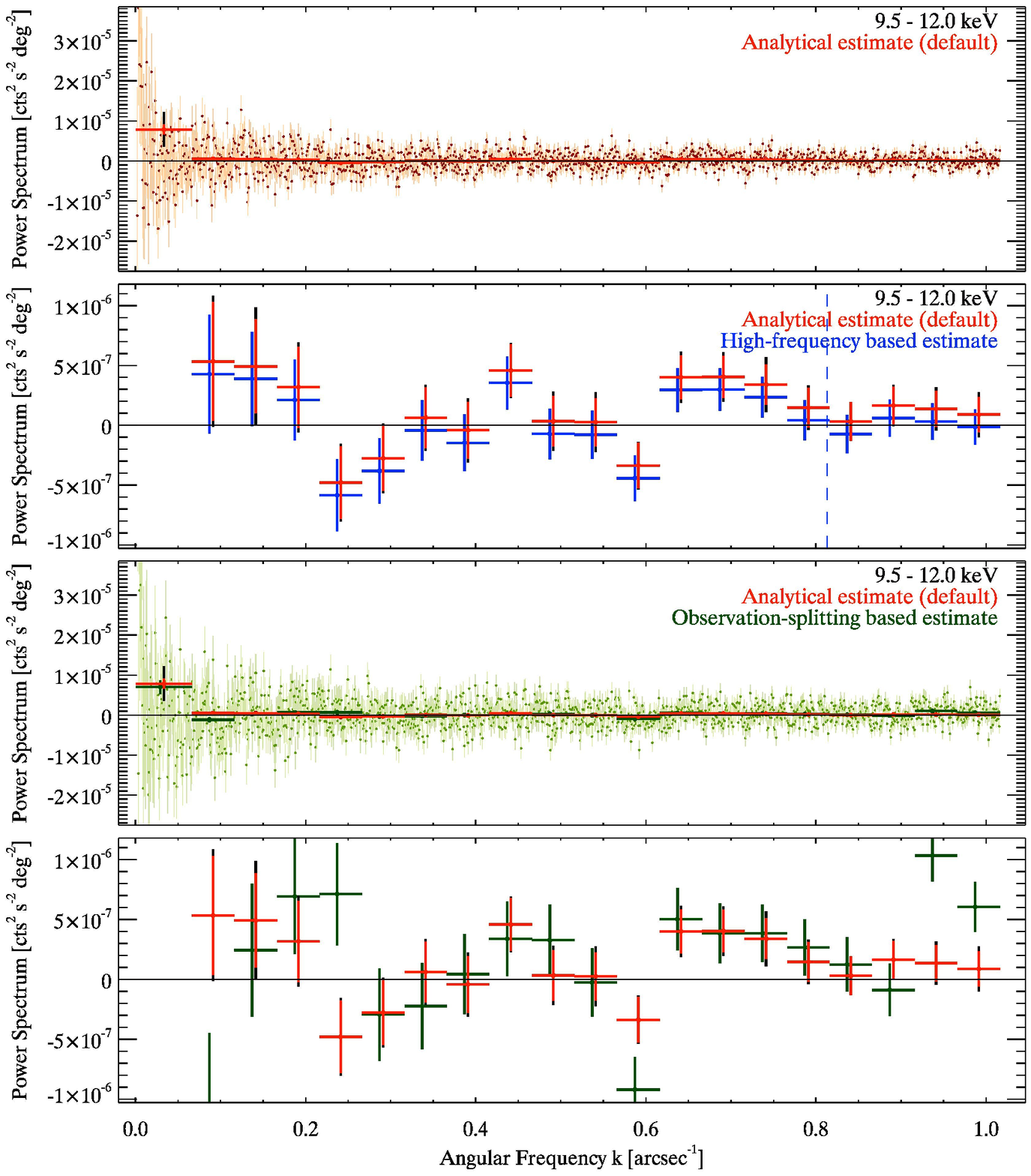}}
							\caption{\label{fig:PS_Signal_ePart}%
								Photon-shot-noise-subtracted \PoSp{} $P(k)$ (\Eref{eq:DefPS}) in the \ePartB{}.
								\Mc{Red crosses in all panels were obtained using our default method of analytical shot-noise estimate  (\Eref{eq:AnaSNoise}).
								Blue crosses in the two upper panels were computed using the high-frequency based  estimate  (\Sref{app:HFnoise}).
								Green crosses in the two lower panels were obtained using the observation-splitting based estimate  (\Sref{app:EOnoise}). 
								The colored error-bars show the standard deviation due to error propagation of all binned frequencies.
								The black error-bars show  the standard deviation of the sample mean of all binned frequencies.}
								Vertical blue dashed line shows the  lower limit $k^\mathrm{(HF)}_\mathrm{min}$ used to compute the high-frequency based  estimate.
							}	
						\end{center}
					\end{figure}

			The \PoSp{} in the \ePartB{} is ideal for evaluating the photon shot-noise estimators. 
			\Mc{Since the effective area of \chandra{} mirrors is virtually zero at these energies}%
				\footnote{\url{http://cxc.harvard.edu/proposer/POG/html/chap4.html#tth_sEc4.2.2}},
			\Mc{all  registered events are due to the instrumental background.
			At small angular scales the power spectrum of the instrumental background is flat and consists of the photon shot-noise only. 
			This fact can be used to evaluate and compare our shot-noise estimators.}

			\Mc{The photon shot-noise subtracted power spectra are plotted in \Fref{fig:PS_Signal_ePart}. As one can see, there are quite significant fluctuations in the binned spectra. They are caused by the correlation between nearby Fourier modes caused by the mask effect. This correlations is not accounted for in computing the errors and leads to seemingly statistically significant deviations from the zero in the binned power spectrum.  In \Fref{fig:PS_Signal_ePart} we also show in black colour  the RMS of power in individual Fourier modes  computed within the broad frequency bins.  They somewhat higher than the theoretical error bars computed as described above (\Eref{eq:PSErr}).}
									
			\Mc{The lowest  frequency bin  contains  a real signal due to the slightly  non-uniformity of the instrumental background  at large scales, which also can be seen in the power spectrum of the  stowed background map (\Sref{app:ss:Bkg}, \Fref{fig:PS_eSoft_Extra03}).
			Therefore, we exclude the frequencies $<0.02\InvArcSec$ from the following evaluation.}

			To compare different estimates we compute the sample mean and RMS deviation for the quantity $Z(k) = P(k)/\sigma_{P(k)}$.  Since the unbinned \PoSp{} has more than $1000$ Fourier frequencies, the average of $Z(k)$ follows Gaussian distribution and, in the ideal case, has the mean zero with a RMS deviation of one.
			We measure a RMS deviation of $\approx1.14$ for all shot-noise estimates and the  sample mean of $0.05\pm0.03$, $0.00\pm0.03$ and $0.08\pm0.03$ for the analytical, high-frequency based, and observation-splitting based estimates, respectively.
			\Mc{This calculation shows somewhat inferior quality of the  observation-splitting based estimate which is also confirmed by the visual inspection of Figs. \ref{fig:PS_Signal_ePart} and \ref{fig:PS_Signal_eSoft}. We choose the analytical estimator as our default one for the analysis in this paper.}
			
			\subsubsection{The \protect\eSoftB{}} \label{app:sss:eSoft}
					\begin{figure}
						\begin{center}
							\resizebox{\hsize}{!}{\includegraphics{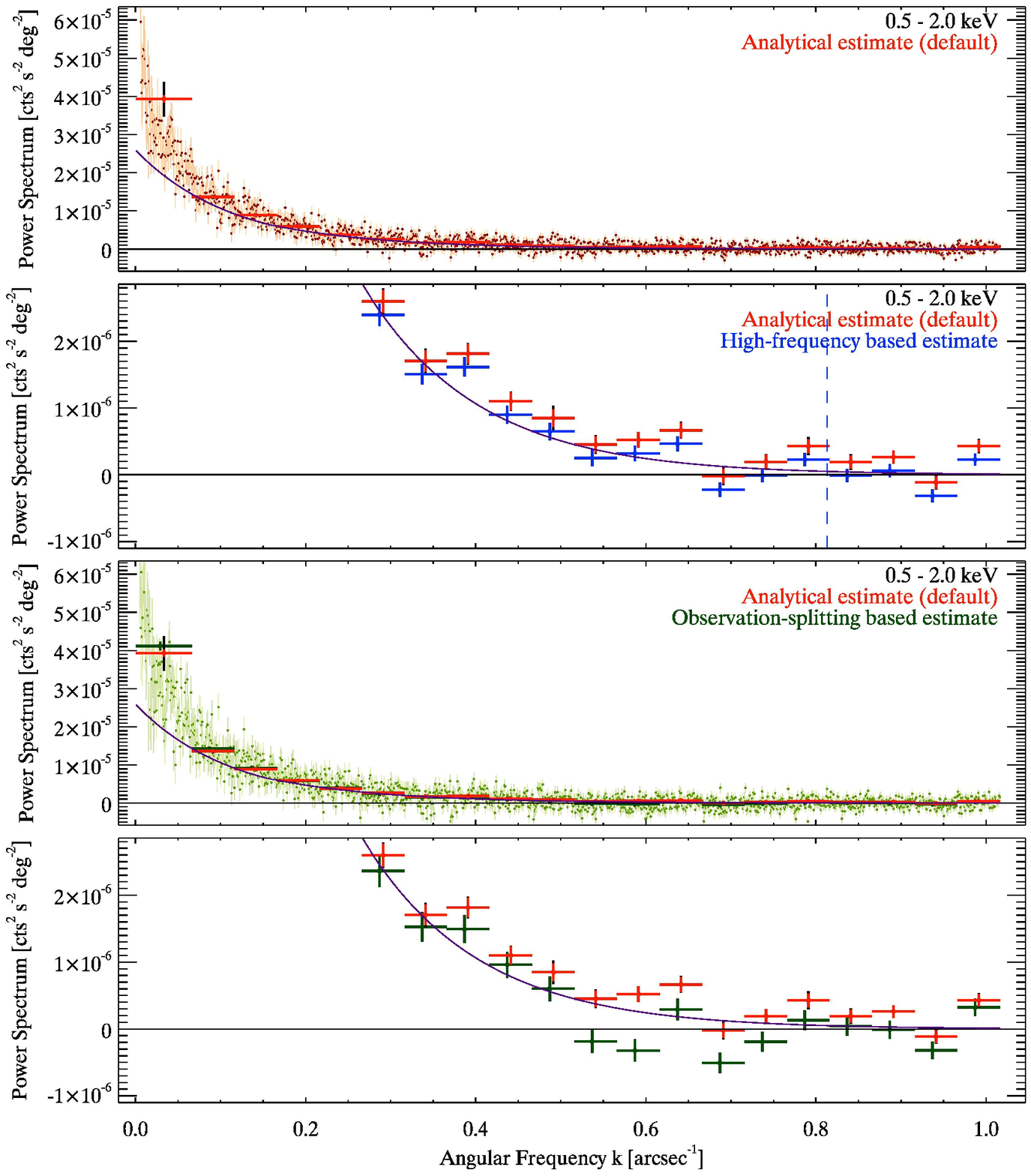}}
							\caption{\label{fig:PS_Signal_eSoft}%
								Same as \Fref{fig:PS_Signal_ePart} but for the \eSoftB{}.
								Purple solid curve shows  AGN shot-noise model (\Sref{ss:shot_noise}) multiplied by the PSF-smearing model (\Sref{app:PSFsmear}).
							}	
						\end{center}
					\end{figure}
				
			\Mc{In \Fref{fig:PS_Signal_eSoft} we show the photon shot-noise subtracted \PoSp{} $P(k)$ for the \eSoftB{} for our three shot-noise estimators.
			We also plot the AGN shot-noise model (\Sref{ss:shot_noise}), multiplied with the PSF-smearing model (\Sref{app:PSFsmear}). One can see good agreement of the data with the model in the entire frequency range. 
			This plot justifies  neglecting the contribution of the signal for the frequencies of $> k_\mathrm{Ny} \times 0.80 \approx 0.81\InvArcSec$, which were used  to compute the high-frequency based estimate. It also demonstrates that our AGN shot-noise model combined with the PSF-smearing model is able to describe the measured \PoSp{} up to the highest frequencies.}
			
			\Mc{Finally, we note that the differences between different photon shot-noise estimators  are significant only at highest frequencies. For the analysis presented in this paper, which is focused at frequencies $\lesssim 0.3\InvArcSec$ these differences are unimportant.}
		
	\section{Other systematic effects} \label{app:Sys}
		\Mc{Below we discuss several less significant systematic effects having impact on the resulting power spectrum.}
		
		\subsection{Mask effect} \label{app:ss:MCS}
				\begin{figure}
					\resizebox{\hsize}{!}{\includegraphics{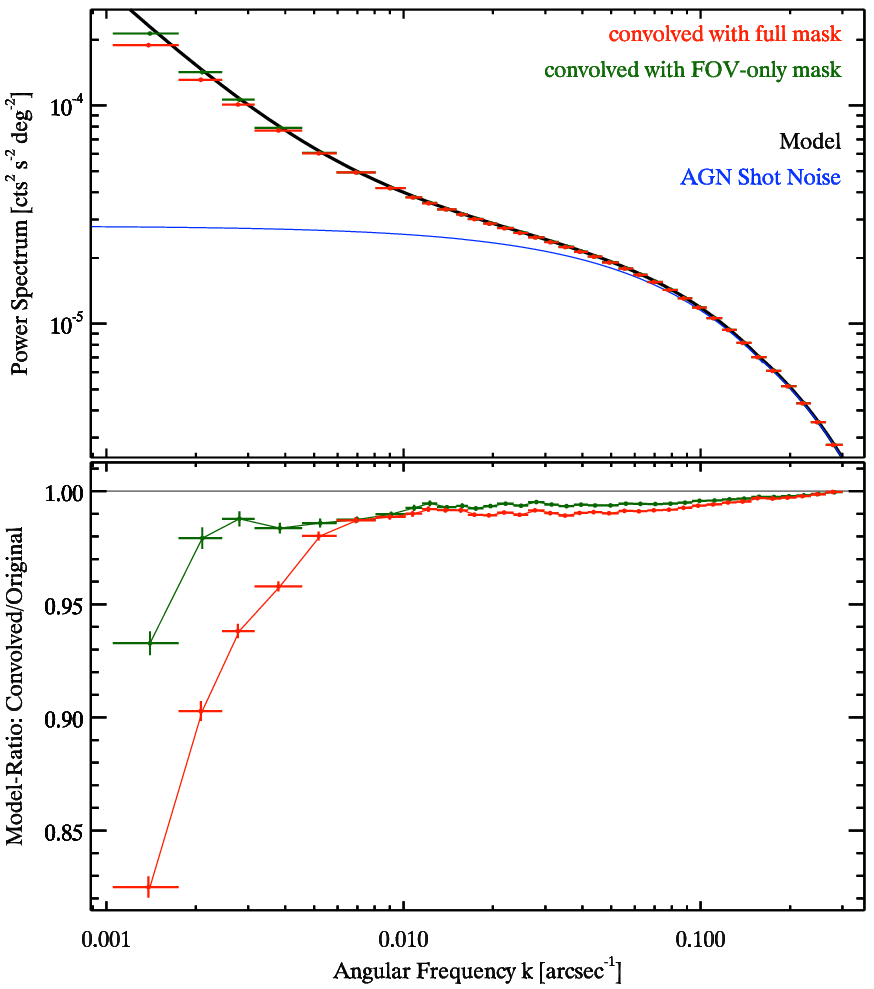}}
					\caption{\label{fig:MaskEff}%
						\Mc{The impact of the mask effect on the power spectrum. 
						\emph{Top:}
							\Ac{The simulated power spectra after two different masks were applied to the simulated images, which are based on our input model (black curve).}
							Also shown is the \Ac{AGN shot noise} model (blue curve). 
						\emph{Bottom:} 
							The ratio of the simulated power spectra to \Ac{our input} model.
							The data points of different color show the effect of two different masks:
							the FOV-only mask (green) and the full mask taking into account the resolved sources and CCD gaps (red).} 
					}	
				\end{figure}
		
		\Mc{The measured \PoSp{} is  a convolution of the true \PoSp{} with a that  of the window function, because the analysed image has a finite size and is modified by the mask excluding the regions corresponding  to resolved sources, CCD gaps and bad pixels.}
		This  alters the \PoSp{} and leads to correlation of adjacent Fourier modes. We refer to  these modifications with the term \emph{mask effect}. Here, we estimate how this  effect distorts the source \PoSp{} in the considered frequency range of $0.001 - 0.300\,\InvArcSec$
		($\sim3\arcsec - 17\arcmin$). 
		
		\Mc{To compute the mask effect we perform simulations as follows. We construct a model power spectrum consisting of the shot-noise of unresolved AGN and a power law component $P\propto k^{-1.3}$ 
		corresponding to the LSS signal, both multiplied by the PSF-smearing function. From this power spectrum we create a two-dimensional Fourier-image assuming random and uniformly distributed phases and perform an inverse Fourier transform to compute the corresponding image. The resulting  image is multiplied by the mask. This is repeated for each Chandra observation, each time using the real mask for that observation. From the produced set of images we compute the average power spectrum following same procedures as were applied to the real data.
		This procedure was repeated $50$ times and the result was averaged to produce the final power spectrum which was then compared with the input model. }
		
		\Mc{The result of these simulations  is shown  in \Fref{fig:MaskEff}.
		One can  see that due the mask effect the \PoSp{} is suppressed by less than $\sim20\,\%$ at the lowest considered frequency bin and at the  frequencies of $\gtrsim4\times10^{-3}\,\InvArcSec$ it is  suppressed by less than $\sim2\,\%$.
		This means that the measured power spectrum is a reasonably good representation of the true power spectrum of the signal. Due to the complexity of the inverse problem, we  chose not to correct the power spectrum for the mask effect. Instead, these effects can be straightforwardly  taken into account when fitting the measured spectrum with theoretical models. }

		\subsection{Instrumental background} \label{app:ss:Bkg}		
\begin{figure}
				\begin{center}
					\resizebox{\hsize}{!}{\includegraphics{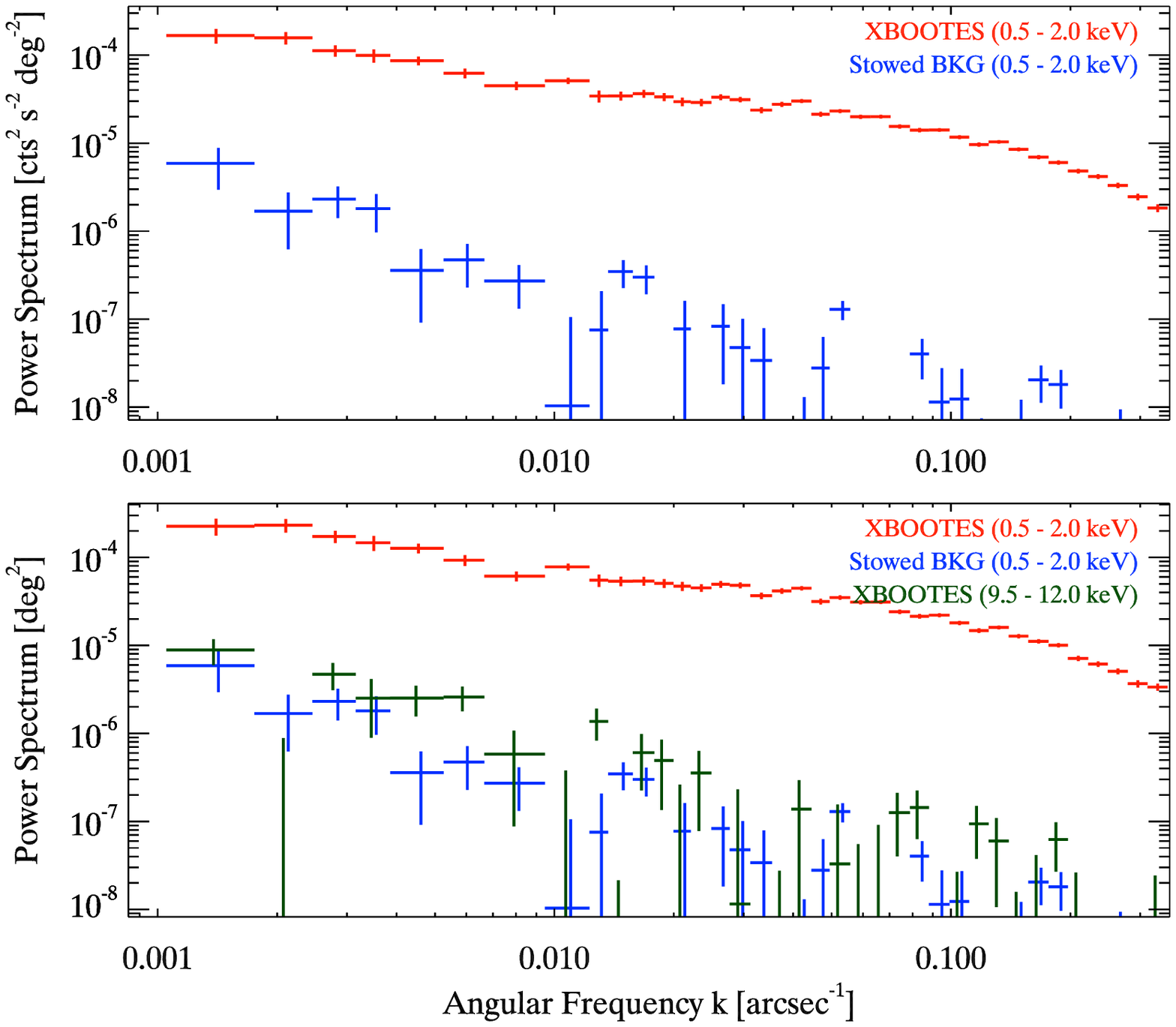}}
					\caption{\label{fig:PS_eSoft_Extra03}%
					\Mc{
					\emph{Top:}
						Photon-shot-noise-subtracted \PoSpa{} of the total-count map,  $\mathbf{C}^\mathrm{Total}$ in the terminology of \Sref{ss:QuiBKG}, (red crosses)
						and of the \acisi{} stowed-background map $\mathbf{C}^\mathrm{Stow}$ (blue crosses) in the \eSoftB.
					\emph{Bottom:} 
						Same as top panel but in the units of squared fractional RMS \PoSp{} $Q(k)$ (\Eref{eq:PS_FluxNorm}).
						Also shown in the bottom panel is the \PoSp{} for the total-count map $\mathbf{C}^\mathrm{Total}$ in the \ePartB{} (green crosses).
					}
					}
				\end{center}
			\end{figure}

		\Mc{At the flux limit of XBOOTES survey, the instrumental background accounts for $\sim 2/3$ of  the unresolved emission, therefore its non-uniformity can significantly contaminate the power spectrum.
		In addition to the non-uniform distribution of the instrumental background over the detector, its surface brightness distribution may be  affected by the large scale non-uniformity of the detector efficiency (if any).
		Thus, investigation of the stowed background data can be also used to constrain the amplitude of the latter.
		The power spectrum of the stowed background ($\mathbf{C}^\mathrm{Stow}$  in the terminology of  \Sref{ss:QuiBKG})  is shown in \Fref{fig:PS_eSoft_Extra03} (blue crosses),
		along with the power spectrum of the  unresolved background in XBOOTES data ($\mathbf{C}_X^\mathrm{Total}$, red crosses)).
		The upper panel shows the power spectra in flux units, to characterize the absolute contamination of the measured power spectrum by the instrumental background.
		In the lower panel, we show the power spectra in the units of squared fractional variations (i.e. divided by the square of the mean flux).
		The latter characterizes the combined effect of the instrumental background variations and the spatial on-uniformity of the detector efficiency, thus placing an upper limit on the latter.
		As we can see from  \Fref{fig:PS_eSoft_Extra03}, in both case the power spectrum of the XBOOTES data exceeds the contamination by more than an order of magnitude at all frequencies.
		In the lower panel we also show the flux normalized power spectrum of the XBOOTES data in the \ePartB{} (green crosses).
		It is similar to the power spectrum of the  stowed background data, as it should be expected as both are due to the instrumental background.
		They are not identical though, as they are computed for different energy bands.}
					
\begin{figure}
				\begin{center}
					\resizebox{\hsize}{!}{\includegraphics{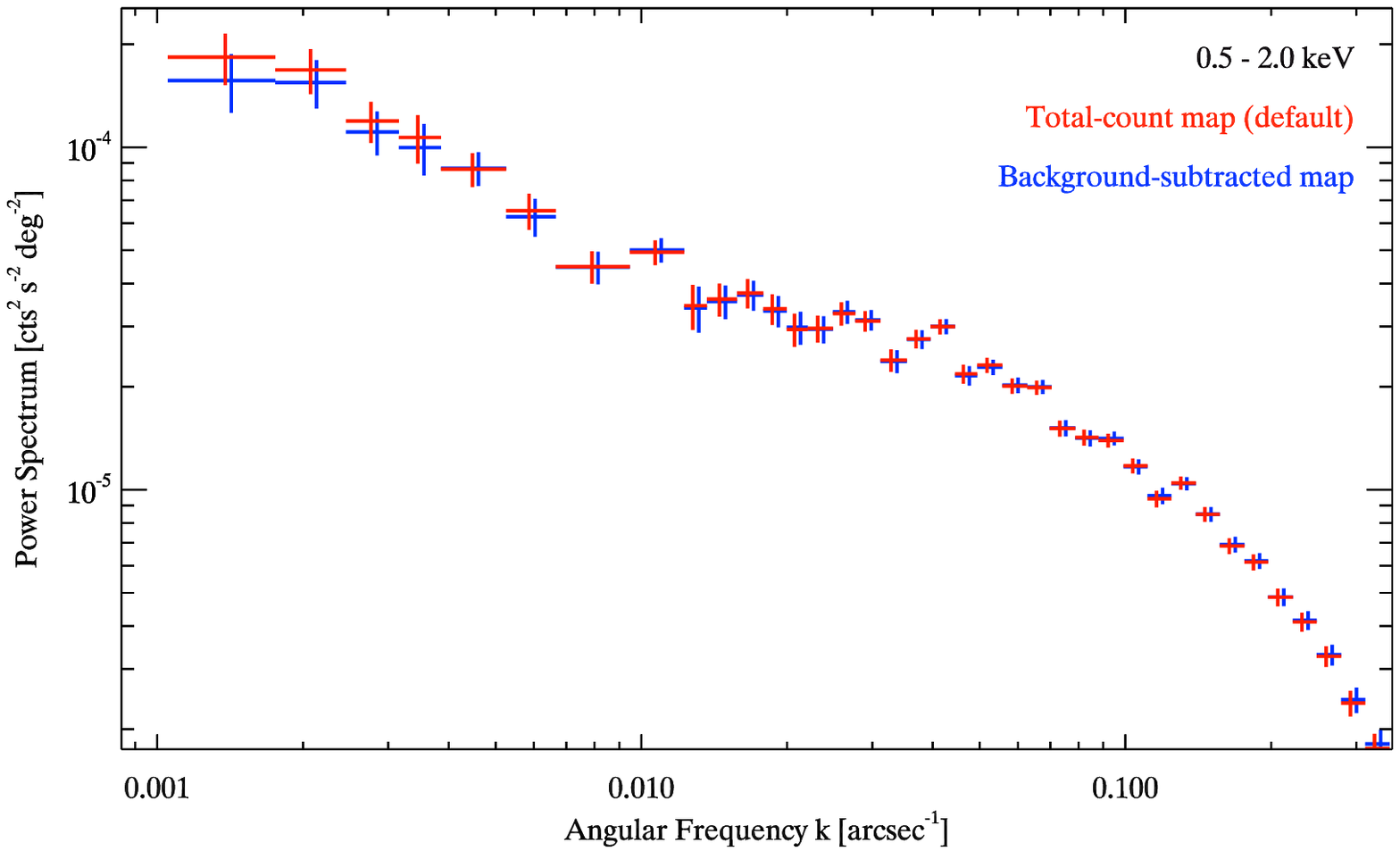}}
					\caption{\label{fig:PS_eSoft_Extra02}%
						Photon-shot-noise-subtracted \PoSp{} in the \eSoftB{}
						computed from the total-count map $\mathbf{C}_X^\mathrm{Total}$ (red)
						and the background-subtracted map $\mathbf{C}_X^\mathrm{CXB}$ (blue, \Eref{eq:NetCtsMap}, \Sref{ss:QuiBKG}).						}	
				\end{center}
			\end{figure}

		\Mc{Given the rather good uniformity of the  instrumental background (as compared to the unresolved CXB),  subtraction of the actual stowed background map from the data (\Sref{ss:QuiBKG}) is unnecessary.
		Instead, from each XBOOTES image we subtract a constant corresponding to the mean unresolved  flux in the given observation (\Eref{eq:FlucMap}).
		This simplifies the error propagation, removing correlations between fluctuation maps of different observations.
		In  \Fref{fig:PS_eSoft_Extra02} we compare our default power spectrum produced from total-count map $\mathbf{C}_X^\mathrm{Total}$ (\Sref{ss:QuiBKG}) with that of the background-subtracted map $\mathbf{C}_X^\mathrm{CXB}$ (\Eref{eq:NetCtsMap}).
		The difference between the two power spectra is insignificant for the purpose of the analysis  presented here.}
	
\newcommand*\aap{A\&A}
\let\astap=\aap
\newcommand*\aapr{A\&A~Rev.}
\newcommand*\aaps{A\&AS}
\newcommand*\actaa{Acta Astron.}
\newcommand*\aj{AJ}
\newcommand*\ao{Appl.~Opt.}
\let\applopt\ao
\newcommand*\apj{ApJ}
\newcommand*\apjl{ApJ}
\let\apjlett\apjl
\newcommand*\apjs{ApJS}
\let\apjsupp\apjs
\newcommand*\aplett{Astrophys.~Lett.}
\newcommand*\apspr{Astrophys.~Space~Phys.~Res.}
\newcommand*\apss{Ap\&SS}
\newcommand*\araa{ARA\&A}
\newcommand*\azh{AZh}
\newcommand*\baas{BAAS}
\newcommand*\bac{Bull. astr. Inst. Czechosl.}
\newcommand*\bain{Bull.~Astron.~Inst.~Netherlands}
\newcommand*\caa{Chinese Astron. Astrophys.}
\newcommand*\cjaa{Chinese J. Astron. Astrophys.}
\newcommand*\fcp{Fund.~Cosmic~Phys.}
\newcommand*\gca{Geochim.~Cosmochim.~Acta}
\newcommand*\grl{Geophys.~Res.~Lett.}
\newcommand*\iaucirc{IAU~Circ.}
\newcommand*\icarus{Icarus}
\newcommand*\jcap{J. Cosmology Astropart. Phys.}
\newcommand*\jcp{J.~Chem.~Phys.}
\newcommand*\jgr{J.~Geophys.~Res.}
\newcommand*\jqsrt{J.~Quant.~Spec.~Radiat.~Transf.}
\newcommand*\jrasc{JRASC}
\newcommand*\memras{MmRAS}
\newcommand*\memsai{Mem.~Soc.~Astron.~Italiana}
\newcommand*\mnras{MNRAS}
\newcommand*\na{New A}
\newcommand*\nar{New A Rev.}
\newcommand*\nat{Nature}
\newcommand*\nphysa{Nucl.~Phys.~A}
\newcommand*\pasa{PASA}
\newcommand*\pasj{PASJ}
\newcommand*\pasp{PASP}
\newcommand*\physrep{Phys.~Rep.}
\newcommand*\physscr{Phys.~Scr}
\newcommand*\planss{Planet.~Space~Sci.}
\newcommand*\pra{Phys.~Rev.~A}
\newcommand*\prb{Phys.~Rev.~B}
\newcommand*\prc{Phys.~Rev.~C}
\newcommand*\prd{Phys.~Rev.~D}
\newcommand*\pre{Phys.~Rev.~E}
\newcommand*\prl{Phys.~Rev.~Lett.}
\newcommand*\procspie{Proc.~SPIE}
\newcommand*\qjras{QJRAS}
\newcommand*\rmxaa{Rev. Mexicana Astron. Astrofis.}
\newcommand*\skytel{S\&T}
\newcommand*\solphys{Sol.~Phys.}
\newcommand*\sovast{Soviet~Ast.}
\newcommand*\ssr{Space~Sci.~Rev.}
\newcommand*\zap{ZAp}			
			
\bibliographystyle{mn2e} 
	

\label{lastpage}
\end{document}